\documentclass[%
 reprint,
superscriptaddress,
longbibliography,
 amsmath,amssymb,
prl,
]{revtex4-1}

\usepackage{lipsum}
\usepackage{graphicx}
\usepackage{dcolumn}
\usepackage{bm}
\usepackage{hyperref}
\usepackage[mathlines]{lineno}
\usepackage{url}

\newcommand{\expct}[1]{\langle #1 \rangle}
\newcommand{\figref}[1]{Fig.~\ref{#1}}
\renewcommand{\eqref}[1]{Eq.~(\ref{#1})}
\newcommand{\tbref}[1]{Table~\ref{#1}}
\newcommand{\pref}[1]{(\ref{#1})}
\renewcommand{\(}{\left(}
\renewcommand{\)}{\right)}
\newcommand{\diff}[2]{\frac{\mathrm{d} #1}{\mathrm{d} #2}}

\newcommand{\unit}[1]{~\mathrm{#1}}
\newcommand{\degc}{\unit{{}^\circ{}C}}

\newcommand{\XCD}{X^\mathrm{CD}}
\newcommand{\XCDth}{X^\mathrm{CD,th}}
\newcommand{\vth}{v^\mathrm{th}}
\newcommand{\mean}{\mathrm{mean}}
\newcommand{\std}{\mathrm{std}}

\newcommand{\numori}{\#\mathrm{ori}}

\DeclareMathOperator{\Std}{Std}

\begin{document}

\title{Scale invariance of cell size fluctuations in starving bacteria}

\author{Takuro Shimaya}
\email{t.shimaya@noneq.phys.s.u-tokyo.ac.jp}
\affiliation{Department of Physics, Graduate School of Science, University of Tokyo, Tokyo 113-0033, Japan}
\author{Reiko Okura} 
\affiliation{Department of Basic Science, Graduate School of Arts and Sciences, University of Tokyo, Tokyo 153-8902, Japan}
\author{Yuichi Wakamoto}
\affiliation{Department of Basic Science, Graduate School of Arts and Sciences, University of Tokyo, Tokyo 153-8902, Japan}
\author{Kazumasa A. Takeuchi}
\email{kat@kaztake.org}
\affiliation{Department of Physics, Graduate School of Science, University of Tokyo, Tokyo 113-0033, Japan}
\affiliation{Department of Physics, School of Science, Tokyo Institute of Technology, Tokyo 152-8551, Japan}

\date{\today}
\maketitle

\section{Abstract}
In stable environments, cell size fluctuations are thought to be governed by simple physical principles, as suggested by recent findings of scaling properties.
Here, by developing a microfluidic device and using {\sl E.~coli}, we investigate the response of cell size fluctuations against starvation.
By abruptly switching to non-nutritious medium, we find that the cell size distribution changes but satisfies scale invariance: the rescaled distribution is kept unchanged and determined by the growth condition before starvation.
These findings are underpinned by a model based on cell growth and cell cycle.
Further, we numerically determine the range of validity of the scale invariance over various characteristic times of the starvation process, and find the violation of the scale invariance for slow starvation.
Our results, combined with theoretical arguments, suggest the relevance of the multifork replication, which helps retaining information of cell cycle states and may thus result in the scale invariance.

\section{Introduction}
Recent studies on microbes in the steady growth phase
 suggested that the cellular body size fluctuations may be governed by simple physical principles.
For instance, Giometto \textit{et al.} \cite{Giometto2013} proposed that size fluctuations of various eukaryotic cells are governed by a common distribution function, if the cell sizes of a given species are normalized by their mean value (see also \cite{Zaoli2019}).
In other words, the distribution of cell volumes $v$, $p(v)$, can be described
 as follows:
\begin{equation}
 p(v) = v^{-1}F(v/V),  \label{eq:1}
\end{equation}
 with a function $F(\cdot)$ and
 $V = \expct{v}$ being the mean cell volume.
This property of distribution is often called scale invariance.
Interestingly, this finding can account for power laws of community size distributions, i.e.,
 the size distribution of all individuals regardless of species,
 which were observed in various natural ecosystems \cite{Camacho2001, Marquet2005}.
Scale invariance akin to \eqref{eq:1} was also found for bacteria
 \cite{Iyer-Biswas2014pnas, Kennard2016} for each cell age, 
 and the function $F(\cdot)$ was shown to be robust against changes
 in growth conditions such as the temperature.

Those results, as well as theoretical models proposed in this context
 \cite{Giometto2013, Iyer-Biswas2014prl, Amir2014},
 have been obtained under steady environments,
 for which our understanding of single-cell growth statistics
 has also been significantly deepened recently \cite{Ho2018, Jun2018, Cadart2019}.
By contrast, it is unclear whether such a simple concept
 as scale invariance is valid under time-dependent conditions,
 where different regulations of cell cycle kinetics may come into play in response to environmental variations.
In particular, when bacterial cells enter the stationary phase from the exponential growth phase,
 they undergo reductive division,
 during which both the typical cell size and the amount of DNA per cell
 decrease \cite{Nystrom2004, Kaprelyants1993, Arias2012, Gray2019}.
Although this behavior itself is commonly observed in batch cultivation,
 little is known about single-cell statistical properties during the transient.
The bacterial reductive division is therefore an important stepping stone
 for studying cell size statistics under time-dependent environments
 and understanding the robustness of the scale invariance against environmental changes.

To investigate size distributions of large bacterial populations under time-dependent growth conditions,
 we should care about experimental methods.
A microfluidic device called the mother machine \cite{Wang2010} consists of many small separate chambers of cells supplied with medium, and thus allows for tracking of bacteria trapped therein.
Although this type of device has also been used for time-dependent problems as well \cite{Arnoldini2014, Kaiser2018, Julou2020, MarcoCosentino2020,Bakshi2021}, for our purpose involving size fluctuations of a large population of cells, it is not obvious if they are equivalent to those of a collection of many independent small populations.
More precisely, since the size of a cell depends strongly on its age, it is reasonable to use large enough chambers so that the chamber size may not affect the age distribution.
This led us to develop another system that can deal with large enough populations in each chamber and uniformly control non-steady environments without delicate optimization.

In this study, we develop a microfluidic device, which we name the ``extensive microperfusion system'' (EMPS).
This device can culture cells uniformly by supplying fresh medium through a porous membrane, similarly to previously reported systems \cite{Inoue2001,Charvin2008, Ducret2009}, but here we realize wide quasi-two-dimensional traps of dense bacteria in such a system.
We confirm that bacteria can freely swim and grow inside,
 and evaluate the uniformity and the switching efficiency of the culture condition.
Then we use this system for quantitative observations of
 bacterial reductive division processes, triggered by abrupt switching to non-nutrious medium.
We observe {\sl Escherichia coli} cells and find that the distribution of cell volumes, collected irrespective of cell ages,
 maintain the scale invariance as in \eqref{eq:1} at each time, with the mean cell size that gradually decreases.
On the other hand, the rescaled distribution function $F$ is found to depend on the growth condition before starvation, slightly but significantly.
To obtain theoretical insights on these experimental findings,
 we devise a cell cycle model describing reductive division processes,
 by extending the Cooper-Helmstetter model and its variants \cite{Cooper1968, Witz2019, Ho2015}
 for steady growth environments.
We numerically find that this model indeed shows the scale invariance under starvation conditions,
 confirming the robustness of this property.
We also provide theoretical descriptions on the time evolution of the cell size distribution,
 and propose a condition for the scale invariance.
Finally, we numerically show the range of validity of the scale invariance over various characteristic times of the starvation process, revealing the number of multifork replications may be important for the scale invariance.

\begin{figure}[t]
       \includegraphics[width=\hsize]{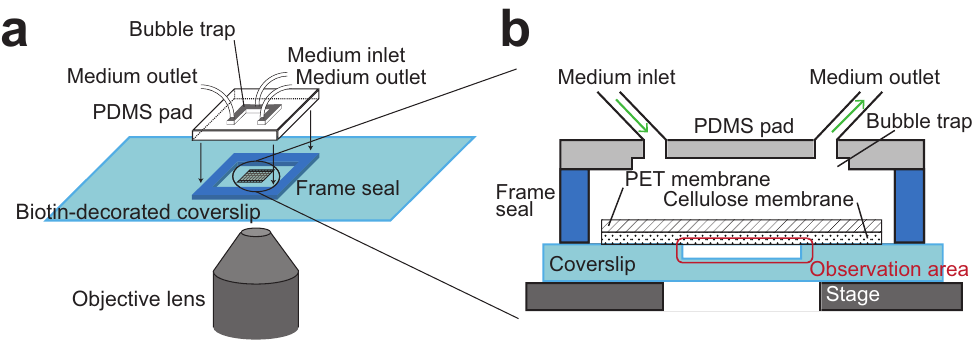}
       \caption{
       \label{fig-1}
       Sketch of the extensive microperfusion system (EMPS).
       {\bf a} Entire view of the device.
       Microwells are created on a glass coverslip.
       We attach a polydimethylsiloxane (PDMS) pad on the coverslip with a square frame seal
        to fill the system with liquid medium.
       {\bf b} Cross-sectional view inside the PDMS pad.
       A polyethylene terephthalate (PET)-cellulose bilayer porous membrane is attached via the biotin-streptavidin bonding.
       Note that there are two outlets as in ({\bf a}).
       }
\end{figure}

\section{Results}
\subsection{Development of the extensive microperfusion system}
To achieve uniformly controlled environments with dense bacterial suspensions, 
we adopt a perfusion system, which supplies fresh medium through a porous membrane attached over the observation area. 
Among several existing devices of this kind \cite{Inoue2001, Charvin2008, Ducret2009}, here we choose the one developed in ref.~\cite{Inoue2001} as a prototype. In this device, bacteria are confined in microwells made on a coverslip, covered by a cellulose porous membrane attached to the coverslip via biotin-streptavidin bonding.
Note that cellulose cannot be metabolized by {\sl E. coli} strains common for laboratory use \cite{Gao2015}, which we confirmed explicitly with MG1655 (Supplementary~Fig.~1g).
The pore size of the membrane is chosen so that it can confine bacteria and also that it can exchange nutrients and waste substances across the membrane.
To continuously perfuse the system with fresh medium, a polydimethylsiloxane (PDMS) pad with a bubble trap is attached above the membrane by a two-sided frame seal (\figref{fig-1}a and Supplementary~Fig.~1a).
This setup can maintain a spatially homogeneous environment
 for cell populations in each microwell, in particular if the microwells are sufficiently shallow so that all cells remain near the membrane.
However, because the soft cellulose membrane may droop and adhere to the bottom for wide and shallow microwells, the horizontal size of such quasi-two-dimensional microwells has been limited up to a few tens of micrometers, preventing from characterization of the instantaneous cell size distribution.

By the EMPS, we overcome this problem and realize quasi-two-dimensional wells sufficiently large for statistical characterization of cell populations.
This is made possible by introducing a bilayer membrane, where the cellulose membrane is sustained by a polyethylene terephthalate (PET) porous membrane via biotin-streptavidin bonding (\figref{fig-1}b, Supplementary~Fig.~1b and Methods).
Because the PET membrane is more rigid than the cellulose membrane, we can realize extended area without bending of the membrane.

Here, we conducted several experiments to evaluate how well the experimental condition inside the EMPS can be controlled (see also Supplementary~Note~2).
We first examined the flatness of the observation area using motile bacteria.
If a cellulose membrane alone is used, it is bent and adheres to the bottom of the well (Supplementary~Fig.~1c,d and Supplementary~Movie~1).
However, if it is replaced by our PET-cellulose bilayer membrane, it keeps flat enough so that bacteria can freely swim in the shallow well (Supplementary~Fig.~1e,f and Supplementary~Movie~2).
We also show that, using non-motile bacteria, growth rate of the bacteria is spatially uniform (Supplementary~Fig.~2 and Supplementary~Movie~3,4).
The doubling time of the cell population was 59 $\pm$ 10 min, which is comparable to that in the previous system without the PET membrane \cite{Inoue2001, Wakamoto2005, Hashimoto2016}.
Furthermore, similarly to other microfluidic devices, we can also switch the culture condition by changing the medium to supply.
We evaluate how efficiently the medium in the well is exchanged, by using fluorescent dye and non-motile bacteria.
We found that the medium exchange was almost completed within $5\unit{min}$ (Supplementary~Fig.~3), much shorter than the length of the bacterial cell cycle.
On the other hand, since medium exchange in EMPS relies on diffusion of molecules through the membrane, other devices that can replace medium more directly, such as mother machines in which medium is poured to a main channel connected to observed growth channels \cite{Arnoldini2014, Kaiser2018, Julou2020, MarcoCosentino2020, Bakshi2021}, may be advantageous in this respect.
Since this difference may affect, e.g., the amount of molecules that can remain on the surface of the observation area, cellular states may also change differently after an environmental switch.
In this respect, advantages of EMPS in environmental changes may be in the fact that (i) we can control the environment without hydrodynamic perturbations and, simultaneously, (ii) we can observe cells in large space under a uniform and time-dependent environment without mechanical trapping.
The absence of hydrodynamic perturbations can be seen from Brownian motion of non-motile cells in Supplementary~Movie~5,6, which is hardly affected by relatively strong medium flow above the membrane used to switch the medium.
Therefore, EMPS is indeed able to change the growth condition for cells in large space uniformly, without noticeable fluid flow perturbations, which is a unique strength of our device.

\begin{figure*}[t]
    \includegraphics[width=\hsize]{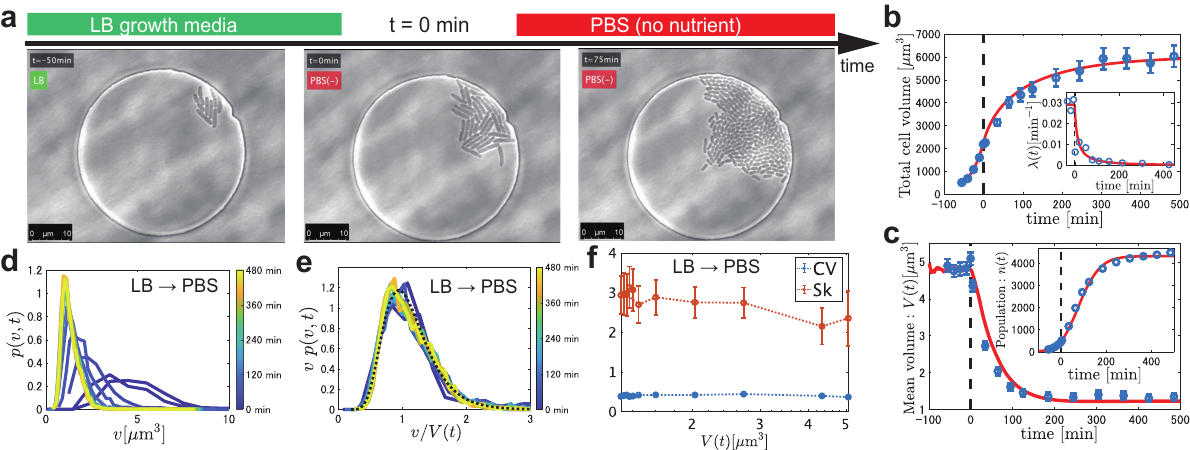}
    \caption{
    \label{fig-2}
    Results from the observations of reductive division.
    {\bf a} Snapshots taken during the reductive division process of {\sl E. coli} MG1655 in the EMPS.
    The medium is switched from LB broth to phosphate buffered saline (PBS) at $t=0$.
    See also Supplementary~Movie~7.
    {\bf b,c} Experimental data (blue symbols) for the total cell volume $V_\mathrm{tot}(t)$ ({\bf b}), the growth rate $\lambda(t)$ ({\bf b}, Inset, see also Supplementary~Fig.~9a showing the same data in logarithmic scale), the mean cell volume $V(t)$ ({\bf c}) and the number of the cells $n(t)$ ({\bf c}, Inset) in the case of LB $\to$ PBS, compared with the simulation results (red curves).
    The error bars indicate segmentation uncertainty in the image analysis (see Methods).
    $t=0$ is the time at which PBS enters the device (black dashed line).
    The data were collected from 15 wells recorded in a single experiment.
    {\bf d} Time evolution of the cell size distributions during starvation
    in the case of LB $\to$ PBS at
     $t=0,5,30,60,90,120,180,240,300,360,420,480\unit{min}$ from right to left.
     The sample size is $n(t)$ for each distribution (see {\bf c}, Inset).
    {\bf e} Rescaling of the data in ({\bf d}). The overlapped curves indicate the function $F(v/V(t))$ in \eqref{eq:massdist}.
    The dashed line represents the fitted log-normal distribution ($\sigma=0.34(2)$).
    {\bf f} The coefficient of variation (CV) and the skewness (Sk) [\eqref{eq:CVSk}] against $V(t)=\expct{v}$.
    The error bars were estimated by the bootstrap method with 1000 realizations.
    }
\end{figure*}

\begin{figure}[t]
       \includegraphics[width=\hsize]{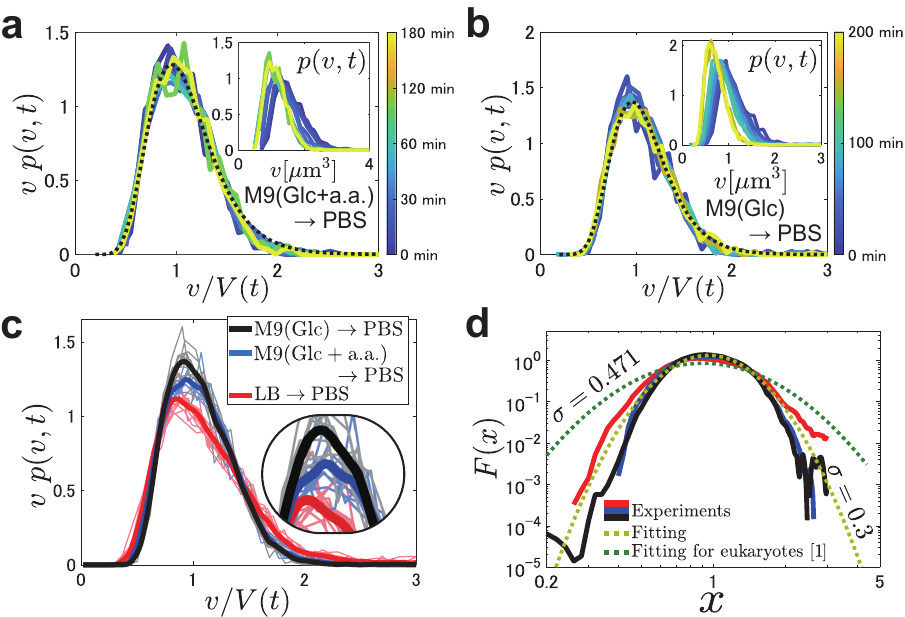}
       \caption{
       \label{fig-3}
       Rescaled cell size distributions.
       {\bf a} The results for M9(glucose (Glc) + amino acids (a.a.)) $\to$ PBS.
       The dashed line represents the fitted log-normal distribution ($\sigma=0.31(2)$).
       The data were taken from 17 wells recorded in a single experiment.
       The sample size ranges from $n(0)=685$ to $n(180)=1260$ (see Supplementary~Fig.~4).
       (Inset) Time evolution of the non-rescaled cell size distributions at
       $t=0,10,20,30,40,50,60,90,120,180\unit{min}$.
       {\bf b} The results for M9(glucose (Glc)) $\to$ PBS.
       The dashed line represents the fitted log-normal distribution ($\sigma=0.29(2)$).
       The data were taken from 26 wells recorded in a single experiment.
       The sample size ranges from $n(0)=836$ to $n(200)=2160$ (see Supplementary~Fig.~4).
       (Inset) Time evolution of the non-rescaled cell size distributions at $t=0,10,20,30,40,50,60,80,100,150,200\unit{min}$.
       {\bf c} Experimental results of $F(v/V(t)) = vp(v,t)$ for the three cases studied in this work.
       The raw data obtained at different times are shown by thin lines with relatively light colors, and the time-averaged data are shown by the bold lines. 
       The time-averaged distributions (bold lines) are found to be slightly but significantly different among the three cases.
       The difference can also be seen in the instantaneous distributions (thin lines; see the inset for enlargement near the peak).
       {\bf d} Fitting of the experimentally obtained $F(x)$ (solid lines; time-averaged data in ({\bf c}) are shown) to the log-normal distribution (yellow dotted line). 
       Also shown is the fitting result by Giometto \textit{et al.} \cite{Giometto2013} for unicellular eukaryotes (green dotted line).
       $\sigma$ is the standard deviation parameter of the log-normal distribution (see text).
       }
\end{figure}

\subsection{Characterization of bacterial reductive division by EMPS}
Now we observe the reductive division of {\sl E. coli} MG1655 in the EMPS,
 triggering starvation by switching from nutritious medium to non-nutritious buffer.
In the beginning, a few cells are trapped in a quasi-two-dimensional well
 (diameter $55\unit{\mu{}m}$ and depth $0.8\unit{\mu{}m}$)
 and grown in nutritious medium,
 until a microcolony composed of approximately 100 cells appear.
We then quickly switch the medium to a non-nutritious buffer,
 which is continuously supplied until the end of the observation
 (see Methods for more details).
By doing so, we not only trigger cell starvation, but also intend to remove various substances secreted by cells, such as autoinducers for quorum sensing and waste products, to reduce their effects on cell growth \cite{Carbonell2002, Bruger2016, Haeaar2018, Maier2015}.
Note that $5\unit{min}$ required to exchange the medium in the trap is sufficiently shorter than the typical length of the cell cycle of \textit{E.~coli}, which is several tens of minutes.
This implies that starvation is triggered abruptly for cells.
Throughout this experiment, the well is entirely recorded by phase contrast microscopy.
We then measure the length and the width of all cells in the well,
 to obtain the volume $v$ of each cell by assuming the spherocylindrical shape,
 at different times before and after the medium switch.
Here we mainly show the results for the case where the medium is switched from LB broth to phosphate buffered saline (PBS)
 (denoted by LB $\to$ PBS) in \figref{fig-2},
 while the results for M9 medium with glucose (Glc) and 12 amino acids (a.a.) $\to$ PBS,
 M9 medium with glucose (Glc) $\to$ PBS,
 M9 medium with glycerol (Glyc) $\to$ PBS,
 and M9 medium with glucose (Glc) $\to$ M9 medium with $\alpha$-methyl-D-glucoside ($\alpha$MG),
 a glucose analog which cannot be metabolized \cite{Chou1994}, are also presented in Supplementary~Fig.~4.
We observe that, after switching to the non-nutritious buffer, the growth of the total volume decelerates, and the mean cell volume rapidly decays because of excessive cell divisions (Supplementary~Movie~7-11, \figref{fig-2}a,b, Supplementary~Fig.~4), until cells eventually stop growing and dividing.
Note that, unlike other cases (Supplementary~Fig.~4), cell growth did not stop completely in the case of LB $\to$ PBS, but we consider that this will not affect our analysis because the ultimately remaining growth rate $\sim 10^{-4}\unit{min^{-1}}$ (Supplementary~Fig.~9a) was sufficiently low compared to the time scale of the volume reduction $\sim 10^2\unit{min}$ (\figref{fig-2}c) and all other time scales relevant in this study.
Concerning the volume reduction, it is mostly due to the decrease of the cell length, 
 while we notice that the mean cell width may also change slightly (Supplementary~Fig.~5).
We consider that this is not due to osmotic shock \cite{Rojas2014},
 because then the cell width would increase when the osmotic pressure is decreased,
 which is contradictory to our observation for LB $\to$ PBS (Supplementary~Table~1 and Supplementary~Fig.~5).
Such a change in cell widths was also reported for a transition between
 two different growth conditions \cite{HARRIS2016}.
In any case, \figref{fig-2}d shows how the distribution of the cell volumes $v$,
 $p(v,t)$, changes over time;
 as the mean volume decreases, the histograms shift leftward and become sharper.
However, when we take the ratio $v/V(t)$,
 with $V(t) = \expct{v}$ being the mean cell volume at each time $t$,
 and plot $vp(v,t)$ instead,
 we find that all those histograms overlap onto a single curve (\figref{fig-2}e).
In other words, we find that the time-dependent cell size distribution during the reductive division
 maintains the following scale-invariant form all the time: 
\begin{equation}
       p(v,t) = v^{-1}F(v/V(t)).
       \label{eq:massdist}
\end{equation}
This is analogous to \eqref{eq:1} previously reported for the steady growth condition,
 but here importantly the mean volume $V(t)$ changes over time significantly (\figref{fig-2}c).
The scale invariance also holds for the length distribution (Supplementary~Fig.~6); this is expected because the length changes are dominant in the studied volume changes.

To further test the scale invariance of the distribution,
 we evaluate the coefficient of variation (CV) and the skewness (Sk) defined by
\begin{equation}
 \mathrm{CV} \equiv \frac{\sqrt{\expct{\delta v^2}}}{\expct{v}}, \quad
 \mathrm{Sk} \equiv \frac{\expct{\delta v^3}}{\expct{\delta v^2}^{3/2}},  \label{eq:CVSk}
\end{equation}
 with $\delta v \equiv v - \expct{v}$.
Both quantities measure the shape of the distribution function of $v/V(t)$ and not affected by variation of $V(t)$.
The results in \figref{fig-2}f indeed confirm that both CV and Sk remain essentially constant, so that the function $F(\cdot)$ remains unchanged and the scale invariance holds during the reductive division.
Remarkably, we reach the same conclusion for all combinations of the growth and starving conditions that we test, as shown in \figref{fig-3}a,b,c and Supplementary~Fig.~7 (see \figref{fig-2}f and Supplementary~Fig.~4 for the results of CV and Sk).
Our results therefore indicate that the scale invariance as in \eqref{eq:massdist},
 which has been observed for steady conditions \cite{Giometto2013, Zaoli2019},
 also holds in non-steady reductive division processes of {\sl E. coli}
 rather robustly.

In addition to the robustness of the scaling relation \pref{eq:massdist},
 the functional form of the scale-invariant distribution, i.e., that of $F(x)$, is of interest.
We detect weak dependence of $F(x)$ on the choice of the medium in the growth phase (\figref{fig-3}c).
More specifically, we find the trend that the fluctuations of the rescaled cell volumes are larger for richer growth conditions (Supplementary~Fig.~7h,i), consistently with a past observation in ref.\cite{Gangan2017}.
The lower the nutrient level of the growth medium is, the sharper the function $F(x)$ becomes, and therefore, the smaller the variance is.
We have also confirmed that the variation in $F(x)$ among different sets of media is more significant than that among biological replicates (Supplementary~Fig.~7h,i).
This is somewhat unexpected in view of the past studies reporting the robustness of cell size fluctuations against varying temperatures and other environmental factors \cite{Giometto2013, Iyer-Biswas2014pnas, Iyer-Biswas2014prl}.

Moreover, we find that our observations for {\sl E. coli} are significantly different from those for unicellular eukaryotes reported by Giometto \textit{et al.} \cite{Giometto2013} (\figref{fig-3}d).
More precisely, they showed that the rescaled cell size distribution for unicellular eukaryotes is well fitted by the log-normal distribution,
 $\propto (1/x)\exp(-(\log x-m)^2/2\sigma^2)$ with $m=-\sigma^2/2$ (due to the normalization $\expct{x}=1$), and obtained $\sigma=0.471(3)$.
We find that our data for {\sl E. coli} can also be fitted by the log-normal distribution (\figref{fig-2}e, \figref{fig-3}a,b,d and Supplementary~Fig.~7),
 but here the value of $\sigma$, evaluated by the standard deviation of $\log x$, is found to be around $\sigma=0.3$ ($\sigma=0.34(2)$ for LB $\to$ PBS, $\sigma=0.30(2)$ for M9(Glc+a.a.) $\to$ PBS, and $\sigma=0.29(2)$ for M9(Glc) $\to$ PBS), much lower than $\sigma=0.471(3)$ for the unicellular eukaryotes.
In the literature, a previous study on {\sl B. subtilis} \cite{Wakita2010} reported values of $\sigma$ from $0.24$ to $0.26$,
 which are comparable to our results for {\sl E. coli}.
Compared to this substantial difference between bacteria and unicellular eukaryotes, the dependence on the environmental factors seems to be much weaker (\figref{fig-3}d).

\begin{figure*}[t]
    \includegraphics[width=\hsize]{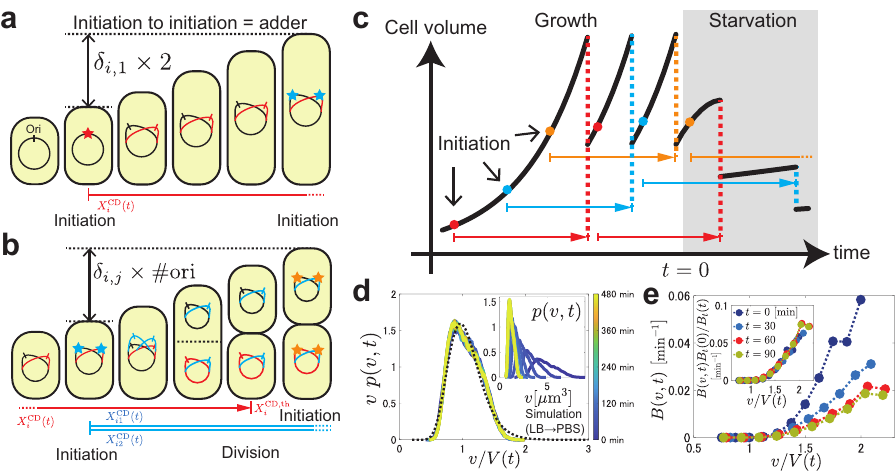}
    \caption{
    \label{fig-4}
    Model of reductive division and simulation results.
    {\bf a,b} Single ({\bf a}) and multifork ({\bf b}, where $\#\mathrm{ori}=4$) intracellular cycle processes.
    See \eqref{eq:initiation} for the criterion that triggers the initiation.
    Progress of each cycle is represented by a coordinate $\XCD_i(t)$, which increases at speed $\mu_i(t)$ and ends at $\XCD_i(t)=\XCDth_i$ by triggering cell division.
    {\bf c} Illustration of cell cycles in this model.
    Each colored arrow represents a single intracellular cycle process.
    {\bf d} Overlapping of the rescaled cell size distributions during starvation
    in the model for LB $\to$ PBS.
    The dashed line represents the fitted log-normal distribution ($\sigma=0.25(2)$).
    (Inset) The non-rescaled cell size distributions
    at $t=0,5,30,60,90,120,180,240,300,360,420,480\unit{min}$ from right to left.
    {\bf e} Numerically measured division rate, $B(v,t)$, in the model for LB $\to$ PBS.
    See Supplementary~Note~4.B for the measurement method.
    (Inset) Test of the condition of \eqref{eq:separability}.
    Here $B_t(0)/B_t(t)$ is evaluated by $B_t(0)/B_t(t) = \int B(xV(0),0)dx / \int B(xV(t),t) dx$, with $x$ running in the range $0 \leq x \leq 1.8$.
    Overlapping of the data demonstrates that \eqref{eq:separability} indeed holds in our model. 
    }
\end{figure*}

\subsection{Modeling the reductive division process}
To obtain theoretical insights on the experimentally observed scale invariance
 of the cell size distributions,
 we construct a simple cell cycle model for the bacterial reductive division.
For the steady growth conditions, a large number of studies on {\sl E. coli} have been carried out
 to clarify what aspect of cells triggers the division event \cite{Jun2018, Ho2018}.
Significant advances have been made recently to provide molecular-level understanding
 \cite{Ho2018, Ho2015, HARRIS2016, Wallden2016, SI2017, MICALI2018, Micalieaau2018, SI2019}.
Here we extend such a model to describe the starvation process.

One of the most established models in this context is the Cooper-Helmstetter (CH) model \cite{Cooper1968,Wang2009}, 
 which consists of cellular volume growth and multifork DNA replication.
The multifork replication is the phenomenon that a cell replicates its DNA not only for its daughters but also for its granddaughters, before the birth of the daughter cells (\figref{fig-4}a,b) -- a phenomenon well known for fast growing bacteria such as {\sl E. coli} and {\sl B. subtilis} \cite{Wang2009,Cooper1968}.
In the CH model, completion of the DNA replication triggers the cell division,
 and this gives a homeostatic balance between the DNA amount and the cell volume.
An unknown factor of the CH model is how DNA replication is initiated,
 and a few studies attempted to fill this gap to complement the CH model \cite{Ho2015, Wallden2016}.
Ho and Amir \cite{Ho2015} assumed that replication is initiated
 when a critical amount of ``initiators'' accumulate at the origin of replication.
In the presence of a constant concentration of autorepressors, expressed together with the initiators, this assumption means that the cellular volume increases by a fixed amount between two initiation events, regardless of the absolute volume at the initiation.
This ``adder'' principle between initiations is now supported by several observations \cite{SI2019,MICALI2018, Micalieaau2018}.
By the initiation considered above, the cell starts the C period of the bacterial cell cycle, which is followed by the D period, and eventually the cell divides \cite{Wang2009,Cooper1968}.
While Ho and Amir assumed that a constant time is needed to complete the C+D period (``timer'' principle) \cite{Ho2015}, further experimental investigations by Witz \textit{et al.} clarified that the model assuming the adder principle for the C+D period captured single-cell behavior better \cite{Witz2019}.
Clarifying the mechanism of cell division control is currently a target of intensive studies and different models have also been proposed \cite{SI2019,MICALI2018, Micalieaau2018}.
In the present work, 
 we choose to extend Witz \textit{et al.}'s model \cite{Witz2019} to cope with the switch to the non-nutritious condition, and measure the cell size fluctuations during the reductive division process.
We also checked that our main conclusions do not change if we use instead Ho and Amir's model \cite{Ho2015} as the starting point.

The model consists of two processes that proceed simultaneously,
 namely the volume growth and the intracellular cycle.
The volume of each cell (indexed by $i$), $v_i(t)$, grows as $\diff{v_i}{t}=\lambda(t)v_i(t)$,
 with a time-dependent growth rate $\lambda(t)$.
Following Witz \textit{et al.}'s model \cite{Ho2015},
 we assume that the volume growth is coupled to the intracellular cycle as follows.
To begin with the simplest case, suppose that a newborn cell $i$ has a single origin of replication in its chromosome, and that the replication starts at some point in time (\figref{fig-4}a, red star).
By this initiation of replication, the cell starts to have two origins of replication.
Then, the next initiation is triggered when the cell volume $v_i(t)$ increases by a fixed amount $\delta_{i,1}$ per origin, i.e., when $v_i(t)$ increases by $\delta_{i,1} \times 2$, since the last initiation (\figref{fig-4}a, blue stars).
Note that this criterion does not change whether a cell divides or not before the initiation; if a cell divides and produces daughter cells $i_1$ and $i_2$, the initiation in the daughter cells occurs when $v_{i_1}(t) + v_{i_2}(t) - v_i(t_\mathrm{init}) = \delta_{i,1} \times 2$, where $t_\mathrm{init}$ is the time at which the last initiation occurred.
Similarly, if multifork replication takes place in a single cell (i.e., $\#\mathrm{ori} = 2^j$ with $j \geq 2$), the threshold for the added volume is given by $\delta_{i,j} \times \#\mathrm{ori}$ (see \figref{fig-4}b, for an example with $\numori=4$).
The criterion therefore reads:
\begin{equation}
 \(\sum_{i' \in \text{offspring of $i$}} v_{i'}(t) \) - v_i(t_\mathrm{init}) = \delta_{i,j} \times \#\mathrm{ori}.  \label{eq:initiation}
\end{equation}
Following the experimental results by Si \textit{et al.} \cite{SI2017}, we assume that $\delta_{i,j}$ does not depend on environmental conditions.
On the other hand, to take into account stochastic nature of division events,
 we generate $\delta_{i,j}$ randomly from the Gaussian distribution
 with mean $\expct{\delta_{i,j}} = \delta_\mean$
 and standard deviation $\Std[\delta_{i,j}] = \delta_\std$.

After an initiation, the cell undergoes the C+D period and finally divides.
Here, for the sake of simplicity, the progression of the C and D period is collectively expressed by a coordinate $\XCD_i(t)$,
 which starts form zero and increases at time-dependent speed $\mu_i(t)$, $\diff{\XCD_i}{t}=\mu_i(t)$.
When $\XCD_i(t)$ reaches a threshold $\XCDth_i$,
 the cell divides (\figref{fig-4}b), leaving two daughter cells of volumes
 $v_{i_1}(t) = x^\mathrm{sep}v_i(t)$ and $v_{i_2}(t) = (1-x^\mathrm{sep})v_i(t)$.
Here, $x^\mathrm{sep}$ is randomly drawn
 from the Gaussian distribution with mean $0.5$ and standard deviation $0.0325$, the latter value being deduced from experimental observations (see Methods and Supplementary~Fig.~8).
To deal with the multifork replication, the index $i$ of $\XCD_i(t)$ denotes the cell to divide by the considered cell cycle progression.
Therefore, if $\#\mathrm{ori} = 2$ when the initiation is triggered, a pair of cell cycles for the future daughter cells, represented by $\XCD_{i_1}(t)$ and $\XCD_{i_2}(t)$, start and run simultaneously (\figref{fig-4}c).
Similarly to $\delta_{i,j}$, we also assume that $\XCDth_i$ is a Gaussian random variable, with $\expct{\XCDth_i} = 1$ and $\Std[\XCDth_i] = \XCDth_\std$.

Now we are left to determine the two time-dependent rates,
 $\lambda(t)$ and $\mu_i(t)$.
Here we consider the situation where growth medium is switched
 to non-nutritious buffer at $t=0$;
 therefore, $t$ denotes time passed since the switch to the non-nutritious condition.
First, we set the volume growth rate $\lambda(t)$
 on the basis of the Monod equation \cite{Monod1949},
 assuming that substrates in each cell are simply diluted by volume growth
 and consumed at a constant rate, without uptake
 because of the non-nutritious condition considered here.
As a result, we obtain
\begin{equation}
       \lambda(t) = \lambda_0\frac{1-A}{e^{ct}-A},
       \label{eq:growthrate}
\end{equation}
 with constant parameters $A$ and $c$, and the growth rate $\lambda_0 (=\lambda(0))$ in the exponential growth phase (see Supplementary~Note~3.A for details).

For the cycle progression speed $\mu_i(t)$, we propose a functional form that conforms with the type of the principle assumed in the original model for the C+D period in steady conditions, i.e., the adder principle for Witz \textit{et al.}'s model and the timer principle for Ho and Amir's model.
We first note that the C+D period mainly consists of DNA replication, followed by its segregation and the septum formation \cite{Wang2009}.
Most parts of those processes involve biochemical reactions of substrates, such as deoxynucleotide triphosphates for the DNA synthesis.
Here we can consider different molecular mechanisms for the cycle progression, depending on the type of the principle to adopt.
For the case of the adder principle (\`a la Witz \textit{et al.}), we can assume that division occurs when a given amount of relevant molecules, such as DNA, is produced.
We assume that such relevant molecules are synthesized from substrates through enzyme catalyses, according to the Michaelis-Menten equation.
Considering dilution due to the volume growth too, we obtain $\mu_i(t) \propto [\mathrm{S_{C+D}}]v_i(t)/(K + [\mathrm{S_{C+D}}])$, with $[\mathrm{S_{C+D}}]$ being the concentration of the corresponding substrates and $K$ an ajustable parameter (see Supplementary~Note~3 for details).
For simplicity, here we assume that $[\mathrm{S_{C+D}}]$ is common to all cells.
Note that, since $[\mathrm{S_{C+D}}]$ is constant in steady conditions, $\mu_i(t) \propto v_i$ and this results in the adder principle as considered in Witz \textit{et al.}'s model.
For the starvation process, we consider that $[\mathrm{S_{C+D}}]$ decreases by dilution due to volume growth, degredation, and consumption.
Those are assumed to be independent of $\#\mathrm{ori}$, based on the experimental results that the duration of the C+D period is independent of $\#\mathrm{ori}$ in steady environments \cite{SI2017}.
From those considerations, we finally obtain the following equation for the cycle progression speed:
\begin{equation}
       \mu_i(t) = \frac{\mu_0}{v_0}\frac{k+1}{k\exp(t/\tau)+1}v_i(t),
       \label{eq:mu}
\end{equation}
 with parameters $k$ and $\tau$, the mean cycle progression speed $\mu_0$ and the mean cell size $v_0$ in the exponential phase (see Supplementary~Note~3.A).
In the case of the timer principle for the C+D period in steady environments (\`a la Ho and Amir), we consider instead that assembly processes of molecules such as deoxynucleotide triphosphates control the cycle progression speed. As a result, we obtain a formula of $\mu_i(t)$ without $v_i$ dependence (see Supplementary~Note~3).
In the following, however, we mostly present results from the model \`a la Witz \textit{et al.} unless otherwise stipulated, while we checked that the main conclusions did not change if the model \`a la Ho and Amir was used instead.

The parameter values are determined from the experimentally measured total cell volume and the cell number, which our simulations turn out to reproduce very well (\figref{fig-2}b,c and Supplementary~Fig.~4a,b), with the aid of relations reported by Wallden \textit{et al.} \cite{Wallden2016} for some of the parameters (see \tbref{tb-1} for the parameter values used in the simulations, and Methods for the estimation method).
With the parameters fixed thereby, we measure the cell size fluctuations at different times and find the scale invariance similar to that revealed experimentally (\figref{fig-4}d and Supplementary~Fig.~9b,e).
The constancy of CV and Sk is also confirmed (Supplementary~Fig.~9c,f).
Interestingly, the scale invariance emerges despite the existence of characteristic scales in the model definition, such as the typical volume added between initiations, $\delta_\mean$.
To check the robustness of those results, we also extended Ho and Amir's model along the same line (see Supplementary~Note~3.B for details) and confirmed the scale invariance of similar quality (Supplementary~Fig.~9g,h).
These findings suggest the existence of a statistical principle underlying the scale invariance, which is not influenced by details of the model.

\subsection{Theoretical conditions for the scale invariance}
To seek for a possible mechanism leading to the scale invariance, here we describe, theoretically, the time dependence of the cell size distribution in a time-dependent process.
Suppose $N(v,t)dv$ is the number of the cells whose volume is larger than $v$ and smaller than $v+dv$.
If we assume, for simplicity, that a cell of volume $v$ can divide to two cells of volume $v/2$, at probability $B(v,t)$, we obtain the following time evolution equation:
\begin{align}
       \frac{\partial N(v,t)}{\partial t} = &-\frac{\partial}{\partial v}[\lambda(t)vN(v,t)]\notag \\
        & -B(v,t)N(v,t)+4B(2v,t)N(2v,t).
       \label{eq:balanceeq}
\end{align}
Note that this equation has been studied by numerous past studies for understanding stable distributions in steady conditions \cite{Sinko1971, Diekmann1983, TYSON1986, Robert2014, Giometto2013, Hosoda2011, TAHERIARAGHI2015}, but here we explicitly include the time dependence of the division rate, $B(v,t)$, for describing the transient dynamics.
To clarify a condition for this equation to have a scale-invariant solution, here we assume the scale invariant form, \eqref{eq:massdist}, where $p(v,t) = N(v,t) / n(t)$ and $n(t)$ is the total number of the cells, and obtain the following self-consistent equation (see Supplementary~Note~4.A for derivation):
\begin{equation}
       F(x) = -x\frac{\partial F(x)}{\partial x} - \frac{B(v,t)}{\bar{B}(t)}F(x)
        + 2\frac{B(2v,t)}{\bar{B}(t)}F(2x).
       \label{eq:Fconsistent}
\end{equation}
Here, $x=v/V(t)$ and $\bar{B}(t)=\int dv B(v,t)p(v,t)$.
For the scale invariance, \eqref{eq:Fconsistent} should hold at any time $t$.
This is fulfilled if $B(v,t)$ can be expressed in the following form (see Supplementary~Note~4.A):
\begin{equation}
       B(v,t) = B_v(v/V(t))B_t(t).
       \label{eq:separability}
\end{equation}
This is a sufficient condition for the cell size distribution to maintain the scale invariant form, \eqref{eq:massdist}, during the reductive division.
Note that ref.~\cite{Kennard2016} proposed a similar scale-invariant form of the division rate for the steady environment.
It is also important to remark that, as opposed to \eqref{eq:balanceeq}, \eqref{eq:Fconsistent} does not include the growth rate $\lambda(t)$ explicitly.
The scale-invariant distribution $F(x)$ is therefore completely characterized by the division rate $B(v,t)$ in this framework.

To test whether the condition of \eqref{eq:separability} is satisfied in our model, we measure the division rate $B(v,t)$ in our simulations (\figref{fig-4}e).
The data overlap if $B(v,t) B_t(0)/B_t(t)$ is plotted against $v/V(t)$, 
demonstrating that \eqref{eq:separability} indeed holds here.
On the other hand, our theory does not seem to account for the functional form $F(x)$ of the scale-invariant distribution; the right hand side of \eqref{eq:Fconsistent} differs significantly from the left hand side, if the numerically obtained $B(v,t)$ is used together with the function $F(x)$ from the simulations or the experiments (Supplementary~Fig.~10).
The disagreement did not improve by taking into account the effect of septum fluctuations (see Supplementary~Note~4.C).
The lack of quantitative precision is probably not surprising given the simplicity of the theoretical description, which incorporates all effects of intracellular cycles into the simple division rate function $B(v,t)$.
The virtue of this theory is that it clarifies that the intracellular cycle seems to have important relevance in the scale invariance and the functional form of the cell size distribution. 
The significant difference in $F(x)$ identified between bacteria and unicellular eukaryotes (\figref{fig-3}d) may be originated from the different replication mechanisms that the two taxonomic domains adopt.

\subsection{Violation of the scale invariance}

Here we investigate the robustness of the scale invariance during the reductive division.
In particular, we aim to clarify whether it breaks down for other starvation conditions, and if so, what the condition is for the scale invariance to hold.
As shown in \figref{fig-3}c and Supplementary~Fig.~7, the form of $F(x)$ obtained by our experiments depends on the growth environment before starvation.
This suggests that $F(x)$ may change if one switches between two growth media in a quasistatic manner, i.e., the scale invariance may break down in this case.
Motivated by this hypothesis, we numerically investigate whether there is a lower bound on the relaxation speed of the cellular state, below which the scale invariance breaks down.
For simplicity, we consider that the environment starts to change at $t=0$, and the volume growth rate $\lambda(t)$ and the cycle progression speed $\mu_i(t)$ decrease as follows:
\begin{align}
       \lambda(t) & = \lambda_0\exp(-t/\tau_\lambda), \label{eq:lamdecrease} \\
       \mu_i(t) & = \frac{\mu_0}{v_0}\exp(-t/\tau_\mu)v_i(t).
       \label{eq:mudecrease}
\end{align}
We regard $\tau_\lambda$, $\tau_\mu$ and $\lambda_0$ as free parameters, while the parameters $\mu_0$ and $v_0$ were set as follows (see also Supplementary~Note~3.C.2 for details). For $\mu_0$, we determined it from $\lambda_0$ using empirical relation reported by Wallden \textit{et al.} \cite{Wallden2016} for steady environments. For $v_0$, we set its value self-consistently, so that the mean cell volume $\expct{v_i}$ obtained numerically in the exponential phase with $\mu_i = (\mu_0/v_0)v_i$ falls within 1\% error from the given value of $v_0$.
The number of cells is set to be approximately $50,000$ at $t=0$ and kept constant afterward during the starvation process, by eliminating one of the daughter cells produced by division (see Methods for details).

\begin{figure}[t]
       \includegraphics[width=\hsize]{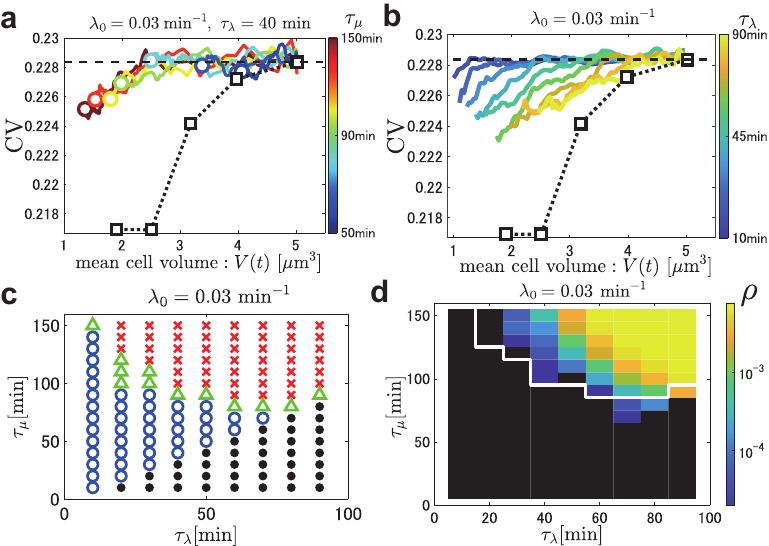}
       \caption{
       \label{fig-5}
       Numerical results on the range of validity of the scale invariance.
       The initial growth rate is fixed at $\lambda(0)=\lambda_0=0.03\unit{min^{-1}}$ unless otherwise stipulated.
       {\bf a} Trajectories in the $\expct{v}$-CV space for different $\tau_\mu$ (from $50$ to $150\unit{min}$), with $\tau_\lambda=40\unit{min}$ fixed.
       The endpoint of each trajectory is indicated by a colored open circle.
       The black squares represent the states in steady growth conditions with the growth rate $\lambda_0$ ranging from $0.01$ to $0.03\unit{min^{-1}}$.
       The dashed plateau indicates the initial CV at $\lambda_0=0.03\unit{min^{-1}}$.
       {\bf b} The master curves of the $\expct{v}$-CV trajectories for different $\tau_\lambda$. Those are obtained by taking the average of the CV values at each $\expct{v}$ over different $\tau_\mu (> \tau_\lambda)$.
       $\tau_\lambda$ ranges from $10$ to $90\unit{min}$.
       {\bf c} Phase diagram. $\times$: the scale invariance breaks down.
        Blue $\circ$: the scale invariance holds.
        Green $\triangle$: near the boundary.
        Black dots: the scale invariance holds but the mean volume increases.
       {\bf d} Pseudocolor plot of $\rho$ for different $\tau_\lambda$ and $\tau_\mu$. See the main text for the definition of $\rho$. 
       The black region indicates $\rho=0$.
       The white line represents the transition line obtained from ({\bf c}).
       The boundaries ($\triangle$) are not included in the region where the scale invariance breaks down.
       }
\end{figure}

First we evaluate the mean cell volume $\expct{v}$ and the coefficient of variation, $\mathrm{CV} = \sqrt{\expct{\delta v^2}}/\expct{v}$, in the exponential growth phase under steady conditions, by varying the growth rate $\lambda_0$ from $0.01$ to $0.03\unit{min^{-1}}$.
As shown by the black squares in \figref{fig-5}a,b, lower growth rates (smaller mean volumes) lead to lower CVs.
This is consistent with our experimental results (\figref{fig-3}c and Supplementary~Fig.~7).
We then investigate how the mean volume $\expct{v}$ and CV change during the starvation process, starting at $t=0$ from the growth phase with $\lambda_0=0.03\unit{min^{-1}}$ (\figref{fig-5}a, the top right black square).
As expected, our data showed that the mean cell volume decreases if $\tau_\mu > \tau_\lambda$ and increases otherwise; therefore, in the following, we deal with the case of $\tau_\mu > \tau_\lambda$, which corresponds to the reductive division.
The color curves in \figref{fig-5} show trajectories in the $\expct{v}$-CV space during the starvation process, each curve corresponding to a different $\tau_\mu (> \tau_\lambda)$ with $\tau_\lambda$ fixed at $\tau_\lambda = 40 \unit{min}$.
Remarkably, these trajectories overlap to a single curve with an extended plateau region, which indicates that CV is kept constant, i.e., the scale invariance.
Each curve stops in the middle of the master curve, the location of the endpoint (at $t\to\infty$, shown by the open circles) being determined by $\tau_\mu$.
Importantly, for small $\tau_\mu$, the trajectories stop in the plateau region, so that the scale invariance holds during the entire process.
By contrast, for large $\tau_\mu$, the trajectories go over the plateau and CV decreases abruptly; in other words the scale invariance breaks down.
Next, \figref{fig-5}b shows the master curves for different $\tau_\lambda$, each constructed by using the trajectories for all $\tau_\mu$ greater than $\tau_\lambda$.
We find that the smaller $\tau_\lambda$ is, the more extended the plateau region is.
Finally, we show the phase diagram for various combinations of $\tau_\mu$ and $\tau_\lambda$ in \figref{fig-5}c.
This clearly shows a region in which the scale invariance is maintained during the entire starvation process, bordered by a transition line over which the scale invariance breaks down.
Note that the scale-invariant region becomes narrower for larger $\tau_\lambda$ and $\tau_\mu$ and seem to disappear eventually; this is consistent with our expectation described at the beginning, that the scale invariance does not hold for quasistatic changes.
All those results were also confirmed when the extension of Ho and Amir's model was used instead (Supplementary~Fig.~11; see Supplementary~Note~2B for the model definition).

To understand what triggers the violation of the scale invariance, we focus on the state of the multifork replications, since our theory suggested the importance of the division rate, which is controlled by the state of the cell cycle.
As illustrated in \figref{fig-4}c, the first few divisions after the onset of starvation are tied to the initiation that occurred in the exponential growth phase.
We may expect that these division events retaining ``memories'' from the growth phase are less affected by the starvation, and therefore may not violate the scale invariance.
Based on this expectation, we investigate the state of the cell cycle as follows.
First, we observe that the change in the number of origins of replication ($\#\mathrm{ori}$) during the growth phase is rather stable, doubling (by initiation) and decreasing (by division) between $\#\mathrm{ori} = 2^{j-1}$ and $2^j$ with a fixed $j$ for the majority of cells ($j=3$ in the case of \figref{fig-4}c; note that $j-1$ and $j$ correspond to the numbers of parallel arrows therein).
This number is maintained for a while in the starvation process, but eventually it may decrease, because a cell may divide without initiating a new replication during the life.
We therefore measure the fraction of such cells, $\rho$.
To be precise, with $J$ being $j$ of each cell in the growth phase, $\rho$ is the fraction of cells such that the C+D period with $j<J$ is initiated during the lifetime, and that this C+D period ends and triggers a cell division afterward, before the cell cycle progression completely stops (note that, since $\mu_i(t) \to 0$, not all cell cycles complete).
It is measured at the final time point of the simulations (specifically $t = 600\unit{min}$) and shown in Figure~\ref{fig-5}d for $\lambda_0=0.03\unit{min^{-1}}$.
Intuitively, $\rho$ corresponds to the fraction of cells that lost memories from the growth phase.
Here we find $\rho=0$ indeed in most part of the scale-invariant region, while $\rho > 0$ when the scale invariance breaks down.
We therefore consider that the state of the multifork replications may be a key factor that determines whether the scale invariance holds or not during the reductive division.
Note that, for gradual environmental changes, actual cells are known to emit signals such as ppGpp \cite{MAGNUSSON2005, Ferullo2008} that control growth and cycle progression, which are not taken into account in our model.
Investigating the effect of such signals in this problem is an interesting problem left for future studies.

\section{Discussion}
In this work, we developed a membrane-based microfluidic device that we named the extensive microperfusion system (EMPS).
Advantages of this device are that we can realize a uniformly controlled environment for wide-area observations of microbes, and can switch it without hydrodynamic perturbations.
Those advantages may be useful for applications in a wide range of problems with dense cellular populations, including living active matter systems \cite{Baer2019,Be'er2019} and biofilm growth \cite{HallStoodley2004,Boudarel2018, Fuquae2019}.
In this work, we focused on statistical characterizations of single cell morphology during the reductive division of {\sl E. coli}.
Thanks to the EMPS, we recorded the time-dependent distribution of cell size fluctuations and revealed that the rescaled distribution is scale-invariant and robust against the abrupt environmental change, despite the decrease of the mean cell size (e.g., \figref{fig-2}).
We confirmed the robustness of the result against different combinations of the growth and non-nutritious media, while we also found that the shape of the rescaled distribution does depend on the choice of the growth medium before the switch (\figref{fig-3}).
Moreover, those findings were successfully reproduced by simulations of a model based on the CH model \cite{Ho2015, Witz2019}, which we propose as an extension for dealing with time-dependent environments (\figref{fig-4}).
We further inspected theoretical mechanism behind this scale invariance and found the significance of the division rate function $B(v,t)$.
We obtained a sufficient condition for the scale invariance, \eqref{eq:separability}, which was indeed confirmed in our numerical data.
Finally, we numerically clarified the range of validity of the scale invariance during the reductive division, showing that the state of the multifork replications may play a crucial role (\figref{fig-5}).

Notably, our experiments (on the growth condition dependence) and simulations suggest that the scale invariance breaks down for slow starvation.
Further investigations of cell size fluctuations in such cases, both experimentally and theoretically, will be an important step toward clarifying what determines the critical time scale of environmental changes for the violation of the scale invariance.
Elucidating the $(\tau_\lambda, \tau_\mu)$ phase diagram is particularly important, because it may serve as a further test of the two cell cycle models used here, which predicted significantly different diagrams (\figref{fig-5}c and Supplementary~Fig.~11b).
It is a challenge experimentally, but may also be possible with EMPS, by combining a technique to control the progression speed of the C+D period, such as the one developed in \cite{SI2017}.

It is also worth noting that the cell size distribution we measured is that of the entire population, which is given by $p(v) = \int p(v|a) p_\text{age}(a) da$ with the size distribution $p(v|a)$ of cells at a given age $a$ and the age distribution $p_\text{age}(a)$ of the population.
Since those distributions have also been studied in the past for steady conditions (e.g., \cite{Iyer-Biswas2014pnas, Kennard2016} for $p(v|a)$, \cite{Hashimoto2016} for $p_\text{age}(a)$), it is an important future work to understand how these distributions change under time-dependent conditions and how they contribute to the scale invariance.
It is also important to understand the dependence on the population size, which will be a crucial point to consider an analogous experiment in the mother machine.

In the context of possible follow-up experiments using the mother machine, another aspect that deserves attention is the way nutrients are delivered to cells and removed.
As we have mentioned, EMPS relies on diffusion of molecules and is therefore prone to have a slight amount of residual nutrients in the observation area.
While this may better correspond to natural conditions, in which the surrounding medium is not necessarily replaced by a strong flow as in the mother machine \cite{Arnoldini2014, Kaiser2018, Julou2020, MarcoCosentino2020, Bakshi2021}, we cannot exclude the possibility that the slightly remaining nutrients might affect the cellular state after starvation in EMPS.
Therefore, it would be interesting to investigate whether the scale invariance, which our study has shown to be robust in various starvation conditions in EMPS and in models, can also be verified in the mother machine.

After all, our results backed by the cell cycle model suggest that mechanism of intracellular replication processes may have direct impact on the scale-invariant distribution, which may account for the significant difference we identified between bacteria and eukaryotes (\figref{fig-3}d).
Since the number of species studied in each taxonomic domain is rather limited ({\sl E. coli} (this work) and {\sl B. subtilis} \cite{Wakita2010} for bacteria, 13 protist species for eukaryotes \cite{Giometto2013}), it is of crucial importance to test the distribution trend further in each taxonomic domain, and to clarify how and to what extent the cell size distribution is determined by the intracellular replication dynamics.
The influence of cell-to-cell interactions, e.g., quorum sensing \cite{Bruger2016, Haeaar2018}, may also exist.
Theoretical approaches, such as models considering the cellular age \cite{Grilli2017}, knowledge from the universal protein number fluctuations \cite{Furusawa2005, Salman2012, Brenner2015}, and renormalization group approaches for living cell tissues \cite{Rulands2018}, may also be useful.
We hope that our understanding of the population-level response against nutrient starvation will be further refined by future experimental and theoretical investigations.

\section{Methods}
\subsection{Strains and culture media}
We used wild-type {\sl E. coli} strains (MG1655 and RP437) and a mutant strain (W3110 $\Delta$fliC $\Delta$flu $\Delta$fimA)
 in this study.
Culture media and buffer are listed in Supplementary~Table~1.
The osmotic pressure of each medium was measured by the freezing-point depression method
 by the OSMOMAT 030 (Genotec, Berlin Germany).
Details on the strains and culture conditions in each experiment
 are provided below (see also Supplementary~Note~1).

\subsection{Fabrication of the EMPS}
The EMPS consists of a microfabricated glass coverslip,
a bilayer porous membrane and a PDMS pad.
The microfabricated coverslip and the PDMS pad were prepared
according to ref. \cite{Inoue2001, Hashimoto2016}.
We fabricated the bilayer porous membrane by combining a streptavidin decorated cellulose membrane
and a biotin decorated polyethylene-terephthalate (PET) membrane.
The streptavidin decoration of the cellulose membrane (Spectra/Por 7, Repligen, Waltham Massachusetts, molecular weight cut-off $25000$) was realized by the method described in ref. \cite{Inoue2001, Hashimoto2016}.
The PET membrane (Transwell 3450, Corning, Corning New York, nominal pore size $0.4\unit{\mu{}m}$) was decorated with biotin as follows.
We soaked a PET membrane in 1 wt\% solution of
3-(2-aminoethyl aminopropyl) trimethoxysilane (Shinetsu Kagaku Kogyo, Tokyo Japan) for 45 min,
dried it at 125${}^\circ$C for 25 min and washed it by ultrasonic cleaning in Milli-Q water for 5 min.
This preprocessed PET membrane was stored in a desiccator at room temperature, until it was used to assemble the EMPS.

The EMPS was assembled as follows.
The preprocessed PET membrane was cut into $5\unit{mm} \times 5\unit{mm}$ squares, soaked in the biotin solution for 4 hours and dried on filter paper.
The biotin decorated PET membrane was attached with a streptavidin decorated cellulose membrane, cut to the size of the PET membrane, by sandwiching them between agar pads (M9 medium with $2\unit{wt\%}$ agarose).
In the meantime, a $1\unit{\mu{}l}$ droplet of bacterial suspension was inoculated on a biotin decorated coverslip
(see also details below).
We then took the bilayer membrane from the agar pad, air-dried for tens of seconds,
and carefully put on the coverslip on top of the bacterial suspension.
The bilayer membrane was then attached to the coverslip via streptavidin-biotin binding as shown in Supplementary~Fig.~1b.
We then air-dried the membrane for a minute and attached a PDMS pad on the coverslip by a double-sided tape.

\subsection{Observation of the bacterial reductive division}
We used a wild-type strain MG1655.
Before the time-lapse observation,
 we inoculated the strain from a glycerol stock into $2\unit{ml}$ growth medium in a test tube.
The same medium as for the main observation was used (LB broth, M9(Glc+a.a.) or M9(Glc)).
After shaking it overnight at $37\degc$, we transferred $20\unit{\mu{}l}$ of the incubated suspension to $2\unit{ml}$ fresh medium and cultured it until the OD at $600\unit{nm}$ wavelength reached $0.1$-$0.5$.
The bacterial suspension was finally diluted to $\mathrm{OD}=0.05$ before it was inoculated on the coverslip.

For this experiment, we used a substrate with wells of $55\unit{\mu{}m}$ diameter and $0.8\unit{\mu{}m}$ depth.
The well diameter was chosen so that all cells in the well can be recorded.
The device was placed on the microscope stage, in the incubation box maintained at $37\degc$.
The microscope we used was Leica DMi8, equipped with a 100x (N.A. 1.30) oil immersion objective and operated by Leica LasX.
To fill the device with growth medium, we injected fresh medium stored at $37\degc$ from the inlet (Supplementary~Fig.~1),
 at the rate of $60\unit{ml~hr^{-1}}$ for $5\unit{min}$ by a syringe pump (NE-1000, New Era Pump Systems).

In the beginning of the observation, growth medium was constantly supplied at the rate of $2\unit{ml~hr^{-1}}$
 (flow speed approximately $0.2\unit{mm~sec^{-1}}$ above the membrane).
When a microcolony composed of approximately 100 cells appeared, 
 we quickly switched the medium to a non-nutritious buffer (PBS or M9 medium with $\alpha$-methyl-D-glucoside ($\alpha$MG), see Supplementary~Table~1) stored at $37\degc$, by exchanging the syringe.
The flow rate was set to be $60\unit{ml~hr^{-1}}$ for the first 5 minutes, then returned to $2\unit{ml~hr^{-1}}$.
Throughout the experiment, the device and the media were always in the microscope incubation box, maintained at $37\degc$.
Cells were observed by phase contrast microscopy and recorded at the time interval of $5\unit{min}$.
The data for obtaining each distribution are taken from several wells (stated in the figure captions) in a single experiment.

The cell volumes were evaluated as follows.
We determined the major axis and the minor axis of each cell, manually, by using a painting software.
By measuring the axis lengths, we obtained the set of the length $L_i$ and the width $w_i$ for all cells (indexed by $i$).
We estimated the uncertainty in manual segmentation at $\pm 0.15\unit{\mu{}m}$.
However, the measurement of the individual cell widths is less accurate than that of the lengths, essentially because the width depends on the choice of the section of the cell.
To estimate the cell volume, therefore, we neglected the fluctuation of the width among the cells as follows.
We measured the width $w_i$ at the center of each cell and took the ensemble average $\expct{w_i}$.
Together with the cell length $L_i$, we obtained the volume of this cell, $v_i$, by $v_i = \frac{4\pi}{3}\( \frac{\expct{w_i}}{2}\)^3 + \pi\( \frac{\expct{w_i}}{2}\)^2(L_i-\expct{w_i})$.
Note that the scale invariance holds for the length distribution as well (Supplementary~Fig.~6), which suggests that neglecting the width fluctuation does not affect the main finding of the paper.

Finally, let us note that there may be some technical limitation specific to the combination of the LB medium and the EMPS.
When we used LB medium in the growth phase, the bacteria continued growing, albeit very slowly, even long time after the medium was switched to PBS (see Supplementary~Movie~7), while they stopped growing completely in all cases where we used chemically defined medium before the switch (Supplementary~Fig.~4 and Supplementary~Movie~8-11).
This may be because some nutrient molecules specific to LB might remain on the well surface or inside the membrane.
However, since we are focusing on the earlier stage of the starvation process, in which the typical cell sizes change most significantly, we believe that this remaining slow growth in the case LB $\to$ PBS does not affect our main results.
More quantitatively, from Supplementary~Fig.~9a, we can evaluate the rate of this remaining cell growth observed in the case LB $\to$ PBS to be nearly $10^{-4}\unit{min^{-1}}$ or eventually even less. Because the corresponding time scale $\gtrsim 10^4\unit{min}$ is much longer than the time scales relevant to the scale invariance we found, which are around $100\unit{min}$, there is a clear scale separation, from which we can expect that the remaining slow cell growth will not affect our main finding.
We also noticed relatively poor reproducibility of experiments using LB medium (Supplementary~Fig.~7), which may be attributed to its chemical undefinedness \cite{Sezonov2007}, since the experiments using defined media recorded good reproducibility (see the same figure).

\begin{table*}[t]\centering
    \caption{
    \label{tb-1}
    Parameters used for the simulations.}
    \begin{tabular}{lc|rr}
        & Parameters & LB$\to$PBS & M9(Glc+a.a.)$\to$PBS \\
       \hline
       Parameters on & $\lambda_0$ & $0.029$ $\mathrm{min^{-1}}$ & $0.010$ $\mathrm{min^{-1}}$ \tabularnewline
       the exponential & $\mu_0^{-1}$ & $1.3\lambda_0^{-0.84}+42\simeq67$ $\mathrm{min}$ & $1.3\lambda_0^{-0.84}+42\simeq104$ $\mathrm{min}$  \tabularnewline
       growth phase & $v_0$ & $4.9$ $\mu\mathrm{m}^3$ & $1.8$ $\mu\mathrm{m}^3$  \tabularnewline
        & $\delta_\mean$ or $\vth_\mean$ & $\delta_\mean=0.275$ $\mu\mathrm{m}^3$ & $\delta_\mean=0.25$ $\mu\mathrm{m}^3$ \tabularnewline
        & $\delta_\std$ or $\vth_\std$ & $0.1\times \delta_\mean=0.025$ $\mu\mathrm{m}^3$ & $0.1\times \delta_\mean=0.0225$ $\mu\mathrm{m}^3$ \tabularnewline
        & $X^\mathrm{CD,th}_\mathrm{std}$ & $0.05\times\expct{X^\mathrm{CD,th}_i}=0.05$ & $0.05\times\expct{X^\mathrm{CD,th}_i}=0.05$ \tabularnewline
        & $x^\mathrm{sep}_\mathrm{std}$ & $0.0325$ & $0.0325$ \tabularnewline
        \hline
       Time-dependent rates & $\lambda(t)=\lambda_0\dfrac{1-A}{e^{ct}-A}$ & $A=0.93$, $c=0.0059$ $\mathrm{min^{-1}}$ & $A=0.84$, $c=0.011$ $\mathrm{min^{-1}}$ \tabularnewline
        & $\mu_i(t)=\dfrac{\mu_0}{v_0}\dfrac{k+1}{ke^{t/\tau}+1}v_i$ & $k=0.01$, $\tau=40$ $\mathrm{min}$ & $k=0.01$, $\tau=16$ $\mathrm{min}$ \tabularnewline
       \hline
       \end{tabular}
\end{table*}

\subsection{Simulation}
The parameters used in the simulations were evaluated as follows. 
First, from the observations of the exponential growth phase, we determined the growth rate $\lambda_0$ and the mean cell size $v_0$ directly.
This allowed us to set the cycle progression speed $\mu_0$ too, by using the relation $\mu_0^{-1}\simeq (1.3\lambda_0^{-0.84}+42)$ proposed by Wallden \textit{et al.} \cite{Wallden2016} (the values of $\lambda_0$ and $\mu_0$ in the unit of $\mathrm{min^{-1}}$ are used here).
Concerning the volume threshold for initiating the replication, we found such a value of $\delta_\mean$ (or $\vth_\mean$) that reproduced the experimentally observed mean cell volume in the growth phase.
The standard deviation $\delta_\std$ (or $\vth_\std$) was set to be $10\%$ of the mean $\delta_\mean$ ($\vth_\mean$), based on the relation on the initiation volume found by Wallden \textit{et al.} \cite{Wallden2016}.
They also measured the fluctuations of the time length of the C+D period; this led us to estimate $\XCDth_\std$ at 5\% of $\expct{\XCDth}$, i.e., $\XCDth_\std=0.05$.
On the septum positions, we measured their fluctuations and found little difference in $x^\mathrm{sep}_\std$ among the different growth conditions we used, and also in the non-nutritious case (Supplementary~Fig.~8).
We therefore used a single value $x^\mathrm{sep}_\std=0.0325$ for all simulations.
Note that, without this stochastic asymmetric division, the cell size distribution exhibited periodic oscillations, presumably because cellular states between siblings were strongly correlated then.

In the following, we describe how the remaining parameters were evaluated and how the simulations were carried out for each set of the simulations presented in this work.

\subsubsection{Methods for the results that reproduced the experimental observations}

We evaluated the time-dependent rates $\lambda(t)$ and $\mu_i(t)$ as follows.
The growth rate $\lambda(t)$ can be determined independently of the cell divisions, because the total volume $V_\mathrm{tot}(t) = \sum_i v_i(t)$ grows as $V_\mathrm{tot}(t) = V_\mathrm{tot}(0) \exp(\int^t_0 \lambda(t)dt)$.
With $\lambda(t)$ given by \eqref{eq:growthrate}, we compared $V_\mathrm{tot}(t)$ with experimental data and determined the values of $A$ and $c$ (\figref{fig-3}c).
Finally, only $k$ and $\tau$ in \eqref{eq:mu} remained as free parameters.
We tuned them so that the mean cell volume $V(t)$ and the number of the cells $n(t)$ observed in the simulations reproduced those from the experiments (\figref{fig-3}d).
The parameter values determined thereby are summarized in \tbref{tb-1}, for the simulations for LB$\to$PBS and M9(Glc+a.a.)$\to$PBS.

We started the simulations from 10 cells with
 volumes in the range of $0.07$-$1.13\unit{\mu{}m^3}$, randomly generated from the uniform distribution.
The cells grew in the exponential phase
 (with the constant growth rate $\lambda_0$ and the cycle progression speed $\mu_i = (\mu_0/v_0)v_i$)
 until the number of cells reached 100,000.
We then randomly picked up 10 cells from this ``precultured'' sample and grew them until the number of cells exceeded 500.
Those cells were kept growing for 1000 minutes to sufficiently mix cell cycle progressions in the population.
During this process, we kept the number of cells constant, by eliminating one of the daughter cells after each division.
We then used them as the initial population of each simulation.
To precisely compare the number of cells in simulations with the experimentally obtained population $n_\mathrm{exp}(t)$ (\figref{fig-2}c and Supplementary~Fig.~4b),
 the numerically obtained population $n_\mathrm{sim}(t)$ is rescaled by multiplying $n_\mathrm{exp}(0)/n_\mathrm{sim}(0)$. 

\subsubsection{Methods for the results on violation of the scale invariance}

The functional forms of $\lambda(t)$ and $\mu_i(t)$ were given by Eqs.~\pref{eq:lamdecrease} and \pref{eq:mudecrease}, with variable parameters $\tau_\lambda$ and $\tau_\mu$.
For $\mu_0$, we determined it from $\lambda_0$ using the empirical relation reported by Wallden \textit{et al.} \cite{Wallden2016}.
For $v_0$, we set its value self-consistently, so that the mean cell volume $\expct{v_i(0)}$ obtained numerically in the exponential phase with $\mu_i = (\mu_0/v_0)v_i$ falls within 1\% error from the given value of $v_0$.
As a result, we obtained $v_0 = 1.9,2.5,3.2,4.0,5.0$ $\unit{\mu{}m}^3$ for $\lambda = 0.01, 0.015, 0.02, 0.025, 0.03\unit{min^{-1}}$, respectively. 
These values satisfy the growth law, i.e., the mean cell size increases exponentially with the growth rate \cite{TAHERIARAGHI2015}.
The other parameters were fixed at $\delta_\mean = 0.25 \unit{\mu{}m^3}$, $\delta_\std = 0.025 \unit{\mu m^3}$, $\XCDth_\std = 0.05$, and $x^\mathrm{sep}_\std = 0.0325$.
We started the simulations from 50 cells with
 volumes in the range of $0.07$-$1.13\unit{\mu{}m^3}$, randomly generated from the uniform distribution.
The cells grew in the exponential phase until the number of cells reached 500,000.
We then randomly picked up 50 cells and grew them until the number of cells exceeded 50,000.
Those cells were kept growing for 1,000 minutes to sufficiently mix cell cycle progressions in the population, with the number of cells kept constant by eliminating one of the daughter cells produced by division.
Using them as the initial population (at $t=0$), we started the simulations for $t \geq 0$, with the number of cells still kept constant by the same method.
Strictly, this situation with a constant number of cells is different from the experimental setting, but we confirmed that this change did not influence the validity of the scale invariance and had only a minor effect on the value of CV, at least for the situation shown in Fig.~4c. Therefore, for the results on the violation of the scale invariance (Fig.~5), we carried out simulations with a constant number of cells as described above, to reduce the computation time.

\section*{Acknowledgments}
\begin{acknowledgments}
We are grateful to Y. Himeoka for motivating us to compare the abrupt and quasi-static changes, which led to the results presented in \figref{fig-5}.
We also acknowledge useful discussions with H. Chat\'{e}, Y. Furuta, T. Hiraiwa, Y. Kitahara, H. Nakaoka, and D. Nishiguchi.
We thank I. Naguro for letting us use the OSMOMAT 030.
This work is supported by KAKENHI from Japan Society for the Promotion of Science (JSPS) (No. 16H04033, No. 19H05800), a Grant-in-Aid for JSPS Fellows (No. 20J10682) and by the grants associated with the ``Planting Seeds for Research'' program and Suematsu Award from Tokyo Tech.
\end{acknowledgments}

\section*{Competing interests}
The authors declare that no competing interests exist.

\section*{Author contributions}
T.S. and K.A.T. designed research.
T.S., R.O., Y.W., and K.A.T. developed the extensive microperfusion system.
T.S. performed all bacterial experiments and analyzed data.
T.S. and K.A.T. did the modeling, and T.S. wrote the codes for the simulations.
T.S. performed the theoretical calculations.
T.S. and K.A.T. wrote the manuscript, and all authors worked for revision.

\section*{Data availability}
The data that support the findings of this study are available at \url{https://github.com/shimasaan/bacterial_rd}.
\section*{Code availability}
The codes used in this study are available at \url{https://github.com/shimasaan/bacterial_rd}.

\section*{References}
\bibliography{EMPSpaper}

\providecommand{\noopsort}[1]{}\providecommand{\singleletter}[1]{#1}%
\begin{thebibliography}{65}%
\makeatletter
\providecommand \@ifxundefined [1]{%
 \@ifx{#1\undefined}
}%
\providecommand \@ifnum [1]{%
 \ifnum #1\expandafter \@firstoftwo
 \else \expandafter \@secondoftwo
 \fi
}%
\providecommand \@ifx [1]{%
 \ifx #1\expandafter \@firstoftwo
 \else \expandafter \@secondoftwo
 \fi
}%
\providecommand \natexlab [1]{#1}%
\providecommand \enquote  [1]{``#1''}%
\providecommand \bibnamefont  [1]{#1}%
\providecommand \bibfnamefont [1]{#1}%
\providecommand \citenamefont [1]{#1}%
\providecommand \href@noop [0]{\@secondoftwo}%
\providecommand \href [0]{\begingroup \@sanitize@url \@href}%
\providecommand \@href[1]{\@@startlink{#1}\@@href}%
\providecommand \@@href[1]{\endgroup#1\@@endlink}%
\providecommand \@sanitize@url [0]{\catcode `\\12\catcode `\$12\catcode
  `\&12\catcode `\#12\catcode `\^12\catcode `\_12\catcode `\%12\relax}%
\providecommand \@@startlink[1]{}%
\providecommand \@@endlink[0]{}%
\providecommand \url  [0]{\begingroup\@sanitize@url \@url }%
\providecommand \@url [1]{\endgroup\@href {#1}{\urlprefix }}%
\providecommand \urlprefix  [0]{URL }%
\providecommand \Eprint [0]{\href }%
\providecommand \doibase [0]{http://dx.doi.org/}%
\providecommand \selectlanguage [0]{\@gobble}%
\providecommand \bibinfo  [0]{\@secondoftwo}%
\providecommand \bibfield  [0]{\@secondoftwo}%
\providecommand \translation [1]{[#1]}%
\providecommand \BibitemOpen [0]{}%
\providecommand \bibitemStop [0]{}%
\providecommand \bibitemNoStop [0]{.\EOS\space}%
\providecommand \EOS [0]{\spacefactor3000\relax}%
\providecommand \BibitemShut  [1]{\csname bibitem#1\endcsname}%
\let\auto@bib@innerbib\@empty
\bibitem [{\citenamefont {Giometto}\ \emph {et~al.}(2013)\citenamefont
  {Giometto}, \citenamefont {Altermatt}, \citenamefont {Carrara}, \citenamefont
  {Maritan},\ and\ \citenamefont {Rinaldo}}]{Giometto2013}%
  \BibitemOpen
  \bibfield  {author} {\bibinfo {author} {\bibfnamefont {A.}~\bibnamefont
  {Giometto}}, \bibinfo {author} {\bibfnamefont {F.}~\bibnamefont {Altermatt}},
  \bibinfo {author} {\bibfnamefont {F.}~\bibnamefont {Carrara}}, \bibinfo
  {author} {\bibfnamefont {A.}~\bibnamefont {Maritan}}, \ and\ \bibinfo
  {author} {\bibfnamefont {A.}~\bibnamefont {Rinaldo}},\ }\bibfield  {title}
  {\enquote {\bibinfo {title} {Scaling body size fluctuations},}\ }\href
  {\doibase 10.1073/pnas.1301552110} {\bibfield  {journal} {\bibinfo  {journal}
  {Proc. Natl. Acad. Sci. USA}\ }\textbf {\bibinfo {volume} {110}},\ \bibinfo
  {pages} {4646--4650} (\bibinfo {year} {2013})}\BibitemShut {NoStop}%
\bibitem [{\citenamefont {Zaoli}\ \emph {et~al.}(2019)\citenamefont {Zaoli},
  \citenamefont {Giometto}, \citenamefont {Mara{\~n}{\'o}n}, \citenamefont
  {Escrig}, \citenamefont {Meibom}, \citenamefont {Ahluwalia}, \citenamefont
  {Stocker}, \citenamefont {Maritan},\ and\ \citenamefont
  {Rinaldo}}]{Zaoli2019}%
  \BibitemOpen
  \bibfield  {author} {\bibinfo {author} {\bibfnamefont {S.}~\bibnamefont
  {Zaoli}}, \bibinfo {author} {\bibfnamefont {A.}~\bibnamefont {Giometto}},
  \bibinfo {author} {\bibfnamefont {E.}~\bibnamefont {Mara{\~n}{\'o}n}},
  \bibinfo {author} {\bibfnamefont {S.}~\bibnamefont {Escrig}}, \bibinfo
  {author} {\bibfnamefont {A.}~\bibnamefont {Meibom}}, \bibinfo {author}
  {\bibfnamefont {A.}~\bibnamefont {Ahluwalia}}, \bibinfo {author}
  {\bibfnamefont {R.}~\bibnamefont {Stocker}}, \bibinfo {author} {\bibfnamefont
  {A.}~\bibnamefont {Maritan}}, \ and\ \bibinfo {author} {\bibfnamefont
  {A.}~\bibnamefont {Rinaldo}},\ }\bibfield  {title} {\enquote {\bibinfo
  {title} {Generalized size scaling of metabolic rates based on single-cell
  measurements with freshwater phytoplankton},}\ }\href {\doibase
  10.1073/pnas.1906762116} {\bibfield  {journal} {\bibinfo  {journal} {Proc.
  Natl. Acad. Sci. USA}\ }\textbf {\bibinfo {volume} {116}},\ \bibinfo {pages}
  {17323--17329} (\bibinfo {year} {2019})}\BibitemShut {NoStop}%
\bibitem [{\citenamefont {Camacho}\ and\ \citenamefont
  {Sol\'e}(2001)}]{Camacho2001}%
  \BibitemOpen
  \bibfield  {author} {\bibinfo {author} {\bibfnamefont {J.}~\bibnamefont
  {Camacho}}\ and\ \bibinfo {author} {\bibfnamefont {R.~V.}\ \bibnamefont
  {Sol\'e}},\ }\bibfield  {title} {\enquote {\bibinfo {title} {Scaling in
  ecological size spectra},}\ }\href {\doibase 10.1209/epl/i2001-00347-0}
  {\bibfield  {journal} {\bibinfo  {journal} {EPL}\ }\textbf {\bibinfo {volume}
  {55}},\ \bibinfo {pages} {774--780} (\bibinfo {year} {2001})}\BibitemShut
  {NoStop}%
\bibitem [{\citenamefont {Marquet}\ \emph {et~al.}(2005)\citenamefont
  {Marquet}, \citenamefont {Qui{\~n}ones}, \citenamefont {Abades},
  \citenamefont {Labra}, \citenamefont {Tognelli}, \citenamefont {Arim},\ and\
  \citenamefont {Rivadeneira}}]{Marquet2005}%
  \BibitemOpen
  \bibfield  {author} {\bibinfo {author} {\bibfnamefont {P.~A.}\ \bibnamefont
  {Marquet}}, \bibinfo {author} {\bibfnamefont {R.~A.}\ \bibnamefont
  {Qui{\~n}ones}}, \bibinfo {author} {\bibfnamefont {S.}~\bibnamefont
  {Abades}}, \bibinfo {author} {\bibfnamefont {F.}~\bibnamefont {Labra}},
  \bibinfo {author} {\bibfnamefont {M.}~\bibnamefont {Tognelli}}, \bibinfo
  {author} {\bibfnamefont {M.}~\bibnamefont {Arim}}, \ and\ \bibinfo {author}
  {\bibfnamefont {M.}~\bibnamefont {Rivadeneira}},\ }\bibfield  {title}
  {\enquote {\bibinfo {title} {Scaling and power-laws in ecological systems},}\
  }\href {\doibase 10.1242/jeb.01588} {\bibfield  {journal} {\bibinfo
  {journal} {J. Exp. Biol.}\ }\textbf {\bibinfo {volume} {208}},\ \bibinfo
  {pages} {1749--1769} (\bibinfo {year} {2005})}\BibitemShut {NoStop}%
\bibitem [{\citenamefont {Iyer-Biswas}\ \emph
  {et~al.}(2014{\natexlab{a}})\citenamefont {Iyer-Biswas}, \citenamefont
  {Wright}, \citenamefont {Henry}, \citenamefont {Lo}, \citenamefont {Burov},
  \citenamefont {Lin}, \citenamefont {Crooks}, \citenamefont {Crosson},
  \citenamefont {Dinner},\ and\ \citenamefont {Scherer}}]{Iyer-Biswas2014pnas}%
  \BibitemOpen
  \bibfield  {author} {\bibinfo {author} {\bibfnamefont {S.}~\bibnamefont
  {Iyer-Biswas}}, \bibinfo {author} {\bibfnamefont {C.~S.}\ \bibnamefont
  {Wright}}, \bibinfo {author} {\bibfnamefont {J.~T.}\ \bibnamefont {Henry}},
  \bibinfo {author} {\bibfnamefont {K.}~\bibnamefont {Lo}}, \bibinfo {author}
  {\bibfnamefont {S.}~\bibnamefont {Burov}}, \bibinfo {author} {\bibfnamefont
  {Y.}~\bibnamefont {Lin}}, \bibinfo {author} {\bibfnamefont {G.~E.}\
  \bibnamefont {Crooks}}, \bibinfo {author} {\bibfnamefont {S.}~\bibnamefont
  {Crosson}}, \bibinfo {author} {\bibfnamefont {A.~R.}\ \bibnamefont {Dinner}},
  \ and\ \bibinfo {author} {\bibfnamefont {N.~F.}\ \bibnamefont {Scherer}},\
  }\bibfield  {title} {\enquote {\bibinfo {title} {Scaling laws governing
  stochastic growth and division of single bacterial cells},}\ }\href {\doibase
  10.1073/pnas.1403232111} {\bibfield  {journal} {\bibinfo  {journal} {Proc.
  Natl. Acad. Sci. USA}\ }\textbf {\bibinfo {volume} {111}},\ \bibinfo {pages}
  {15912--15917} (\bibinfo {year} {2014}{\natexlab{a}})}\BibitemShut {NoStop}%
\bibitem [{\citenamefont {Kennard}\ \emph {et~al.}(2016)\citenamefont
  {Kennard}, \citenamefont {Osella}, \citenamefont {Javer}, \citenamefont
  {Grilli}, \citenamefont {Nghe}, \citenamefont {Tans}, \citenamefont
  {Cicuta},\ and\ \citenamefont {Cosentino~Lagomarsino}}]{Kennard2016}%
  \BibitemOpen
  \bibfield  {author} {\bibinfo {author} {\bibfnamefont {A.~S.}\ \bibnamefont
  {Kennard}}, \bibinfo {author} {\bibfnamefont {M.}~\bibnamefont {Osella}},
  \bibinfo {author} {\bibfnamefont {A.}~\bibnamefont {Javer}}, \bibinfo
  {author} {\bibfnamefont {J.}~\bibnamefont {Grilli}}, \bibinfo {author}
  {\bibfnamefont {P.}~\bibnamefont {Nghe}}, \bibinfo {author} {\bibfnamefont
  {S.~J.}\ \bibnamefont {Tans}}, \bibinfo {author} {\bibfnamefont
  {P.}~\bibnamefont {Cicuta}}, \ and\ \bibinfo {author} {\bibfnamefont
  {M.}~\bibnamefont {Cosentino~Lagomarsino}},\ }\bibfield  {title} {\enquote
  {\bibinfo {title} {Individuality and universality in the growth-division laws
  of single \textit{E. coli} cells},}\ }\href {\doibase
  10.1103/PhysRevE.93.012408} {\bibfield  {journal} {\bibinfo  {journal} {Phys.
  Rev. E}\ }\textbf {\bibinfo {volume} {93}},\ \bibinfo {pages} {012408}
  (\bibinfo {year} {2016})}\BibitemShut {NoStop}%
\bibitem [{\citenamefont {Iyer-Biswas}\ \emph
  {et~al.}(2014{\natexlab{b}})\citenamefont {Iyer-Biswas}, \citenamefont
  {Crooks}, \citenamefont {Scherer},\ and\ \citenamefont
  {Dinner}}]{Iyer-Biswas2014prl}%
  \BibitemOpen
  \bibfield  {author} {\bibinfo {author} {\bibfnamefont {S.}~\bibnamefont
  {Iyer-Biswas}}, \bibinfo {author} {\bibfnamefont {G.~E.}\ \bibnamefont
  {Crooks}}, \bibinfo {author} {\bibfnamefont {N.~F.}\ \bibnamefont {Scherer}},
  \ and\ \bibinfo {author} {\bibfnamefont {A.~R.}\ \bibnamefont {Dinner}},\
  }\bibfield  {title} {\enquote {\bibinfo {title} {Universality in stochastic
  exponential growth},}\ }\href {\doibase 10.1103/PhysRevLett.113.028101}
  {\bibfield  {journal} {\bibinfo  {journal} {Phys. Rev. Lett.}\ }\textbf
  {\bibinfo {volume} {113}},\ \bibinfo {pages} {028101} (\bibinfo {year}
  {2014}{\natexlab{b}})}\BibitemShut {NoStop}%
\bibitem [{\citenamefont {Amir}(2014)}]{Amir2014}%
  \BibitemOpen
  \bibfield  {author} {\bibinfo {author} {\bibfnamefont {A}~\bibnamefont
  {Amir}},\ }\bibfield  {title} {\enquote {\bibinfo {title} {Cell size
  regulation in bacteria},}\ }\href {\doibase 10.1103/PhysRevLett.112.208102}
  {\bibfield  {journal} {\bibinfo  {journal} {Phys. Rev. Lett.}\ }\textbf
  {\bibinfo {volume} {112}},\ \bibinfo {pages} {208102} (\bibinfo {year}
  {2014})}\BibitemShut {NoStop}%
\bibitem [{\citenamefont {Ho}\ \emph {et~al.}(2018)\citenamefont {Ho},
  \citenamefont {Lin},\ and\ \citenamefont {Amir}}]{Ho2018}%
  \BibitemOpen
  \bibfield  {author} {\bibinfo {author} {\bibfnamefont {P.}~\bibnamefont
  {Ho}}, \bibinfo {author} {\bibfnamefont {J.}~\bibnamefont {Lin}}, \ and\
  \bibinfo {author} {\bibfnamefont {A.}~\bibnamefont {Amir}},\ }\bibfield
  {title} {\enquote {\bibinfo {title} {Modeling cell size regulation: From
  single-cell-level statistics to molecular mechanisms and population-level
  effects},}\ }\href {\doibase 10.1146/annurev-biophys-070317-032955}
  {\bibfield  {journal} {\bibinfo  {journal} {Annu. Rev. Biophys.}\ }\textbf
  {\bibinfo {volume} {47}},\ \bibinfo {pages} {251--271} (\bibinfo {year}
  {2018})}\BibitemShut {NoStop}%
\bibitem [{\citenamefont {Jun}\ \emph {et~al.}(2018)\citenamefont {Jun},
  \citenamefont {Si}, \citenamefont {Pugatch},\ and\ \citenamefont
  {Scott}}]{Jun2018}%
  \BibitemOpen
  \bibfield  {author} {\bibinfo {author} {\bibfnamefont {S.}~\bibnamefont
  {Jun}}, \bibinfo {author} {\bibfnamefont {F.}~\bibnamefont {Si}}, \bibinfo
  {author} {\bibfnamefont {R.}~\bibnamefont {Pugatch}}, \ and\ \bibinfo
  {author} {\bibfnamefont {M.}~\bibnamefont {Scott}},\ }\bibfield  {title}
  {\enquote {\bibinfo {title} {Fundamental principles in bacterial
  physiology-history, recent progress, and the future with focus on cell size
  control: a review},}\ }\href {\doibase 10.1088/1361-6633/aaa628} {\bibfield
  {journal} {\bibinfo  {journal} {Rep. Prog. Phys.}\ }\textbf {\bibinfo
  {volume} {81}},\ \bibinfo {pages} {056601} (\bibinfo {year}
  {2018})}\BibitemShut {NoStop}%
\bibitem [{\citenamefont {Cadart}\ \emph {et~al.}(2019)\citenamefont {Cadart},
  \citenamefont {Venkova}, \citenamefont {Recho}, \citenamefont {Lagomarsino},\
  and\ \citenamefont {Piel}}]{Cadart2019}%
  \BibitemOpen
  \bibfield  {author} {\bibinfo {author} {\bibfnamefont {C.}~\bibnamefont
  {Cadart}}, \bibinfo {author} {\bibfnamefont {L.}~\bibnamefont {Venkova}},
  \bibinfo {author} {\bibfnamefont {P.}~\bibnamefont {Recho}}, \bibinfo
  {author} {\bibfnamefont {M.~C.}\ \bibnamefont {Lagomarsino}}, \ and\ \bibinfo
  {author} {\bibfnamefont {M.}~\bibnamefont {Piel}},\ }\bibfield  {title}
  {\enquote {\bibinfo {title} {{The physics of cell-size regulation across
  timescales}},}\ }\href {\doibase 10.1038/s41567-019-0629-y} {\bibfield
  {journal} {\bibinfo  {journal} {Nat. Phys.}\ }\textbf {\bibinfo {volume}
  {15}},\ \bibinfo {pages} {993--1004} (\bibinfo {year} {2019})}\BibitemShut
  {NoStop}%
\bibitem [{\citenamefont {Nystr\"{o}m}(2004)}]{Nystrom2004}%
  \BibitemOpen
  \bibfield  {author} {\bibinfo {author} {\bibfnamefont {T.}~\bibnamefont
  {Nystr\"{o}m}},\ }\bibfield  {title} {\enquote {\bibinfo {title}
  {Stationary-phase physiology},}\ }\href {\doibase
  10.1146/annurev.micro.58.030603.123818} {\bibfield  {journal} {\bibinfo
  {journal} {Annu. Rev. Microbiol.}\ }\textbf {\bibinfo {volume} {58}},\
  \bibinfo {pages} {161--181} (\bibinfo {year} {2004})}\BibitemShut {NoStop}%
\bibitem [{\citenamefont {Kaprelyants}\ and\ \citenamefont
  {Kell}(1993)}]{Kaprelyants1993}%
  \BibitemOpen
  \bibfield  {author} {\bibinfo {author} {\bibfnamefont {A.~S.}\ \bibnamefont
  {Kaprelyants}}\ and\ \bibinfo {author} {\bibfnamefont {D.~B.}\ \bibnamefont
  {Kell}},\ }\bibfield  {title} {\enquote {\bibinfo {title} {Dormancy in
  stationary-phase cultures of micrococcus luteus: Flow cytometric analysis of
  starvation and resuscitation},}\ }\href {\doibase
  10.1128/aem.59.10.3187-3196.1993} {\bibfield  {journal} {\bibinfo  {journal}
  {Appl. Environ. Microbiol.}\ }\textbf {\bibinfo {volume} {59}},\ \bibinfo
  {pages} {3187--3196} (\bibinfo {year} {1993})}\BibitemShut {NoStop}%
\bibitem [{\citenamefont {Arias}\ \emph {et~al.}(2012)\citenamefont {Arias},
  \citenamefont {LaFrentz}, \citenamefont {Cai},\ and\ \citenamefont
  {Olivares-Fuster}}]{Arias2012}%
  \BibitemOpen
  \bibfield  {author} {\bibinfo {author} {\bibfnamefont {C.~R.}\ \bibnamefont
  {Arias}}, \bibinfo {author} {\bibfnamefont {S.}~\bibnamefont {LaFrentz}},
  \bibinfo {author} {\bibfnamefont {W.}~\bibnamefont {Cai}}, \ and\ \bibinfo
  {author} {\bibfnamefont {O.}~\bibnamefont {Olivares-Fuster}},\ }\bibfield
  {title} {\enquote {\bibinfo {title} {{Adaptive response to starvation in the
  fish pathogen Flavobacterium columnare: cell viability and ultrastructural
  changes}},}\ }\href {\doibase 10.1186/1471-2180-12-266} {\bibfield  {journal}
  {\bibinfo  {journal} {BMC Microbiol.}\ }\textbf {\bibinfo {volume} {12}},\
  \bibinfo {pages} {266} (\bibinfo {year} {2012})}\BibitemShut {NoStop}%
\bibitem [{\citenamefont {Gray}\ \emph {et~al.}(2019)\citenamefont {Gray},
  \citenamefont {Dugar}, \citenamefont {Gamba}, \citenamefont {Strahl},
  \citenamefont {Jonker},\ and\ \citenamefont {Hamoen}}]{Gray2019}%
  \BibitemOpen
  \bibfield  {author} {\bibinfo {author} {\bibfnamefont {D.~A.}\ \bibnamefont
  {Gray}}, \bibinfo {author} {\bibfnamefont {G.}~\bibnamefont {Dugar}},
  \bibinfo {author} {\bibfnamefont {P.}~\bibnamefont {Gamba}}, \bibinfo
  {author} {\bibfnamefont {H.}~\bibnamefont {Strahl}}, \bibinfo {author}
  {\bibfnamefont {M.~J.}\ \bibnamefont {Jonker}}, \ and\ \bibinfo {author}
  {\bibfnamefont {L.~W.}\ \bibnamefont {Hamoen}},\ }\bibfield  {title}
  {\enquote {\bibinfo {title} {{Extreme slow growth as alternative strategy to
  survive deep starvation in bacteria}},}\ }\href {\doibase
  10.1038/s41467-019-08719-8} {\bibfield  {journal} {\bibinfo  {journal} {Nat.
  Commun.}\ }\textbf {\bibinfo {volume} {10}},\ \bibinfo {pages} {890}
  (\bibinfo {year} {2019})}\BibitemShut {NoStop}%
\bibitem [{\citenamefont {Wang}\ \emph {et~al.}(2010)\citenamefont {Wang},
  \citenamefont {Robert}, \citenamefont {Pelletier}, \citenamefont {Dang},
  \citenamefont {Taddei}, \citenamefont {Wright},\ and\ \citenamefont
  {Jun}}]{Wang2010}%
  \BibitemOpen
  \bibfield  {author} {\bibinfo {author} {\bibfnamefont {P.}~\bibnamefont
  {Wang}}, \bibinfo {author} {\bibfnamefont {L.}~\bibnamefont {Robert}},
  \bibinfo {author} {\bibfnamefont {J.}~\bibnamefont {Pelletier}}, \bibinfo
  {author} {\bibfnamefont {W.~L.}\ \bibnamefont {Dang}}, \bibinfo {author}
  {\bibfnamefont {F.}~\bibnamefont {Taddei}}, \bibinfo {author} {\bibfnamefont
  {A.}~\bibnamefont {Wright}}, \ and\ \bibinfo {author} {\bibfnamefont
  {S.}~\bibnamefont {Jun}},\ }\bibfield  {title} {\enquote {\bibinfo {title}
  {Robust growth of \textit{Escherichia coli}},}\ }\href {\doibase
  10.1016/j.cub.2010.04.045} {\bibfield  {journal} {\bibinfo  {journal} {Curr.
  Biol.}\ }\textbf {\bibinfo {volume} {20}},\ \bibinfo {pages} {1099--1103}
  (\bibinfo {year} {2010})}\BibitemShut {NoStop}%
\bibitem [{\citenamefont {Arnoldini}\ \emph {et~al.}(2014)\citenamefont
  {Arnoldini}, \citenamefont {Vizcarra}, \citenamefont {Pe\~{n}a Miller},
  \citenamefont {Stocker}, \citenamefont {Diard}, \citenamefont {Vogel},
  \citenamefont {Beardmore}, \citenamefont {Hardt},\ and\ \citenamefont
  {Ackermann}}]{Arnoldini2014}%
  \BibitemOpen
  \bibfield  {author} {\bibinfo {author} {\bibfnamefont {M.}~\bibnamefont
  {Arnoldini}}, \bibinfo {author} {\bibfnamefont {IA}~\bibnamefont {Vizcarra}},
  \bibinfo {author} {\bibfnamefont {R.}~\bibnamefont {Pe\~{n}a Miller}},
  \bibinfo {author} {\bibfnamefont {N.}~\bibnamefont {Stocker}}, \bibinfo
  {author} {\bibfnamefont {M.}~\bibnamefont {Diard}}, \bibinfo {author}
  {\bibfnamefont {V.}~\bibnamefont {Vogel}}, \bibinfo {author} {\bibfnamefont
  {R.~E.}\ \bibnamefont {Beardmore}}, \bibinfo {author} {\bibfnamefont
  {W.}~\bibnamefont {Hardt}}, \ and\ \bibinfo {author} {\bibfnamefont
  {M.}~\bibnamefont {Ackermann}},\ }\bibfield  {title} {\enquote {\bibinfo
  {title} {Bistable expression of virulence genes in salmonella leads to the
  formation of an antibiotic-tolerant subpopulation},}\ }\href {\doibase
  10.1371/journal.pbio.1001928} {\bibfield  {journal} {\bibinfo  {journal}
  {PLOS Biol.}\ }\textbf {\bibinfo {volume} {12}},\ \bibinfo {pages} {e1001928}
  (\bibinfo {year} {2014})}\BibitemShut {NoStop}%
\bibitem [{\citenamefont {Kaiser}\ \emph {et~al.}(2018)\citenamefont {Kaiser},
  \citenamefont {Jug}, \citenamefont {Julou}, \citenamefont {Deshpande},
  \citenamefont {Pfohl}, \citenamefont {Silander}, \citenamefont {Myers},\ and\
  \citenamefont {van Nimwegen}}]{Kaiser2018}%
  \BibitemOpen
  \bibfield  {author} {\bibinfo {author} {\bibfnamefont {M.}~\bibnamefont
  {Kaiser}}, \bibinfo {author} {\bibfnamefont {F.}~\bibnamefont {Jug}},
  \bibinfo {author} {\bibfnamefont {T.}~\bibnamefont {Julou}}, \bibinfo
  {author} {\bibfnamefont {S.}~\bibnamefont {Deshpande}}, \bibinfo {author}
  {\bibfnamefont {T.}~\bibnamefont {Pfohl}}, \bibinfo {author} {\bibfnamefont
  {O.~K.}\ \bibnamefont {Silander}}, \bibinfo {author} {\bibfnamefont
  {G.}~\bibnamefont {Myers}}, \ and\ \bibinfo {author} {\bibfnamefont
  {E.}~\bibnamefont {van Nimwegen}},\ }\bibfield  {title} {\enquote {\bibinfo
  {title} {Monitoring single-cell gene regulation under dynamically
  controllable conditions with integrated microfluidics and software},}\ }\href
  {\doibase 10.1038/s41467-017-02505-0} {\bibfield  {journal} {\bibinfo
  {journal} {Nat. Commun.}\ }\textbf {\bibinfo {volume} {9}},\ \bibinfo {pages}
  {212} (\bibinfo {year} {2018})}\BibitemShut {NoStop}%
\bibitem [{\citenamefont {Julou}\ \emph {et~al.}(2020)\citenamefont {Julou},
  \citenamefont {Zweifel}, \citenamefont {Blank}, \citenamefont {Fiori},\ and\
  \citenamefont {van Nimwegen}}]{Julou2020}%
  \BibitemOpen
  \bibfield  {author} {\bibinfo {author} {\bibfnamefont {T.}~\bibnamefont
  {Julou}}, \bibinfo {author} {\bibfnamefont {L.}~\bibnamefont {Zweifel}},
  \bibinfo {author} {\bibfnamefont {D.}~\bibnamefont {Blank}}, \bibinfo
  {author} {\bibfnamefont {A.}~\bibnamefont {Fiori}}, \ and\ \bibinfo {author}
  {\bibfnamefont {E.}~\bibnamefont {van Nimwegen}},\ }\bibfield  {title}
  {\enquote {\bibinfo {title} {Subpopulations of sensorless bacteria drive
  fitness in fluctuating environments},}\ }\href {\doibase
  10.1371/journal.pbio.3000952} {\bibfield  {journal} {\bibinfo  {journal}
  {PLOS Biol.}\ }\textbf {\bibinfo {volume} {18}},\ \bibinfo {pages} {e3000952}
  (\bibinfo {year} {2020})}\BibitemShut {NoStop}%
\bibitem [{\citenamefont {Panlilio}\ \emph {et~al.}(2021)\citenamefont
  {Panlilio}, \citenamefont {Grilli}, \citenamefont {Tallarico}, \citenamefont
  {Iuliani}, \citenamefont {Sclavi}, \citenamefont {Cicuta},\ and\
  \citenamefont {Cosentino~Lagomarsino}}]{MarcoCosentino2020}%
  \BibitemOpen
  \bibfield  {author} {\bibinfo {author} {\bibfnamefont {M.}~\bibnamefont
  {Panlilio}}, \bibinfo {author} {\bibfnamefont {J.}~\bibnamefont {Grilli}},
  \bibinfo {author} {\bibfnamefont {G.}~\bibnamefont {Tallarico}}, \bibinfo
  {author} {\bibfnamefont {I.}~\bibnamefont {Iuliani}}, \bibinfo {author}
  {\bibfnamefont {B.}~\bibnamefont {Sclavi}}, \bibinfo {author} {\bibfnamefont
  {P.}~\bibnamefont {Cicuta}}, \ and\ \bibinfo {author} {\bibfnamefont
  {M.}~\bibnamefont {Cosentino~Lagomarsino}},\ }\bibfield  {title} {\enquote
  {\bibinfo {title} {Threshold accumulation of a constitutive protein explains
  \textit{E. coli} cell-division behavior in nutrient upshifts},}\ }\href
  {\doibase 10.1073/pnas.2016391118} {\bibfield  {journal} {\bibinfo  {journal}
  {Proc. Natl. Acad. Sci. USA}\ }\textbf {\bibinfo {volume} {118}},\ \bibinfo
  {pages} {e2016391118} (\bibinfo {year} {2021})}\BibitemShut {NoStop}%
\bibitem [{\citenamefont {Bakshi}\ \emph {et~al.}(2021)\citenamefont {Bakshi},
  \citenamefont {Leoncini}, \citenamefont {Baker}, \citenamefont
  {Ca{\~n}as-Duarte}, \citenamefont {Okumus},\ and\ \citenamefont
  {Paulsson}}]{Bakshi2021}%
  \BibitemOpen
  \bibfield  {author} {\bibinfo {author} {\bibfnamefont {S.}~\bibnamefont
  {Bakshi}}, \bibinfo {author} {\bibfnamefont {E.}~\bibnamefont {Leoncini}},
  \bibinfo {author} {\bibfnamefont {C.}~\bibnamefont {Baker}}, \bibinfo
  {author} {\bibfnamefont {S.~J.}\ \bibnamefont {Ca{\~n}as-Duarte}}, \bibinfo
  {author} {\bibfnamefont {B.}~\bibnamefont {Okumus}}, \ and\ \bibinfo {author}
  {\bibfnamefont {J.}~\bibnamefont {Paulsson}},\ }\bibfield  {title} {\enquote
  {\bibinfo {title} {Tracking bacterial lineages in complex and dynamic
  environments with applications for growth control and persistence},}\ }\href
  {\doibase 10.1038/s41564-021-00900-4} {\bibfield  {journal} {\bibinfo
  {journal} {Nat. Microbiol.}\ }\textbf {\bibinfo {volume} {6}},\ \bibinfo
  {pages} {783–791} (\bibinfo {year} {2021})}\BibitemShut {NoStop}%
\bibitem [{\citenamefont {Inoue}\ \emph {et~al.}(2001)\citenamefont {Inoue},
  \citenamefont {Wakamoto}, \citenamefont {Moriguchi}, \citenamefont {Okano},\
  and\ \citenamefont {Yasuda}}]{Inoue2001}%
  \BibitemOpen
  \bibfield  {author} {\bibinfo {author} {\bibfnamefont {I.}~\bibnamefont
  {Inoue}}, \bibinfo {author} {\bibfnamefont {Y.}~\bibnamefont {Wakamoto}},
  \bibinfo {author} {\bibfnamefont {H.}~\bibnamefont {Moriguchi}}, \bibinfo
  {author} {\bibfnamefont {K.}~\bibnamefont {Okano}}, \ and\ \bibinfo {author}
  {\bibfnamefont {K.}~\bibnamefont {Yasuda}},\ }\bibfield  {title} {\enquote
  {\bibinfo {title} {On-chip culture system for observation of isolated
  individual cells},}\ }\href {\doibase 10.1039/B103931H} {\bibfield  {journal}
  {\bibinfo  {journal} {Lab Chip}\ }\textbf {\bibinfo {volume} {1}},\ \bibinfo
  {pages} {50--55} (\bibinfo {year} {2001})}\BibitemShut {NoStop}%
\bibitem [{\citenamefont {Charvin}\ \emph {et~al.}(2008)\citenamefont
  {Charvin}, \citenamefont {Cross},\ and\ \citenamefont
  {Siggia}}]{Charvin2008}%
  \BibitemOpen
  \bibfield  {author} {\bibinfo {author} {\bibfnamefont {G.}~\bibnamefont
  {Charvin}}, \bibinfo {author} {\bibfnamefont {F.~R.}\ \bibnamefont {Cross}},
  \ and\ \bibinfo {author} {\bibfnamefont {E.~D.}\ \bibnamefont {Siggia}},\
  }\bibfield  {title} {\enquote {\bibinfo {title} {A microfluidic device for
  temporally controlled gene expression and long-term fluorescent imaging in
  unperturbed dividing yeast cells},}\ }\href {\doibase
  10.1371/journal.pone.0001468} {\bibfield  {journal} {\bibinfo  {journal}
  {PLOS ONE}\ }\textbf {\bibinfo {volume} {3}},\ \bibinfo {pages} {e1468}
  (\bibinfo {year} {2008})}\BibitemShut {NoStop}%
\bibitem [{\citenamefont {Ducret}\ \emph {et~al.}(2009)\citenamefont {Ducret},
  \citenamefont {Maisonneuve}, \citenamefont {Notareschi}, \citenamefont
  {Grossi}, \citenamefont {Mignot},\ and\ \citenamefont {Dukan}}]{Ducret2009}%
  \BibitemOpen
  \bibfield  {author} {\bibinfo {author} {\bibfnamefont {A.}~\bibnamefont
  {Ducret}}, \bibinfo {author} {\bibfnamefont {E.}~\bibnamefont {Maisonneuve}},
  \bibinfo {author} {\bibfnamefont {P.}~\bibnamefont {Notareschi}}, \bibinfo
  {author} {\bibfnamefont {A.}~\bibnamefont {Grossi}}, \bibinfo {author}
  {\bibfnamefont {T.}~\bibnamefont {Mignot}}, \ and\ \bibinfo {author}
  {\bibfnamefont {S.}~\bibnamefont {Dukan}},\ }\bibfield  {title} {\enquote
  {\bibinfo {title} {A microscope automated fluidic system to study bacterial
  processes in real time},}\ }\href {\doibase 10.1371/journal.pone.0007282}
  {\bibfield  {journal} {\bibinfo  {journal} {PLOS ONE}\ }\textbf {\bibinfo
  {volume} {4}},\ \bibinfo {pages} {e7282} (\bibinfo {year}
  {2009})}\BibitemShut {NoStop}%
\bibitem [{\citenamefont {Cooper}\ and\ \citenamefont
  {Helmstetter}(1968)}]{Cooper1968}%
  \BibitemOpen
  \bibfield  {author} {\bibinfo {author} {\bibfnamefont {S.}~\bibnamefont
  {Cooper}}\ and\ \bibinfo {author} {\bibfnamefont {C.~E.}\ \bibnamefont
  {Helmstetter}},\ }\bibfield  {title} {\enquote {\bibinfo {title} {{Chromosome
  replication and the division cycle of \textit{Escherichia coli} Br}},}\
  }\href {\doibase 10.1016/0022-2836(68)90425-7} {\bibfield  {journal}
  {\bibinfo  {journal} {J. Mol. Biol}\ }\textbf {\bibinfo {volume} {31}},\
  \bibinfo {pages} {519 -- 540} (\bibinfo {year} {1968})}\BibitemShut {NoStop}%
\bibitem [{\citenamefont {Witz}\ \emph {et~al.}(2019)\citenamefont {Witz},
  \citenamefont {van Nimwegen},\ and\ \citenamefont {Julou}}]{Witz2019}%
  \BibitemOpen
  \bibfield  {author} {\bibinfo {author} {\bibfnamefont {G.}~\bibnamefont
  {Witz}}, \bibinfo {author} {\bibfnamefont {E.}~\bibnamefont {van Nimwegen}},
  \ and\ \bibinfo {author} {\bibfnamefont {T.}~\bibnamefont {Julou}},\
  }\bibfield  {title} {\enquote {\bibinfo {title} {Initiation of chromosome
  replication controls both division and replication cycles in \textit{E. coli}
  through a double-adder mechanism},}\ }\href {\doibase 10.7554/eLife.48063}
  {\bibfield  {journal} {\bibinfo  {journal} {eLife}\ }\textbf {\bibinfo
  {volume} {8}},\ \bibinfo {pages} {e48063} (\bibinfo {year}
  {2019})}\BibitemShut {NoStop}%
\bibitem [{\citenamefont {Ho}\ and\ \citenamefont {Amir}(2015)}]{Ho2015}%
  \BibitemOpen
  \bibfield  {author} {\bibinfo {author} {\bibfnamefont {P.}~\bibnamefont
  {Ho}}\ and\ \bibinfo {author} {\bibfnamefont {A.}~\bibnamefont {Amir}},\
  }\bibfield  {title} {\enquote {\bibinfo {title} {Simultaneous regulation of
  cell size and chromosome replication in bacteria},}\ }\href {\doibase
  10.3389/fmicb.2015.00662} {\bibfield  {journal} {\bibinfo  {journal} {Front.
  Microbiol.}\ }\textbf {\bibinfo {volume} {6}},\ \bibinfo {pages} {662}
  (\bibinfo {year} {2015})}\BibitemShut {NoStop}%
\bibitem [{\citenamefont {Gao}\ \emph {et~al.}(2015)\citenamefont {Gao},
  \citenamefont {Luan}, \citenamefont {Wang}, \citenamefont {Liang},\ and\
  \citenamefont {Qi}}]{Gao2015}%
  \BibitemOpen
  \bibfield  {author} {\bibinfo {author} {\bibfnamefont {D.}~\bibnamefont
  {Gao}}, \bibinfo {author} {\bibfnamefont {Y.}~\bibnamefont {Luan}}, \bibinfo
  {author} {\bibfnamefont {Q.}~\bibnamefont {Wang}}, \bibinfo {author}
  {\bibfnamefont {Q.}~\bibnamefont {Liang}}, \ and\ \bibinfo {author}
  {\bibfnamefont {Q.}~\bibnamefont {Qi}},\ }\bibfield  {title} {\enquote
  {\bibinfo {title} {Construction of cellulose-utilizing escherichia coli based
  on a secretable cellulase},}\ }\href {\doibase 10.1186/s12934-015-0349-7}
  {\bibfield  {journal} {\bibinfo  {journal} {Microb. Cell Fact.}\ }\textbf
  {\bibinfo {volume} {14}},\ \bibinfo {pages} {159} (\bibinfo {year}
  {2015})}\BibitemShut {NoStop}%
\bibitem [{\citenamefont {Wakamoto}\ \emph {et~al.}(2005)\citenamefont
  {Wakamoto}, \citenamefont {Ramsden},\ and\ \citenamefont
  {Yasuda}}]{Wakamoto2005}%
  \BibitemOpen
  \bibfield  {author} {\bibinfo {author} {\bibfnamefont {Y.}~\bibnamefont
  {Wakamoto}}, \bibinfo {author} {\bibfnamefont {J.}~\bibnamefont {Ramsden}}, \
  and\ \bibinfo {author} {\bibfnamefont {K.}~\bibnamefont {Yasuda}},\
  }\bibfield  {title} {\enquote {\bibinfo {title} {Single-cell growth and
  division dynamics showing epigenetic correlations},}\ }\href {\doibase
  10.1039/b409860a} {\bibfield  {journal} {\bibinfo  {journal} {The Analyst}\
  }\textbf {\bibinfo {volume} {130}},\ \bibinfo {pages} {311--317} (\bibinfo
  {year} {2005})}\BibitemShut {NoStop}%
\bibitem [{\citenamefont {Hashimoto}\ \emph {et~al.}(2016)\citenamefont
  {Hashimoto}, \citenamefont {Nozoe}, \citenamefont {Nakaoka}, \citenamefont
  {Okura}, \citenamefont {Akiyoshi}, \citenamefont {Kaneko}, \citenamefont
  {Kussell},\ and\ \citenamefont {Wakamoto}}]{Hashimoto2016}%
  \BibitemOpen
  \bibfield  {author} {\bibinfo {author} {\bibfnamefont {M.}~\bibnamefont
  {Hashimoto}}, \bibinfo {author} {\bibfnamefont {T.}~\bibnamefont {Nozoe}},
  \bibinfo {author} {\bibfnamefont {H.}~\bibnamefont {Nakaoka}}, \bibinfo
  {author} {\bibfnamefont {R.}~\bibnamefont {Okura}}, \bibinfo {author}
  {\bibfnamefont {S.}~\bibnamefont {Akiyoshi}}, \bibinfo {author}
  {\bibfnamefont {K.}~\bibnamefont {Kaneko}}, \bibinfo {author} {\bibfnamefont
  {E.}~\bibnamefont {Kussell}}, \ and\ \bibinfo {author} {\bibfnamefont
  {Y.}~\bibnamefont {Wakamoto}},\ }\bibfield  {title} {\enquote {\bibinfo
  {title} {Noise-driven growth rate gain in clonal cellular populations},}\
  }\href {\doibase 10.1073/pnas.1519412113} {\bibfield  {journal} {\bibinfo
  {journal} {Proc. Natl. Acad. Sci. USA}\ }\textbf {\bibinfo {volume} {113}},\
  \bibinfo {pages} {3251--3256} (\bibinfo {year} {2016})}\BibitemShut {NoStop}%
\bibitem [{\citenamefont {Carbonell}\ \emph {et~al.}(2002)\citenamefont
  {Carbonell}, \citenamefont {Corchero}, \citenamefont {Cubars\'{i}},
  \citenamefont {Vila},\ and\ \citenamefont {Villaverde}}]{Carbonell2002}%
  \BibitemOpen
  \bibfield  {author} {\bibinfo {author} {\bibfnamefont {X.}~\bibnamefont
  {Carbonell}}, \bibinfo {author} {\bibfnamefont {J.~L.}\ \bibnamefont
  {Corchero}}, \bibinfo {author} {\bibfnamefont {R.}~\bibnamefont
  {Cubars\'{i}}}, \bibinfo {author} {\bibfnamefont {P.}~\bibnamefont {Vila}}, \
  and\ \bibinfo {author} {\bibfnamefont {A.}~\bibnamefont {Villaverde}},\
  }\bibfield  {title} {\enquote {\bibinfo {title} {Control of
  \textit{Escherichia coli} growth rate through cell density},}\ }\href
  {\doibase 10.1078/0944-5013-00167} {\bibfield  {journal} {\bibinfo  {journal}
  {Microbiol. Res.}\ }\textbf {\bibinfo {volume} {157}},\ \bibinfo {pages} {257
  -- 265} (\bibinfo {year} {2002})}\BibitemShut {NoStop}%
\bibitem [{\citenamefont {Bruger}\ and\ \citenamefont
  {Waters}(2016)}]{Bruger2016}%
  \BibitemOpen
  \bibfield  {author} {\bibinfo {author} {\bibfnamefont {E.~L.}\ \bibnamefont
  {Bruger}}\ and\ \bibinfo {author} {\bibfnamefont {C.~M.}\ \bibnamefont
  {Waters}},\ }\bibfield  {title} {\enquote {\bibinfo {title} {Bacterial quorum
  sensing stabilizes cooperation by optimizing growth strategies},}\ }\href
  {\doibase 10.1128/AEM.01945-16} {\bibfield  {journal} {\bibinfo  {journal}
  {Appl. Environ. Microbiol.}\ }\textbf {\bibinfo {volume} {82}},\ \bibinfo
  {pages} {6498--6506} (\bibinfo {year} {2016})}\BibitemShut {NoStop}%
\bibitem [{\citenamefont {Ha}\ \emph {et~al.}(2018)\citenamefont {Ha},
  \citenamefont {Hauk}, \citenamefont {Cho}, \citenamefont {Eo}, \citenamefont
  {Ma}, \citenamefont {Stephens}, \citenamefont {Cha}, \citenamefont {Jeong},
  \citenamefont {Suh}, \citenamefont {Sintim}, \citenamefont {Bentley},\ and\
  \citenamefont {Ryu}}]{Haeaar2018}%
  \BibitemOpen
  \bibfield  {author} {\bibinfo {author} {\bibfnamefont {J.}~\bibnamefont
  {Ha}}, \bibinfo {author} {\bibfnamefont {P.}~\bibnamefont {Hauk}}, \bibinfo
  {author} {\bibfnamefont {K.}~\bibnamefont {Cho}}, \bibinfo {author}
  {\bibfnamefont {Y.}~\bibnamefont {Eo}}, \bibinfo {author} {\bibfnamefont
  {X.}~\bibnamefont {Ma}}, \bibinfo {author} {\bibfnamefont {K.}~\bibnamefont
  {Stephens}}, \bibinfo {author} {\bibfnamefont {S.}~\bibnamefont {Cha}},
  \bibinfo {author} {\bibfnamefont {M.}~\bibnamefont {Jeong}}, \bibinfo
  {author} {\bibfnamefont {J.}~\bibnamefont {Suh}}, \bibinfo {author}
  {\bibfnamefont {H.~O.}\ \bibnamefont {Sintim}}, \bibinfo {author}
  {\bibfnamefont {W.~E.}\ \bibnamefont {Bentley}}, \ and\ \bibinfo {author}
  {\bibfnamefont {K.}~\bibnamefont {Ryu}},\ }\bibfield  {title} {\enquote
  {\bibinfo {title} {Evidence of link between quorum sensing and sugar
  metabolism in \textit{Escherichia coli} revealed via cocrystal structures of
  lsrk and hpr},}\ }\href {\doibase 10.1126/sciadv.aar7063} {\bibfield
  {journal} {\bibinfo  {journal} {Sci. Adv.}\ }\textbf {\bibinfo {volume}
  {4}},\ \bibinfo {pages} {eaar7063} (\bibinfo {year} {2018})}\BibitemShut
  {NoStop}%
\bibitem [{\citenamefont {Maier}\ and\ \citenamefont
  {Pepper}(2015)}]{Maier2015}%
  \BibitemOpen
  \bibfield  {author} {\bibinfo {author} {\bibfnamefont {R.~M.}\ \bibnamefont
  {Maier}}\ and\ \bibinfo {author} {\bibfnamefont {I.~L.}\ \bibnamefont
  {Pepper}},\ }\bibfield  {title} {\enquote {\bibinfo {title} {Chapter 3 -
  bacterial growth},}\ }in\ \href {\doibase 10.1016/B978-0-12-394626-3.00003-X}
  {\emph {\bibinfo {booktitle} {Environmental Microbiology (Third Edition)}}},\
  \bibinfo {editor} {edited by\ \bibinfo {editor} {\bibfnamefont {I.~L.}\
  \bibnamefont {Pepper}}, \bibinfo {editor} {\bibfnamefont {C.~P.}\
  \bibnamefont {Gerba}}, \ and\ \bibinfo {editor} {\bibfnamefont {T.~J.}\
  \bibnamefont {Gentry}}}\ (\bibinfo  {publisher} {Academic Press},\ \bibinfo
  {address} {San Diego},\ \bibinfo {year} {2015})\ \bibinfo {edition} {third
  edition}\ ed.,\ pp.\ \bibinfo {pages} {37 -- 56}\BibitemShut {NoStop}%
\bibitem [{\citenamefont {Chou}\ \emph {et~al.}(1994)\citenamefont {Chou},
  \citenamefont {Bennett},\ and\ \citenamefont {San}}]{Chou1994}%
  \BibitemOpen
  \bibfield  {author} {\bibinfo {author} {\bibfnamefont {C.}~\bibnamefont
  {Chou}}, \bibinfo {author} {\bibfnamefont {G.~N.}\ \bibnamefont {Bennett}}, \
  and\ \bibinfo {author} {\bibfnamefont {K.}~\bibnamefont {San}},\ }\bibfield
  {title} {\enquote {\bibinfo {title} {Effect of modulated glucose uptake on
  high-level recombinant protein production in a dense \textit{Escherichia
  coli} culture},}\ }\href {\doibase 10.1021/bp00030a009} {\bibfield  {journal}
  {\bibinfo  {journal} {Biotechnol. Prog.}\ }\textbf {\bibinfo {volume} {10}},\
  \bibinfo {pages} {644--647} (\bibinfo {year} {1994})}\BibitemShut {NoStop}%
\bibitem [{\citenamefont {Rojas}\ \emph {et~al.}(2014)\citenamefont {Rojas},
  \citenamefont {Theriot},\ and\ \citenamefont {Huang}}]{Rojas2014}%
  \BibitemOpen
  \bibfield  {author} {\bibinfo {author} {\bibfnamefont {E.}~\bibnamefont
  {Rojas}}, \bibinfo {author} {\bibfnamefont {J.~A.}\ \bibnamefont {Theriot}},
  \ and\ \bibinfo {author} {\bibfnamefont {K.~C.}\ \bibnamefont {Huang}},\
  }\bibfield  {title} {\enquote {\bibinfo {title} {Response of
  \textit{Escherichia coli} growth rate to osmotic shock},}\ }\href {\doibase
  10.1073/pnas.1402591111} {\bibfield  {journal} {\bibinfo  {journal} {Proc.
  Natl. Acad. Sci. USA}\ }\textbf {\bibinfo {volume} {111}},\ \bibinfo {pages}
  {7807--7812} (\bibinfo {year} {2014})}\BibitemShut {NoStop}%
\bibitem [{\citenamefont {Harris}\ and\ \citenamefont
  {Theriot}(2016)}]{HARRIS2016}%
  \BibitemOpen
  \bibfield  {author} {\bibinfo {author} {\bibfnamefont {L.~K.}\ \bibnamefont
  {Harris}}\ and\ \bibinfo {author} {\bibfnamefont {J.~A.}\ \bibnamefont
  {Theriot}},\ }\bibfield  {title} {\enquote {\bibinfo {title} {Relative rates
  of surface and volume synthesis set bacterial cell size},}\ }\href {\doibase
  10.1016/j.cell.2016.05.045} {\bibfield  {journal} {\bibinfo  {journal}
  {Cell}\ }\textbf {\bibinfo {volume} {165}},\ \bibinfo {pages} {1479 -- 1492}
  (\bibinfo {year} {2016})}\BibitemShut {NoStop}%
\bibitem [{\citenamefont {Gangan}\ and\ \citenamefont
  {Athale}(2017)}]{Gangan2017}%
  \BibitemOpen
  \bibfield  {author} {\bibinfo {author} {\bibfnamefont {M.~S.}\ \bibnamefont
  {Gangan}}\ and\ \bibinfo {author} {\bibfnamefont {C.~A.}\ \bibnamefont
  {Athale}},\ }\bibfield  {title} {\enquote {\bibinfo {title} {Threshold effect
  of growth rate on population variability of \textit{Escherichia coli} cell
  lengths},}\ }\href {\doibase 10.1098/rsos.160417} {\bibfield  {journal}
  {\bibinfo  {journal} {R. Soc. Open Sci.}\ }\textbf {\bibinfo {volume} {4}},\
  \bibinfo {pages} {160417} (\bibinfo {year} {2017})}\BibitemShut {NoStop}%
\bibitem [{\citenamefont {Wakita}\ \emph {et~al.}(2010)\citenamefont {Wakita},
  \citenamefont {Kuninaka}, \citenamefont {Matsuyama},\ and\ \citenamefont
  {Matsushita}}]{Wakita2010}%
  \BibitemOpen
  \bibfield  {author} {\bibinfo {author} {\bibfnamefont {J.}~\bibnamefont
  {Wakita}}, \bibinfo {author} {\bibfnamefont {H.}~\bibnamefont {Kuninaka}},
  \bibinfo {author} {\bibfnamefont {T.}~\bibnamefont {Matsuyama}}, \ and\
  \bibinfo {author} {\bibfnamefont {M.}~\bibnamefont {Matsushita}},\ }\bibfield
   {title} {\enquote {\bibinfo {title} {Size distribution of bacterial cells in
  homogeneously spreading disk-like colonies by \textit{Bacillus subtilis}},}\
  }\href {\doibase 10.1143/JPSJ.79.094002} {\bibfield  {journal} {\bibinfo
  {journal} {J. Phys. Soc. Jpn.}\ }\textbf {\bibinfo {volume} {79}},\ \bibinfo
  {pages} {094002} (\bibinfo {year} {2010})}\BibitemShut {NoStop}%
\bibitem [{\citenamefont {Wallden}\ \emph {et~al.}(2016)\citenamefont
  {Wallden}, \citenamefont {Fange}, \citenamefont {Lundius}, \citenamefont
  {Baltekin},\ and\ \citenamefont {Elf}}]{Wallden2016}%
  \BibitemOpen
  \bibfield  {author} {\bibinfo {author} {\bibfnamefont {M.}~\bibnamefont
  {Wallden}}, \bibinfo {author} {\bibfnamefont {D.}~\bibnamefont {Fange}},
  \bibinfo {author} {\bibfnamefont {E.~G.}\ \bibnamefont {Lundius}}, \bibinfo
  {author} {\bibfnamefont {\"{O}.}\ \bibnamefont {Baltekin}}, \ and\ \bibinfo
  {author} {\bibfnamefont {J.}~\bibnamefont {Elf}},\ }\bibfield  {title}
  {\enquote {\bibinfo {title} {The synchronization of replication and division
  cycles in individual \textit{E. coli} cells},}\ }\href {\doibase
  10.1016/j.cell.2016.06.052} {\bibfield  {journal} {\bibinfo  {journal}
  {Cell}\ }\textbf {\bibinfo {volume} {166}},\ \bibinfo {pages} {729 -- 739}
  (\bibinfo {year} {2016})}\BibitemShut {NoStop}%
\bibitem [{\citenamefont {Si}\ \emph {et~al.}(2017)\citenamefont {Si},
  \citenamefont {Li}, \citenamefont {Cox}, \citenamefont {Sauls}, \citenamefont
  {Azizi}, \citenamefont {Sou}, \citenamefont {Schwartz}, \citenamefont
  {Erickstad}, \citenamefont {Jun}, \citenamefont {Li},\ and\ \citenamefont
  {Jun}}]{SI2017}%
  \BibitemOpen
  \bibfield  {author} {\bibinfo {author} {\bibfnamefont {F.}~\bibnamefont
  {Si}}, \bibinfo {author} {\bibfnamefont {D.}~\bibnamefont {Li}}, \bibinfo
  {author} {\bibfnamefont {S.~E.}\ \bibnamefont {Cox}}, \bibinfo {author}
  {\bibfnamefont {John~T.}\ \bibnamefont {Sauls}}, \bibinfo {author}
  {\bibfnamefont {Omid}\ \bibnamefont {Azizi}}, \bibinfo {author}
  {\bibfnamefont {Cindy}\ \bibnamefont {Sou}}, \bibinfo {author} {\bibfnamefont
  {Amy~B.}\ \bibnamefont {Schwartz}}, \bibinfo {author} {\bibfnamefont
  {Michael~J.}\ \bibnamefont {Erickstad}}, \bibinfo {author} {\bibfnamefont
  {Yonggun}\ \bibnamefont {Jun}}, \bibinfo {author} {\bibfnamefont {Xintian}\
  \bibnamefont {Li}}, \ and\ \bibinfo {author} {\bibfnamefont {Suckjoon}\
  \bibnamefont {Jun}},\ }\bibfield  {title} {\enquote {\bibinfo {title}
  {Invariance of initiation mass and predictability of cell size in
  \textit{Escherichia coli}},}\ }\href {\doibase
  https://doi.org/10.1016/j.cub.2017.03.022} {\bibfield  {journal} {\bibinfo
  {journal} {Curr. Biol.}\ }\textbf {\bibinfo {volume} {27}},\ \bibinfo {pages}
  {1278 -- 1287} (\bibinfo {year} {2017})}\BibitemShut {NoStop}%
\bibitem [{\citenamefont {Micali}\ \emph
  {et~al.}(2018{\natexlab{a}})\citenamefont {Micali}, \citenamefont {Grilli},
  \citenamefont {Marchi}, \citenamefont {Osella},\ and\ \citenamefont
  {Lagomarsino}}]{MICALI2018}%
  \BibitemOpen
  \bibfield  {author} {\bibinfo {author} {\bibfnamefont {G.}~\bibnamefont
  {Micali}}, \bibinfo {author} {\bibfnamefont {J.}~\bibnamefont {Grilli}},
  \bibinfo {author} {\bibfnamefont {J.}~\bibnamefont {Marchi}}, \bibinfo
  {author} {\bibfnamefont {M.}~\bibnamefont {Osella}}, \ and\ \bibinfo {author}
  {\bibfnamefont {M.~Cosentino}\ \bibnamefont {Lagomarsino}},\ }\bibfield
  {title} {\enquote {\bibinfo {title} {Dissecting the control mechanisms for
  {DNA} replication and cell division in \textit{E. coli}},}\ }\href {\doibase
  10.1016/j.celrep.2018.09.061} {\bibfield  {journal} {\bibinfo  {journal}
  {Cell Rep.}\ }\textbf {\bibinfo {volume} {25}},\ \bibinfo {pages} {761 --
  771.e4} (\bibinfo {year} {2018}{\natexlab{a}})}\BibitemShut {NoStop}%
\bibitem [{\citenamefont {Micali}\ \emph
  {et~al.}(2018{\natexlab{b}})\citenamefont {Micali}, \citenamefont {Grilli},
  \citenamefont {Osella},\ and\ \citenamefont
  {Cosentino~Lagomarsino}}]{Micalieaau2018}%
  \BibitemOpen
  \bibfield  {author} {\bibinfo {author} {\bibfnamefont {G.}~\bibnamefont
  {Micali}}, \bibinfo {author} {\bibfnamefont {J.}~\bibnamefont {Grilli}},
  \bibinfo {author} {\bibfnamefont {M.}~\bibnamefont {Osella}}, \ and\ \bibinfo
  {author} {\bibfnamefont {M.}~\bibnamefont {Cosentino~Lagomarsino}},\
  }\bibfield  {title} {\enquote {\bibinfo {title} {Concurrent processes set
  \textit{E. coli} cell division},}\ }\href {\doibase 10.1126/sciadv.aau3324}
  {\bibfield  {journal} {\bibinfo  {journal} {Sci. Adv.}\ }\textbf {\bibinfo
  {volume} {4}},\ \bibinfo {pages} {eaau3324} (\bibinfo {year}
  {2018}{\natexlab{b}})}\BibitemShut {NoStop}%
\bibitem [{\citenamefont {Si}\ \emph {et~al.}(2019)\citenamefont {Si},
  \citenamefont {{Le Treut}}, \citenamefont {Sauls}, \citenamefont {Vadia},
  \citenamefont {Levin},\ and\ \citenamefont {Jun}}]{SI2019}%
  \BibitemOpen
  \bibfield  {author} {\bibinfo {author} {\bibfnamefont {F.}~\bibnamefont
  {Si}}, \bibinfo {author} {\bibfnamefont {G.}~\bibnamefont {{Le Treut}}},
  \bibinfo {author} {\bibfnamefont {J.~T.}\ \bibnamefont {Sauls}}, \bibinfo
  {author} {\bibfnamefont {S.}~\bibnamefont {Vadia}}, \bibinfo {author}
  {\bibfnamefont {P.~A.}\ \bibnamefont {Levin}}, \ and\ \bibinfo {author}
  {\bibfnamefont {S.}~\bibnamefont {Jun}},\ }\bibfield  {title} {\enquote
  {\bibinfo {title} {Mechanistic origin of cell-size control and homeostasis in
  bacteria},}\ }\href {\doibase https://doi.org/10.1016/j.cub.2019.04.062}
  {\bibfield  {journal} {\bibinfo  {journal} {Curr. Biol.}\ }\textbf {\bibinfo
  {volume} {29}},\ \bibinfo {pages} {1760 -- 1770.e7} (\bibinfo {year}
  {2019})}\BibitemShut {NoStop}%
\bibitem [{\citenamefont {Wang}\ and\ \citenamefont {Levin}(2009)}]{Wang2009}%
  \BibitemOpen
  \bibfield  {author} {\bibinfo {author} {\bibfnamefont {J.~D.}\ \bibnamefont
  {Wang}}\ and\ \bibinfo {author} {\bibfnamefont {P.~A.}\ \bibnamefont
  {Levin}},\ }\bibfield  {title} {\enquote {\bibinfo {title} {Metabolism, cell
  growth and the bacterial cell cycle},}\ }\href {\doibase 10.1038/nrmicro2202}
  {\bibfield  {journal} {\bibinfo  {journal} {Nat. Rev. Microbiol.}\ }\textbf
  {\bibinfo {volume} {7}},\ \bibinfo {pages} {822--827} (\bibinfo {year}
  {2009})}\BibitemShut {NoStop}%
\bibitem [{\citenamefont {Monod}(1949)}]{Monod1949}%
  \BibitemOpen
  \bibfield  {author} {\bibinfo {author} {\bibfnamefont {J.}~\bibnamefont
  {Monod}},\ }\bibfield  {title} {\enquote {\bibinfo {title} {The growth of
  bacterial cultures},}\ }\href {\doibase 10.1146/annurev.mi.03.100149.002103}
  {\bibfield  {journal} {\bibinfo  {journal} {Annu. Rev. Microbiol.}\ }\textbf
  {\bibinfo {volume} {3}},\ \bibinfo {pages} {371--394} (\bibinfo {year}
  {1949})}\BibitemShut {NoStop}%
\bibitem [{\citenamefont {Sinko}\ and\ \citenamefont
  {Streifer}(1971)}]{Sinko1971}%
  \BibitemOpen
  \bibfield  {author} {\bibinfo {author} {\bibfnamefont {James~W.}\
  \bibnamefont {Sinko}}\ and\ \bibinfo {author} {\bibfnamefont {William}\
  \bibnamefont {Streifer}},\ }\bibfield  {title} {\enquote {\bibinfo {title} {A
  model for population reproducing by fission},}\ }\href
  {http://www.jstor.org/stable/1934592} {\bibfield  {journal} {\bibinfo
  {journal} {Ecology}\ }\textbf {\bibinfo {volume} {52}},\ \bibinfo {pages}
  {330--335} (\bibinfo {year} {1971})}\BibitemShut {NoStop}%
\bibitem [{\citenamefont {Diekmann}\ \emph {et~al.}(1983)\citenamefont
  {Diekmann}, \citenamefont {Lauwerier}, \citenamefont {Aldenberg},\ and\
  \citenamefont {Metz}}]{Diekmann1983}%
  \BibitemOpen
  \bibfield  {author} {\bibinfo {author} {\bibfnamefont {O.}~\bibnamefont
  {Diekmann}}, \bibinfo {author} {\bibfnamefont {H.~A.}\ \bibnamefont
  {Lauwerier}}, \bibinfo {author} {\bibfnamefont {T.}~\bibnamefont
  {Aldenberg}}, \ and\ \bibinfo {author} {\bibfnamefont {J.~A.~J.}\
  \bibnamefont {Metz}},\ }\bibfield  {title} {\enquote {\bibinfo {title}
  {Growth, fission and the stable size distribution},}\ }\href {\doibase
  10.1007/BF00280662} {\bibfield  {journal} {\bibinfo  {journal} {J. Math.
  Biol.}\ }\textbf {\bibinfo {volume} {18}},\ \bibinfo {pages} {135--148}
  (\bibinfo {year} {1983})}\BibitemShut {NoStop}%
\bibitem [{\citenamefont {Tyson}\ and\ \citenamefont
  {Diekmann}(1986)}]{TYSON1986}%
  \BibitemOpen
  \bibfield  {author} {\bibinfo {author} {\bibfnamefont {J.~J.}\ \bibnamefont
  {Tyson}}\ and\ \bibinfo {author} {\bibfnamefont {O.}~\bibnamefont
  {Diekmann}},\ }\bibfield  {title} {\enquote {\bibinfo {title} {Sloppy size
  control of the cell division cycle},}\ }\href {\doibase
  10.1016/S0022-5193(86)80162-X} {\bibfield  {journal} {\bibinfo  {journal} {J.
  Theor. Biol.}\ }\textbf {\bibinfo {volume} {118}},\ \bibinfo {pages} {405 --
  426} (\bibinfo {year} {1986})}\BibitemShut {NoStop}%
\bibitem [{\citenamefont {Robert}\ \emph {et~al.}(2014)\citenamefont {Robert},
  \citenamefont {Hoffmann}, \citenamefont {Krell}, \citenamefont {Aymerich},
  \citenamefont {Robert},\ and\ \citenamefont {Doumic}}]{Robert2014}%
  \BibitemOpen
  \bibfield  {author} {\bibinfo {author} {\bibfnamefont {L.}~\bibnamefont
  {Robert}}, \bibinfo {author} {\bibfnamefont {M.}~\bibnamefont {Hoffmann}},
  \bibinfo {author} {\bibfnamefont {N.}~\bibnamefont {Krell}}, \bibinfo
  {author} {\bibfnamefont {S.}~\bibnamefont {Aymerich}}, \bibinfo {author}
  {\bibfnamefont {J.}~\bibnamefont {Robert}}, \ and\ \bibinfo {author}
  {\bibfnamefont {M.}~\bibnamefont {Doumic}},\ }\bibfield  {title} {\enquote
  {\bibinfo {title} {Division in \textit{Escherichia coli} is triggered by a
  size-sensing rather than a timing mechanism},}\ }\href {\doibase
  10.1186/1741-7007-12-17} {\bibfield  {journal} {\bibinfo  {journal} {BMC
  Biol.}\ }\textbf {\bibinfo {volume} {12}},\ \bibinfo {pages} {17} (\bibinfo
  {year} {2014})}\BibitemShut {NoStop}%
\bibitem [{\citenamefont {Hosoda}\ \emph {et~al.}(2011)\citenamefont {Hosoda},
  \citenamefont {Matsuura}, \citenamefont {Suzuki},\ and\ \citenamefont
  {Yomo}}]{Hosoda2011}%
  \BibitemOpen
  \bibfield  {author} {\bibinfo {author} {\bibfnamefont {K.}~\bibnamefont
  {Hosoda}}, \bibinfo {author} {\bibfnamefont {T.}~\bibnamefont {Matsuura}},
  \bibinfo {author} {\bibfnamefont {H.}~\bibnamefont {Suzuki}}, \ and\ \bibinfo
  {author} {\bibfnamefont {T.}~\bibnamefont {Yomo}},\ }\bibfield  {title}
  {\enquote {\bibinfo {title} {Origin of lognormal-like distributions with a
  common width in a growth and division process},}\ }\href {\doibase
  10.1103/PhysRevE.83.031118} {\bibfield  {journal} {\bibinfo  {journal} {Phys.
  Rev. E}\ }\textbf {\bibinfo {volume} {83}},\ \bibinfo {pages} {031118}
  (\bibinfo {year} {2011})}\BibitemShut {NoStop}%
\bibitem [{\citenamefont {Taheri-Araghi}\ \emph {et~al.}(2015)\citenamefont
  {Taheri-Araghi}, \citenamefont {Bradde}, \citenamefont {Sauls}, \citenamefont
  {Hill}, \citenamefont {Levin}, \citenamefont {Paulsson}, \citenamefont
  {Vergassola},\ and\ \citenamefont {Jun}}]{TAHERIARAGHI2015}%
  \BibitemOpen
  \bibfield  {author} {\bibinfo {author} {\bibfnamefont {S.}~\bibnamefont
  {Taheri-Araghi}}, \bibinfo {author} {\bibfnamefont {S.}~\bibnamefont
  {Bradde}}, \bibinfo {author} {\bibfnamefont {J.~T.}\ \bibnamefont {Sauls}},
  \bibinfo {author} {\bibfnamefont {N.~S.}\ \bibnamefont {Hill}}, \bibinfo
  {author} {\bibfnamefont {P.~A.}\ \bibnamefont {Levin}}, \bibinfo {author}
  {\bibfnamefont {J.}~\bibnamefont {Paulsson}}, \bibinfo {author}
  {\bibfnamefont {M.}~\bibnamefont {Vergassola}}, \ and\ \bibinfo {author}
  {\bibfnamefont {S.}~\bibnamefont {Jun}},\ }\bibfield  {title} {\enquote
  {\bibinfo {title} {Cell-size control and homeostasis in bacteria},}\ }\href
  {\doibase 10.1016/j.cub.2014.12.009} {\bibfield  {journal} {\bibinfo
  {journal} {Curr. Biol.}\ }\textbf {\bibinfo {volume} {25}},\ \bibinfo {pages}
  {385 -- 391} (\bibinfo {year} {2015})}\BibitemShut {NoStop}%
\bibitem [{\citenamefont {Magnusson}\ \emph {et~al.}(2005)\citenamefont
  {Magnusson}, \citenamefont {Farewell},\ and\ \citenamefont
  {Nystr\o{}m}}]{MAGNUSSON2005}%
  \BibitemOpen
  \bibfield  {author} {\bibinfo {author} {\bibfnamefont {L.~U.}\ \bibnamefont
  {Magnusson}}, \bibinfo {author} {\bibfnamefont {A.}~\bibnamefont {Farewell}},
  \ and\ \bibinfo {author} {\bibfnamefont {T.}~\bibnamefont {Nystr\o{}m}},\
  }\bibfield  {title} {\enquote {\bibinfo {title} {{ppGpp}: a global regulator
  in \textit{Escherichia coli}},}\ }\href {\doibase
  https://doi.org/10.1016/j.tim.2005.03.008} {\bibfield  {journal} {\bibinfo
  {journal} {Trends Microbiol.}\ }\textbf {\bibinfo {volume} {13}},\ \bibinfo
  {pages} {236 -- 242} (\bibinfo {year} {2005})}\BibitemShut {NoStop}%
\bibitem [{\citenamefont {Ferullo}\ and\ \citenamefont
  {Lovett}(2008)}]{Ferullo2008}%
  \BibitemOpen
  \bibfield  {author} {\bibinfo {author} {\bibfnamefont {D.~J.}\ \bibnamefont
  {Ferullo}}\ and\ \bibinfo {author} {\bibfnamefont {S.~T.}\ \bibnamefont
  {Lovett}},\ }\bibfield  {title} {\enquote {\bibinfo {title} {The stringent
  response and cell cycle arrest in \textit{Escherichia coli}},}\ }\href
  {\doibase 10.1371/journal.pgen.1000300} {\bibfield  {journal} {\bibinfo
  {journal} {PLoS Genet.}\ }\textbf {\bibinfo {volume} {4}},\ \bibinfo {pages}
  {e1000300} (\bibinfo {year} {2008})}\BibitemShut {NoStop}%
\bibitem [{\citenamefont {B\"{a}r}\ \emph {et~al.}(2019)\citenamefont
  {B\"{a}r}, \citenamefont {Gro{\ss}mann}, \citenamefont {Heidenreich},\ and\
  \citenamefont {Peruani}}]{Baer2019}%
  \BibitemOpen
  \bibfield  {author} {\bibinfo {author} {\bibfnamefont {M.}~\bibnamefont
  {B\"{a}r}}, \bibinfo {author} {\bibfnamefont {R.}~\bibnamefont
  {Gro{\ss}mann}}, \bibinfo {author} {\bibfnamefont {S.}~\bibnamefont
  {Heidenreich}}, \ and\ \bibinfo {author} {\bibfnamefont {F.}~\bibnamefont
  {Peruani}},\ }\bibfield  {title} {\enquote {\bibinfo {title} {Self-propelled
  rods: Insights and perspectives for active matter},}\ }\href {\doibase
  10.1146/annurev-conmatphys-031119-050611} {\bibfield  {journal} {\bibinfo
  {journal} {Annu. Rev. Condens. Matter Phys.}\ }\textbf {\bibinfo {volume}
  {11}},\ \bibinfo {pages} {441--466} (\bibinfo {year} {2019})}\BibitemShut
  {NoStop}%
\bibitem [{\citenamefont {Be{'}er}\ and\ \citenamefont
  {Ariel}(2019)}]{Be'er2019}%
  \BibitemOpen
  \bibfield  {author} {\bibinfo {author} {\bibfnamefont {A.}~\bibnamefont
  {Be{'}er}}\ and\ \bibinfo {author} {\bibfnamefont {G.}~\bibnamefont
  {Ariel}},\ }\bibfield  {title} {\enquote {\bibinfo {title} {{A statistical
  physics view of swarming bacteria}},}\ }\href {\doibase
  10.1186/s40462-019-0153-9} {\bibfield  {journal} {\bibinfo  {journal} {Mov.
  Ecol.}\ }\textbf {\bibinfo {volume} {7}},\ \bibinfo {pages} {9} (\bibinfo
  {year} {2019})}\BibitemShut {NoStop}%
\bibitem [{\citenamefont {Hall-Stoodley}\ \emph {et~al.}(2004)\citenamefont
  {Hall-Stoodley}, \citenamefont {Costerton},\ and\ \citenamefont
  {Stoodley}}]{HallStoodley2004}%
  \BibitemOpen
  \bibfield  {author} {\bibinfo {author} {\bibfnamefont {L.}~\bibnamefont
  {Hall-Stoodley}}, \bibinfo {author} {\bibfnamefont {J.~W.}\ \bibnamefont
  {Costerton}}, \ and\ \bibinfo {author} {\bibfnamefont {P.}~\bibnamefont
  {Stoodley}},\ }\bibfield  {title} {\enquote {\bibinfo {title} {Bacterial
  biofilms: from the natural environment to infectious diseases},}\ }\href
  {\doibase 10.1038/nrmicro821} {\bibfield  {journal} {\bibinfo  {journal}
  {Nat. Rev. Microbiol.}\ }\textbf {\bibinfo {volume} {2}},\ \bibinfo {pages}
  {95--108} (\bibinfo {year} {2004})}\BibitemShut {NoStop}%
\bibitem [{\citenamefont {Boudarel}\ \emph {et~al.}(2018)\citenamefont
  {Boudarel}, \citenamefont {Mathias}, \citenamefont {Blaysat},\ and\
  \citenamefont {Gr\'{e}diac}}]{Boudarel2018}%
  \BibitemOpen
  \bibfield  {author} {\bibinfo {author} {\bibfnamefont {H.}~\bibnamefont
  {Boudarel}}, \bibinfo {author} {\bibfnamefont {J.}~\bibnamefont {Mathias}},
  \bibinfo {author} {\bibfnamefont {B.}~\bibnamefont {Blaysat}}, \ and\
  \bibinfo {author} {\bibfnamefont {M.}~\bibnamefont {Gr\'{e}diac}},\
  }\bibfield  {title} {\enquote {\bibinfo {title} {Towards standardized
  mechanical characterization of microbial biofilms: analysis and critical
  review},}\ }\href {\doibase 10.1038/s41522-018-0062-5} {\bibfield  {journal}
  {\bibinfo  {journal} {NPJ Biofilms Microbiomes}\ }\textbf {\bibinfo {volume}
  {4}},\ \bibinfo {pages} {17} (\bibinfo {year} {2018})}\BibitemShut {NoStop}%
\bibitem [{\citenamefont {Fuqua}\ \emph {et~al.}(2019)\citenamefont {Fuqua},
  \citenamefont {Filloux}, \citenamefont {Ghigo},\ and\ \citenamefont
  {Visick}}]{Fuquae2019}%
  \BibitemOpen
  \bibfield  {author} {\bibinfo {author} {\bibfnamefont {C.}~\bibnamefont
  {Fuqua}}, \bibinfo {author} {\bibfnamefont {A.}~\bibnamefont {Filloux}},
  \bibinfo {author} {\bibfnamefont {J.}~\bibnamefont {Ghigo}}, \ and\ \bibinfo
  {author} {\bibfnamefont {K.~L.}\ \bibnamefont {Visick}},\ }\bibfield  {title}
  {\enquote {\bibinfo {title} {Biofilms 2018: a diversity of microbes and
  mechanisms},}\ }\href {\doibase 10.1128/JB.00118-19} {\bibfield  {journal}
  {\bibinfo  {journal} {J. Bacteriol}\ }\textbf {\bibinfo {volume} {201}},\
  \bibinfo {pages} {e00118--19} (\bibinfo {year} {2019})}\BibitemShut {NoStop}%
\bibitem [{\citenamefont {Grilli}\ \emph {et~al.}(2017)\citenamefont {Grilli},
  \citenamefont {Osella}, \citenamefont {Kennard},\ and\ \citenamefont
  {Cosentino~Lagomarsino}}]{Grilli2017}%
  \BibitemOpen
  \bibfield  {author} {\bibinfo {author} {\bibfnamefont {J.}~\bibnamefont
  {Grilli}}, \bibinfo {author} {\bibfnamefont {M.}~\bibnamefont {Osella}},
  \bibinfo {author} {\bibfnamefont {A.~S.}\ \bibnamefont {Kennard}}, \ and\
  \bibinfo {author} {\bibfnamefont {M.}~\bibnamefont {Cosentino~Lagomarsino}},\
  }\bibfield  {title} {\enquote {\bibinfo {title} {Relevant parameters in
  models of cell division control},}\ }\href {\doibase
  10.1103/PhysRevE.95.032411} {\bibfield  {journal} {\bibinfo  {journal} {Phys.
  Rev. E}\ }\textbf {\bibinfo {volume} {95}},\ \bibinfo {pages} {032411}
  (\bibinfo {year} {2017})}\BibitemShut {NoStop}%
\bibitem [{\citenamefont {Furusawa}\ \emph {et~al.}(2005)\citenamefont
  {Furusawa}, \citenamefont {Suzuki}, \citenamefont {Kashiwagi}, \citenamefont
  {Yomo},\ and\ \citenamefont {Kaneko}}]{Furusawa2005}%
  \BibitemOpen
  \bibfield  {author} {\bibinfo {author} {\bibfnamefont {C.}~\bibnamefont
  {Furusawa}}, \bibinfo {author} {\bibfnamefont {T.}~\bibnamefont {Suzuki}},
  \bibinfo {author} {\bibfnamefont {A.}~\bibnamefont {Kashiwagi}}, \bibinfo
  {author} {\bibfnamefont {T.}~\bibnamefont {Yomo}}, \ and\ \bibinfo {author}
  {\bibfnamefont {K.}~\bibnamefont {Kaneko}},\ }\bibfield  {title} {\enquote
  {\bibinfo {title} {Ubiquity of log-normal distributions in intra-cellular
  reaction dynamics},}\ }\href {\doibase 10.2142/biophysics.1.25} {\bibfield
  {journal} {\bibinfo  {journal} {Biophysics}\ }\textbf {\bibinfo {volume}
  {1}},\ \bibinfo {pages} {25--31} (\bibinfo {year} {2005})}\BibitemShut
  {NoStop}%
\bibitem [{\citenamefont {Salman}\ \emph {et~al.}(2012)\citenamefont {Salman},
  \citenamefont {Brenner}, \citenamefont {Tung}, \citenamefont {Elyahu},
  \citenamefont {Stolovicki}, \citenamefont {Moore}, \citenamefont
  {Libchaber},\ and\ \citenamefont {Braun}}]{Salman2012}%
  \BibitemOpen
  \bibfield  {author} {\bibinfo {author} {\bibfnamefont {H.}~\bibnamefont
  {Salman}}, \bibinfo {author} {\bibfnamefont {N.}~\bibnamefont {Brenner}},
  \bibinfo {author} {\bibfnamefont {C.~K.}\ \bibnamefont {Tung}}, \bibinfo
  {author} {\bibfnamefont {N.}~\bibnamefont {Elyahu}}, \bibinfo {author}
  {\bibfnamefont {E.}~\bibnamefont {Stolovicki}}, \bibinfo {author}
  {\bibfnamefont {L.}~\bibnamefont {Moore}}, \bibinfo {author} {\bibfnamefont
  {A.}~\bibnamefont {Libchaber}}, \ and\ \bibinfo {author} {\bibfnamefont
  {E.}~\bibnamefont {Braun}},\ }\bibfield  {title} {\enquote {\bibinfo {title}
  {Universal protein fluctuations in populations of microorganisms},}\ }\href
  {\doibase 10.1103/PhysRevLett.108.238105} {\bibfield  {journal} {\bibinfo
  {journal} {Phys. Rev. Lett.}\ }\textbf {\bibinfo {volume} {108}},\ \bibinfo
  {pages} {238105} (\bibinfo {year} {2012})}\BibitemShut {NoStop}%
\bibitem [{\citenamefont {Brenner}\ \emph {et~al.}(2015)\citenamefont
  {Brenner}, \citenamefont {Braun}, \citenamefont {Yoney}, \citenamefont
  {Susman}, \citenamefont {Rotella},\ and\ \citenamefont
  {Salman}}]{Brenner2015}%
  \BibitemOpen
  \bibfield  {author} {\bibinfo {author} {\bibfnamefont {N.}~\bibnamefont
  {Brenner}}, \bibinfo {author} {\bibfnamefont {E.}~\bibnamefont {Braun}},
  \bibinfo {author} {\bibfnamefont {A.}~\bibnamefont {Yoney}}, \bibinfo
  {author} {\bibfnamefont {L.}~\bibnamefont {Susman}}, \bibinfo {author}
  {\bibfnamefont {J.}~\bibnamefont {Rotella}}, \ and\ \bibinfo {author}
  {\bibfnamefont {H.}~\bibnamefont {Salman}},\ }\bibfield  {title} {\enquote
  {\bibinfo {title} {Single-cell protein dynamics reproduce universal
  fluctuations in cell populations},}\ }\href {\doibase
  10.1140/epje/i2015-15102-8} {\bibfield  {journal} {\bibinfo  {journal} {Eur.
  Phys. J. E}\ }\textbf {\bibinfo {volume} {38}},\ \bibinfo {pages} {102}
  (\bibinfo {year} {2015})}\BibitemShut {NoStop}%
\bibitem [{\citenamefont {Rulands}\ \emph {et~al.}(2018)\citenamefont
  {Rulands}, \citenamefont {Lescroart}, \citenamefont {Chabab}, \citenamefont
  {Hindley}, \citenamefont {Prior}, \citenamefont {Sznurkowska}, \citenamefont
  {Huch}, \citenamefont {Philpott}, \citenamefont {Blanpain},\ and\
  \citenamefont {Simons}}]{Rulands2018}%
  \BibitemOpen
  \bibfield  {author} {\bibinfo {author} {\bibfnamefont {S.}~\bibnamefont
  {Rulands}}, \bibinfo {author} {\bibfnamefont {F.}~\bibnamefont {Lescroart}},
  \bibinfo {author} {\bibfnamefont {S.}~\bibnamefont {Chabab}}, \bibinfo
  {author} {\bibfnamefont {C.~J.}\ \bibnamefont {Hindley}}, \bibinfo {author}
  {\bibfnamefont {N.}~\bibnamefont {Prior}}, \bibinfo {author} {\bibfnamefont
  {M.~K.}\ \bibnamefont {Sznurkowska}}, \bibinfo {author} {\bibfnamefont
  {M.}~\bibnamefont {Huch}}, \bibinfo {author} {\bibfnamefont {A.}~\bibnamefont
  {Philpott}}, \bibinfo {author} {\bibfnamefont {C.}~\bibnamefont {Blanpain}},
  \ and\ \bibinfo {author} {\bibfnamefont {B.~D.}\ \bibnamefont {Simons}},\
  }\bibfield  {title} {\enquote {\bibinfo {title} {{Universality of clone
  dynamics during tissue development}},}\ }\href {\doibase
  10.1038/s41567-018-0055-6} {\bibfield  {journal} {\bibinfo  {journal} {Nat.
  Phys.}\ }\textbf {\bibinfo {volume} {14}},\ \bibinfo {pages} {469--474}
  (\bibinfo {year} {2018})}\BibitemShut {NoStop}%
\bibitem [{\citenamefont {Sezonov}\ \emph {et~al.}(2007)\citenamefont
  {Sezonov}, \citenamefont {Joseleau-Petit},\ and\ \citenamefont
  {D{\textquoteright}Ari}}]{Sezonov2007}%
  \BibitemOpen
  \bibfield  {author} {\bibinfo {author} {\bibfnamefont {G.}~\bibnamefont
  {Sezonov}}, \bibinfo {author} {\bibfnamefont {D.}~\bibnamefont
  {Joseleau-Petit}}, \ and\ \bibinfo {author} {\bibfnamefont {R.}~\bibnamefont
  {D{\textquoteright}Ari}},\ }\bibfield  {title} {\enquote {\bibinfo {title}
  {\textit{Escherichia coli} physiology in {Luria-Bertani} broth},}\ }\href
  {\doibase 10.1128/JB.01368-07} {\bibfield  {journal} {\bibinfo  {journal} {J.
  Bacteriol}\ }\textbf {\bibinfo {volume} {189}},\ \bibinfo {pages}
  {8746--8749} (\bibinfo {year} {2007})}\BibitemShut {NoStop}%
\end{thebibliography}%


\providecommand{\noopsort}[1]{}\providecommand{\singleletter}[1]{#1}%
\begin{thebibliography}{19}%
\makeatletter
\providecommand \@ifxundefined [1]{%
 \@ifx{#1\undefined}
}%
\providecommand \@ifnum [1]{%
 \ifnum #1\expandafter \@firstoftwo
 \else \expandafter \@secondoftwo
 \fi
}%
\providecommand \@ifx [1]{%
 \ifx #1\expandafter \@firstoftwo
 \else \expandafter \@secondoftwo
 \fi
}%
\providecommand \natexlab [1]{#1}%
\providecommand \enquote  [1]{``#1''}%
\providecommand \bibnamefont  [1]{#1}%
\providecommand \bibfnamefont [1]{#1}%
\providecommand \citenamefont [1]{#1}%
\providecommand \href@noop [0]{\@secondoftwo}%
\providecommand \href [0]{\begingroup \@sanitize@url \@href}%
\providecommand \@href[1]{\@@startlink{#1}\@@href}%
\providecommand \@@href[1]{\endgroup#1\@@endlink}%
\providecommand \@sanitize@url [0]{\catcode `\\12\catcode `\$12\catcode
  `\&12\catcode `\#12\catcode `\^12\catcode `\_12\catcode `\%12\relax}%
\providecommand \@@startlink[1]{}%
\providecommand \@@endlink[0]{}%
\providecommand \url  [0]{\begingroup\@sanitize@url \@url }%
\providecommand \@url [1]{\endgroup\@href {#1}{\urlprefix }}%
\providecommand \urlprefix  [0]{URL }%
\providecommand \Eprint [0]{\href }%
\providecommand \doibase [0]{http://dx.doi.org/}%
\providecommand \selectlanguage [0]{\@gobble}%
\providecommand \bibinfo  [0]{\@secondoftwo}%
\providecommand \bibfield  [0]{\@secondoftwo}%
\providecommand \translation [1]{[#1]}%
\providecommand \BibitemOpen [0]{}%
\providecommand \bibitemStop [0]{}%
\providecommand \bibitemNoStop [0]{.\EOS\space}%
\providecommand \EOS [0]{\spacefactor3000\relax}%
\providecommand \BibitemShut  [1]{\csname bibitem#1\endcsname}%
\let\auto@bib@innerbib\@empty
\bibitem [{\citenamefont {Nakaoka}\ and\ \citenamefont
  {Wakamoto}(2017)}]{Nakaoka2017}%
  \BibitemOpen
  \bibfield  {author} {\bibinfo {author} {\bibfnamefont {H.}~\bibnamefont
  {Nakaoka}}\ and\ \bibinfo {author} {\bibfnamefont {Y.}~\bibnamefont
  {Wakamoto}},\ }\href {\doibase 10.1371/journal.pbio.2001109} {\bibfield
  {journal} {\bibinfo  {journal} {PLoS Biol.}\ }\textbf {\bibinfo {volume}
  {15}},\ \bibinfo {pages} {1} (\bibinfo {year} {2017})}\BibitemShut {NoStop}%
\bibitem [{\citenamefont {Mather}\ \emph {et~al.}(2010)\citenamefont {Mather},
  \citenamefont {Mondrag\'{o}n-Palomino}, \citenamefont {Danino}, \citenamefont
  {Hasty},\ and\ \citenamefont {Tsimring}}]{Mather2010}%
  \BibitemOpen
  \bibfield  {author} {\bibinfo {author} {\bibfnamefont {W.}~\bibnamefont
  {Mather}}, \bibinfo {author} {\bibfnamefont {O.}~\bibnamefont
  {Mondrag\'{o}n-Palomino}}, \bibinfo {author} {\bibfnamefont {T.}~\bibnamefont
  {Danino}}, \bibinfo {author} {\bibfnamefont {J.}~\bibnamefont {Hasty}}, \
  and\ \bibinfo {author} {\bibfnamefont {L.~S.}\ \bibnamefont {Tsimring}},\
  }\href {\doibase 10.1103/PhysRevLett.104.208101} {\bibfield  {journal}
  {\bibinfo  {journal} {Phys. Rev. Lett.}\ }\textbf {\bibinfo {volume} {104}},\
  \bibinfo {pages} {208101} (\bibinfo {year} {2010})}\BibitemShut {NoStop}%
\bibitem [{\citenamefont {Inoue}\ \emph {et~al.}(2001)\citenamefont {Inoue},
  \citenamefont {Wakamoto}, \citenamefont {Moriguchi}, \citenamefont {Okano},\
  and\ \citenamefont {Yasuda}}]{Inoue2001}%
  \BibitemOpen
  \bibfield  {author} {\bibinfo {author} {\bibfnamefont {I.}~\bibnamefont
  {Inoue}}, \bibinfo {author} {\bibfnamefont {Y.}~\bibnamefont {Wakamoto}},
  \bibinfo {author} {\bibfnamefont {H.}~\bibnamefont {Moriguchi}}, \bibinfo
  {author} {\bibfnamefont {K.}~\bibnamefont {Okano}}, \ and\ \bibinfo {author}
  {\bibfnamefont {K.}~\bibnamefont {Yasuda}},\ }\href {\doibase
  10.1039/B103931H} {\bibfield  {journal} {\bibinfo  {journal} {Lab Chip}\
  }\textbf {\bibinfo {volume} {1}},\ \bibinfo {pages} {50} (\bibinfo {year}
  {2001})}\BibitemShut {NoStop}%
\bibitem [{\citenamefont {Hashimoto}\ \emph {et~al.}(2016)\citenamefont
  {Hashimoto}, \citenamefont {Nozoe}, \citenamefont {Nakaoka}, \citenamefont
  {Okura}, \citenamefont {Akiyoshi}, \citenamefont {Kaneko}, \citenamefont
  {Kussell},\ and\ \citenamefont {Wakamoto}}]{Hashimoto2016}%
  \BibitemOpen
  \bibfield  {author} {\bibinfo {author} {\bibfnamefont {M.}~\bibnamefont
  {Hashimoto}}, \bibinfo {author} {\bibfnamefont {T.}~\bibnamefont {Nozoe}},
  \bibinfo {author} {\bibfnamefont {H.}~\bibnamefont {Nakaoka}}, \bibinfo
  {author} {\bibfnamefont {R.}~\bibnamefont {Okura}}, \bibinfo {author}
  {\bibfnamefont {S.}~\bibnamefont {Akiyoshi}}, \bibinfo {author}
  {\bibfnamefont {K.}~\bibnamefont {Kaneko}}, \bibinfo {author} {\bibfnamefont
  {E.}~\bibnamefont {Kussell}}, \ and\ \bibinfo {author} {\bibfnamefont
  {Y.}~\bibnamefont {Wakamoto}},\ }\href {\doibase 10.1073/pnas.1519412113}
  {\bibfield  {journal} {\bibinfo  {journal} {Proc. Natl. Acad. Sci. USA}\
  }\textbf {\bibinfo {volume} {113}},\ \bibinfo {pages} {3251} (\bibinfo {year}
  {2016})}\BibitemShut {NoStop}%
\bibitem [{\citenamefont {Sezonov}\ \emph {et~al.}(2007)\citenamefont
  {Sezonov}, \citenamefont {Joseleau-Petit},\ and\ \citenamefont
  {D{\textquoteright}Ari}}]{Sezonov2007}%
  \BibitemOpen
  \bibfield  {author} {\bibinfo {author} {\bibfnamefont {G.}~\bibnamefont
  {Sezonov}}, \bibinfo {author} {\bibfnamefont {D.}~\bibnamefont
  {Joseleau-Petit}}, \ and\ \bibinfo {author} {\bibfnamefont {R.}~\bibnamefont
  {D{\textquoteright}Ari}},\ }\href {\doibase 10.1128/JB.01368-07} {\bibfield
  {journal} {\bibinfo  {journal} {J. Bacteriol}\ }\textbf {\bibinfo {volume}
  {189}},\ \bibinfo {pages} {8746} (\bibinfo {year} {2007})}\BibitemShut
  {NoStop}%
\bibitem [{\citenamefont {Wakamoto}\ \emph {et~al.}(2005)\citenamefont
  {Wakamoto}, \citenamefont {Ramsden},\ and\ \citenamefont
  {Yasuda}}]{Wakamoto2005}%
  \BibitemOpen
  \bibfield  {author} {\bibinfo {author} {\bibfnamefont {Y.}~\bibnamefont
  {Wakamoto}}, \bibinfo {author} {\bibfnamefont {J.}~\bibnamefont {Ramsden}}, \
  and\ \bibinfo {author} {\bibfnamefont {K.}~\bibnamefont {Yasuda}},\ }\href
  {\doibase 10.1039/b409860a} {\bibfield  {journal} {\bibinfo  {journal} {The
  Analyst}\ }\textbf {\bibinfo {volume} {130}},\ \bibinfo {pages} {311}
  (\bibinfo {year} {2005})}\BibitemShut {NoStop}%
\bibitem [{\citenamefont {Marucci}\ \emph {et~al.}(2006)\citenamefont
  {Marucci}, \citenamefont {Ragnarsson},\ and\ \citenamefont
  {Axelsson}}]{MARUCCI2006}%
  \BibitemOpen
  \bibfield  {author} {\bibinfo {author} {\bibfnamefont {M.}~\bibnamefont
  {Marucci}}, \bibinfo {author} {\bibfnamefont {G.}~\bibnamefont {Ragnarsson}},
  \ and\ \bibinfo {author} {\bibfnamefont {A.}~\bibnamefont {Axelsson}},\
  }\href {\doibase 10.1016/j.jconrel.2006.06.019} {\bibfield  {journal}
  {\bibinfo  {journal} {J. Control. Release.}\ }\textbf {\bibinfo {volume}
  {114}},\ \bibinfo {pages} {369 } (\bibinfo {year} {2006})}\BibitemShut
  {NoStop}%
\bibitem [{\citenamefont {Volfson}\ \emph {et~al.}(2008)\citenamefont
  {Volfson}, \citenamefont {Cookson}, \citenamefont {Hasty},\ and\
  \citenamefont {Tsimring}}]{Volfson2008}%
  \BibitemOpen
  \bibfield  {author} {\bibinfo {author} {\bibfnamefont {D.}~\bibnamefont
  {Volfson}}, \bibinfo {author} {\bibfnamefont {S.}~\bibnamefont {Cookson}},
  \bibinfo {author} {\bibfnamefont {J.}~\bibnamefont {Hasty}}, \ and\ \bibinfo
  {author} {\bibfnamefont {L.~S.}\ \bibnamefont {Tsimring}},\ }\href {\doibase
  10.1073/pnas.0706805105} {\bibfield  {journal} {\bibinfo  {journal} {Proc.
  Natl. Acad. Sci. USA}\ }\textbf {\bibinfo {volume} {105}},\ \bibinfo {pages}
  {15346} (\bibinfo {year} {2008})}\BibitemShut {NoStop}%
\bibitem [{\citenamefont {Boyer}\ \emph {et~al.}(2011)\citenamefont {Boyer},
  \citenamefont {Mather}, \citenamefont {Mondrag\'on-Palomino}, \citenamefont
  {Orozco-Fuentes}, \citenamefont {Danino}, \citenamefont {Hasty},\ and\
  \citenamefont {Tsimring}}]{Boyer2011}%
  \BibitemOpen
  \bibfield  {author} {\bibinfo {author} {\bibfnamefont {D.}~\bibnamefont
  {Boyer}}, \bibinfo {author} {\bibfnamefont {W.}~\bibnamefont {Mather}},
  \bibinfo {author} {\bibfnamefont {O.}~\bibnamefont {Mondrag\'on-Palomino}},
  \bibinfo {author} {\bibfnamefont {S.}~\bibnamefont {Orozco-Fuentes}},
  \bibinfo {author} {\bibfnamefont {T.}~\bibnamefont {Danino}}, \bibinfo
  {author} {\bibfnamefont {J.}~\bibnamefont {Hasty}}, \ and\ \bibinfo {author}
  {\bibfnamefont {L.~S.}\ \bibnamefont {Tsimring}},\ }\href {\doibase
  10.1088/1478-3975/8/2/026008} {\bibfield  {journal} {\bibinfo  {journal}
  {Phys. Biol.}\ }\textbf {\bibinfo {volume} {8}},\ \bibinfo {pages} {026008}
  (\bibinfo {year} {2011})}\BibitemShut {NoStop}%
\bibitem [{\citenamefont {Si}\ \emph {et~al.}(2016)\citenamefont {Si},
  \citenamefont {Li}, \citenamefont {Margolin},\ and\ \citenamefont
  {Sun}}]{Si2015}%
  \BibitemOpen
  \bibfield  {author} {\bibinfo {author} {\bibfnamefont {F.}~\bibnamefont
  {Si}}, \bibinfo {author} {\bibfnamefont {B.}~\bibnamefont {Li}}, \bibinfo
  {author} {\bibfnamefont {W.}~\bibnamefont {Margolin}}, \ and\ \bibinfo
  {author} {\bibfnamefont {S.~X.}\ \bibnamefont {Sun}},\ }\href {\doibase
  10.1038/srep11367} {\bibfield  {journal} {\bibinfo  {journal} {Sci. Rep.}\
  }\textbf {\bibinfo {volume} {5}},\ \bibinfo {pages} {11367} (\bibinfo {year}
  {2016})}\BibitemShut {NoStop}%
\bibitem [{\citenamefont {Chu}\ \emph {et~al.}(2018)\citenamefont {Chu},
  \citenamefont {Kilic}, \citenamefont {Cho}, \citenamefont {Groisman},\ and\
  \citenamefont {Levchenko}}]{Chu2018}%
  \BibitemOpen
  \bibfield  {author} {\bibinfo {author} {\bibfnamefont {E.~K.}\ \bibnamefont
  {Chu}}, \bibinfo {author} {\bibfnamefont {O.}~\bibnamefont {Kilic}}, \bibinfo
  {author} {\bibfnamefont {H.}~\bibnamefont {Cho}}, \bibinfo {author}
  {\bibfnamefont {A.}~\bibnamefont {Groisman}}, \ and\ \bibinfo {author}
  {\bibfnamefont {A.}~\bibnamefont {Levchenko}},\ }\href {\doibase
  10.1038/s41467-018-06552-z} {\bibfield  {journal} {\bibinfo  {journal} {Nat.
  Commun.}\ }\textbf {\bibinfo {volume} {9}},\ \bibinfo {pages} {4087}
  (\bibinfo {year} {2018})}\BibitemShut {NoStop}%
\bibitem [{\citenamefont {Bruger}\ and\ \citenamefont
  {Waters}(2016)}]{Bruger2016}%
  \BibitemOpen
  \bibfield  {author} {\bibinfo {author} {\bibfnamefont {E.~L.}\ \bibnamefont
  {Bruger}}\ and\ \bibinfo {author} {\bibfnamefont {C.~M.}\ \bibnamefont
  {Waters}},\ }\href {\doibase 10.1128/AEM.01945-16} {\bibfield  {journal}
  {\bibinfo  {journal} {Appl. Environ. Microbiol.}\ }\textbf {\bibinfo {volume}
  {82}},\ \bibinfo {pages} {6498} (\bibinfo {year} {2016})}\BibitemShut
  {NoStop}%
\bibitem [{\citenamefont {Ha}\ \emph {et~al.}(2018)\citenamefont {Ha},
  \citenamefont {Hauk}, \citenamefont {Cho}, \citenamefont {Eo}, \citenamefont
  {Ma}, \citenamefont {Stephens}, \citenamefont {Cha}, \citenamefont {Jeong},
  \citenamefont {Suh}, \citenamefont {Sintim}, \citenamefont {Bentley},\ and\
  \citenamefont {Ryu}}]{Haeaar2018}%
  \BibitemOpen
  \bibfield  {author} {\bibinfo {author} {\bibfnamefont {J.}~\bibnamefont
  {Ha}}, \bibinfo {author} {\bibfnamefont {P.}~\bibnamefont {Hauk}}, \bibinfo
  {author} {\bibfnamefont {K.}~\bibnamefont {Cho}}, \bibinfo {author}
  {\bibfnamefont {Y.}~\bibnamefont {Eo}}, \bibinfo {author} {\bibfnamefont
  {X.}~\bibnamefont {Ma}}, \bibinfo {author} {\bibfnamefont {K.}~\bibnamefont
  {Stephens}}, \bibinfo {author} {\bibfnamefont {S.}~\bibnamefont {Cha}},
  \bibinfo {author} {\bibfnamefont {M.}~\bibnamefont {Jeong}}, \bibinfo
  {author} {\bibfnamefont {J.}~\bibnamefont {Suh}}, \bibinfo {author}
  {\bibfnamefont {H.~O.}\ \bibnamefont {Sintim}}, \bibinfo {author}
  {\bibfnamefont {W.~E.}\ \bibnamefont {Bentley}}, \ and\ \bibinfo {author}
  {\bibfnamefont {K.}~\bibnamefont {Ryu}},\ }\href {\doibase
  10.1126/sciadv.aar7063} {\bibfield  {journal} {\bibinfo  {journal} {Sci.
  Adv.}\ }\textbf {\bibinfo {volume} {4}},\ \bibinfo {pages} {eaar7063}
  (\bibinfo {year} {2018})}\BibitemShut {NoStop}%
\bibitem [{\citenamefont {Monod}(1949)}]{Monod1949}%
  \BibitemOpen
  \bibfield  {author} {\bibinfo {author} {\bibfnamefont {J.}~\bibnamefont
  {Monod}},\ }\href {\doibase 10.1146/annurev.mi.03.100149.002103} {\bibfield
  {journal} {\bibinfo  {journal} {Annu. Rev. Microbiol.}\ }\textbf {\bibinfo
  {volume} {3}},\ \bibinfo {pages} {371} (\bibinfo {year} {1949})}\BibitemShut
  {NoStop}%
\bibitem [{\citenamefont {Witz}\ \emph {et~al.}(2019)\citenamefont {Witz},
  \citenamefont {van Nimwegen},\ and\ \citenamefont {Julou}}]{Witz2019}%
  \BibitemOpen
  \bibfield  {author} {\bibinfo {author} {\bibfnamefont {G.}~\bibnamefont
  {Witz}}, \bibinfo {author} {\bibfnamefont {E.}~\bibnamefont {van Nimwegen}},
  \ and\ \bibinfo {author} {\bibfnamefont {T.}~\bibnamefont {Julou}},\ }\href
  {\doibase 10.7554/eLife.48063} {\bibfield  {journal} {\bibinfo  {journal}
  {eLife}\ }\textbf {\bibinfo {volume} {8}},\ \bibinfo {pages} {e48063}
  (\bibinfo {year} {2019})}\BibitemShut {NoStop}%
\bibitem [{\citenamefont {Si}\ \emph {et~al.}(2017)\citenamefont {Si},
  \citenamefont {Li}, \citenamefont {Cox}, \citenamefont {Sauls}, \citenamefont
  {Azizi}, \citenamefont {Sou}, \citenamefont {Schwartz}, \citenamefont
  {Erickstad}, \citenamefont {Jun}, \citenamefont {Li},\ and\ \citenamefont
  {Jun}}]{SI2017}%
  \BibitemOpen
  \bibfield  {author} {\bibinfo {author} {\bibfnamefont {F.}~\bibnamefont
  {Si}}, \bibinfo {author} {\bibfnamefont {D.}~\bibnamefont {Li}}, \bibinfo
  {author} {\bibfnamefont {S.~E.}\ \bibnamefont {Cox}}, \bibinfo {author}
  {\bibfnamefont {J.~T.}\ \bibnamefont {Sauls}}, \bibinfo {author}
  {\bibfnamefont {O.}~\bibnamefont {Azizi}}, \bibinfo {author} {\bibfnamefont
  {C.}~\bibnamefont {Sou}}, \bibinfo {author} {\bibfnamefont {A.~B.}\
  \bibnamefont {Schwartz}}, \bibinfo {author} {\bibfnamefont {M.~J.}\
  \bibnamefont {Erickstad}}, \bibinfo {author} {\bibfnamefont {Y.}~\bibnamefont
  {Jun}}, \bibinfo {author} {\bibfnamefont {X.}~\bibnamefont {Li}}, \ and\
  \bibinfo {author} {\bibfnamefont {S.}~\bibnamefont {Jun}},\ }\href {\doibase
  https://doi.org/10.1016/j.cub.2017.03.022} {\bibfield  {journal} {\bibinfo
  {journal} {Curr. Biol.}\ }\textbf {\bibinfo {volume} {27}},\ \bibinfo {pages}
  {1278 } (\bibinfo {year} {2017})}\BibitemShut {NoStop}%
\bibitem [{\citenamefont {Ho}\ and\ \citenamefont {Amir}(2015)}]{Ho2015}%
  \BibitemOpen
  \bibfield  {author} {\bibinfo {author} {\bibfnamefont {P.}~\bibnamefont
  {Ho}}\ and\ \bibinfo {author} {\bibfnamefont {A.}~\bibnamefont {Amir}},\
  }\href {\doibase 10.3389/fmicb.2015.00662} {\bibfield  {journal} {\bibinfo
  {journal} {Front. Microbiol.}\ }\textbf {\bibinfo {volume} {6}},\ \bibinfo
  {pages} {662} (\bibinfo {year} {2015})}\BibitemShut {NoStop}%
\bibitem [{\citenamefont {Wallden}\ \emph {et~al.}(2016)\citenamefont
  {Wallden}, \citenamefont {Fange}, \citenamefont {Lundius}, \citenamefont
  {Baltekin},\ and\ \citenamefont {Elf}}]{Wallden2016}%
  \BibitemOpen
  \bibfield  {author} {\bibinfo {author} {\bibfnamefont {M.}~\bibnamefont
  {Wallden}}, \bibinfo {author} {\bibfnamefont {D.}~\bibnamefont {Fange}},
  \bibinfo {author} {\bibfnamefont {E.~G.}\ \bibnamefont {Lundius}}, \bibinfo
  {author} {\bibfnamefont {O.}~\bibnamefont {Baltekin}}, \ and\ \bibinfo
  {author} {\bibfnamefont {J.}~\bibnamefont {Elf}},\ }\href {\doibase
  10.1016/j.cell.2016.06.052} {\bibfield  {journal} {\bibinfo  {journal}
  {Cell}\ }\textbf {\bibinfo {volume} {166}},\ \bibinfo {pages} {729 }
  (\bibinfo {year} {2016})}\BibitemShut {NoStop}%
\bibitem [{\citenamefont {Taheri-Araghi}\ \emph {et~al.}(2015)\citenamefont
  {Taheri-Araghi}, \citenamefont {Bradde}, \citenamefont {Sauls}, \citenamefont
  {Hill}, \citenamefont {Levin}, \citenamefont {Paulsson}, \citenamefont
  {Vergassola},\ and\ \citenamefont {Jun}}]{TAHERIARAGHI2015}%
  \BibitemOpen
  \bibfield  {author} {\bibinfo {author} {\bibfnamefont {S.}~\bibnamefont
  {Taheri-Araghi}}, \bibinfo {author} {\bibfnamefont {S.}~\bibnamefont
  {Bradde}}, \bibinfo {author} {\bibfnamefont {J.~T.}\ \bibnamefont {Sauls}},
  \bibinfo {author} {\bibfnamefont {N.~S.}\ \bibnamefont {Hill}}, \bibinfo
  {author} {\bibfnamefont {P.~A.}\ \bibnamefont {Levin}}, \bibinfo {author}
  {\bibfnamefont {J.}~\bibnamefont {Paulsson}}, \bibinfo {author}
  {\bibfnamefont {M.}~\bibnamefont {Vergassola}}, \ and\ \bibinfo {author}
  {\bibfnamefont {S.}~\bibnamefont {Jun}},\ }\href {\doibase
  10.1016/j.cub.2014.12.009} {\bibfield  {journal} {\bibinfo  {journal} {Curr.
  Biol.}\ }\textbf {\bibinfo {volume} {25}},\ \bibinfo {pages} {385 } (\bibinfo
  {year} {2015})}\BibitemShut {NoStop}%
\end{thebibliography}%

\end{document}


\title{Supplementary Information for\\``Scale invariance of cell size fluctuations in starving bacteria''}

\author{Takuro Shimaya}
\email{t.shimaya@noneq.phys.s.u-tokyo.ac.jp}
\affiliation{Department of Physics, Graduate School of Science, University of Tokyo, Tokyo 113-0033, Japan}
\author{Reiko Okura} 
\affiliation{Department of Basic Science, Graduate School of Arts and Sciences, University of Tokyo, Tokyo 153-8902, Japan}
\author{Yuichi Wakamoto}
\affiliation{Department of Basic Science, Graduate School of Arts and Sciences, University of Tokyo, Tokyo 153-8902, Japan}
\author{Kazumasa A. Takeuchi}
\email{kat@kaztake.org}
\affiliation{Department of Physics, Graduate School of Science, University of Tokyo, Tokyo 113-0033, Japan}
\affiliation{Department of Physics, School of Science, Tokyo Institute of Technology, Tokyo 152-8551, Japan}

\maketitle

\section*{Supplementary Note 1: Supplementary experimental methods}
\subsection{Strains and culture media}
We used wild-type {\sl E. coli} strains (MG1655 and RP437) and a mutant strain (W3110 $\Delta$fliC $\Delta$flu $\Delta$fimA)
 in this study.
Culture media and buffer are listed in Supplementary~Table~1.
The osmotic pressure of each medium was measured by the freezing-point depression method
 by the OSMOMAT 030 (Genotec, Berlin Germany).
Details on the strains and culture conditions in each experiment
 are provided below.

\subsection{Fabrication of the PDMS-based device}
We prepared PDMS-based microfluidic devices by following the method
reported in Ref.\cite{Nakaoka2017}.
We adopted a microchannel geometry similar to that in Ref.\cite{Mather2010},
which consists of deep drain channels and shallow U-shape traps (Supplementary~Fig.~2c,d).
In our setup, the drain channels are $25\unit{\mu{}m}$ deep, and 
the U-shape traps are $30\unit{\mu{}m}$ width, $70$-$90\unit{\mu{}m}$ long, and $1.1\unit{\mu{}m}$ deep.
Tygon tubes of 1/16" outer diameter were connected to the inlets and the outlets (Supplementary~Fig.~2d)
via steel tubes (Elveflow, Paris France).
The width of the drains (see Supplementary~Fig.~2d) and the length of tygon tubes were adjusted to realize the desired flow rate in the drain near the U-shape traps.
The tube length was $55\unit{cm}$ for the medium inlet, $40\unit{cm}$ for the medium outlet, and $30\unit{cm}$ for the waste outlet.
The cell inlet was connected to a syringe filled with bacterial suspension via a $10\unit{cm}$ tube, to inject cells at the beginning of observation.
After the intrusion of cells, the syringe with the suspension was fixed,
and no flow was generated around the cell inlet.

\subsection{Fabrication of the EMPS}
The EMPS consists of a microfabricated glass coverslip,
a bilayer porous membrane and a PDMS pad.
The microfabricated coverslip and the PDMS pad were prepared
according to ref. \cite{Inoue2001, Hashimoto2016}.
We fabricated the bilayer porous membrane by combining a streptavidin decorated cellulose membrane
and a biotin decorated polyethylene-terephthalate (PET) membrane.
The streptavidin decoration of the cellulose membrane (Spectra/Por 7, Repligen, Waltham Massachusetts, molecular weight cut-off $25000$) was realized by the method described in ref. \cite{Inoue2001, Hashimoto2016}.
The PET membrane (Transwell 3450, Corning, Corning New York, nominal pore size $0.4\unit{\mu{}m}$) was decorated with biotin as follows.
We soaked a PET membrane in 1 wt\% solution of
3-(2-aminoethyl aminopropyl) trimethoxysilane (Shinetsu Kagaku Kogyo, Tokyo Japan) for 45 min,
dried it at 125${}^\circ$C for 25 min and washed it by ultrasonic cleaning in Milli-Q water for 5 min.
This preprocessed PET membrane was stored in a desiccator at room temperature, until it was used to assemble the EMPS.

The EMPS was assembled as follows.
The preprocessed PET membrane was cut into $5\unit{mm} \times 5\unit{mm}$ squares, soaked in the biotin solution for 4 hours and dried on filter paper.
The biotin decorated PET membrane was attached with a streptavidin decorated cellulose membrane, cut to the size of the PET membrane, by sandwiching them between agar pads (M9 medium with $2\unit{wt\%}$ agarose).
In the meantime, a $1\unit{\mu{}l}$ droplet of bacterial suspension was inoculated on a biotin decorated coverslip
(see also details below).
We then took the bilayer membrane from the agar pad, air-dried for tens of seconds,
and carefully put on the coverslip on top of the bacterial suspension.
The bilayer membrane was then attached to the coverslip via streptavidin-biotin binding as shown in Supplementary~Fig.~1b.
We then air-dried the membrane for a minute and attached a PDMS pad on the coverslip by a double-sided tape.

\subsection{Observation of motile {\sl E. coli} in the EMPS}
We used a wild-type motile strain of {\sl E. coli}, RP437.
First,
 we inoculated the strain from a glycerol stock into $2\unit{ml}$ TB medium (see Supplementary~Table~1 for components) in a test tube.
After shaking it overnight at $37\degc$, we transferred $20\unit{\mu{}l}$ of the incubated suspension to $2\unit{ml}$ fresh TB medium and cultured it until the optical density (OD) at $600\unit{nm}$ wavelength reached $0.1$-$0.5$.
The bacterial suspension was finally diluted to $\mathrm{OD}=0.1$ before it was inoculated on the coverslip of the EMPS.

Regarding the device, here we compared the EMPS with the previous system developed in ref.~\cite{Inoue2001}, whose membrane was composed of a cellulose membrane alone instead of the bilayer one.
In either case, we used a substrate with circular wells of $110$-$210\unit{\mu{}m}$ diameter and $1.1\unit{\mu{}m}$ depth.
After the assembly of the device with the bacterial suspension, it was fixed on the microscope stage inside an incubation box maintained at $37\degc$.
The microscope we used was Leica DMi8, equipped with a 63x (N.A. 1.30) oil immersion objective and operated by Leica LasX.
To fill the device with medium,
 we injected fresh TB medium stored at $37\degc$ from the inlet (Supplementary~Fig.~1),
 at the rate of $60\unit{ml~hr^{-1}}$ for $5\unit{min}$ by a syringe pump (NE-1000, New Era Pump Systems, Farmingdale New York).
 
During the observation, TB medium was constantly supplied from the inlet at the rate of $2\unit{ml~hr^{-1}}$ (flow speed above the membrane was approximately $0.2\unit{mm~sec^{-1}}$) by the syringe pump.
Cells were observed by phase contrast microscopy and recorded at the time interval of $30\unit{sec}$ for the cellulose-membrane device (Supplementary~Fig.~1c,d) and $118\unit{msec}$ for the EMPS (Supplementary~Fig.~1e,f).
The time interval for the former was long, because cells hardly moved in this case (see Supplementary~Movie~1).
We checked that the EMPS can realize quasi-two-dimensional space in which bacteria can freely swim, even for large wells, at least up to $210\unit{\mu{}m}$ in diameter, while we have not investigated the reachable largest diameter.

\subsection{Cell growth measurement in U-shape traps in the PDMS-based device}
We used a non-motile mutant strain W3110 without flagella and pili ($\Delta$fliC $\Delta$flu $\Delta$fimA)
 to prevent cell adhesion to the surface of a coverslip.
Before the time-lapse observation,
 we inoculated the strain from a glycerol stock into $2\unit{ml}$ M9 medium with glucose and amino acids (Glc+a.a.) (see also Supplementary~Table~1) in a test tube.
After shaking it overnight at $37\degc$, we transferred $20\unit{\mu{}l}$ of the incubated suspension to $2\unit{ml}$ fresh M9(Glc+a.a.) medium and cultured it until the OD at $600\unit{nm}$ wavelength reached $0.4$-$0.5$.
We then injected the bacterial suspension into the device from the cell inlet (Supplementary~Fig.~2d)
and left it until a few cells entered the U-shape traps.
The device was placed on the microscope stage, in the incubation box maintained at $37\degc$.
The microscope we used was Leica DMi8, equipped with a 63x (N.A. 1.30) oil immersion objective and operated by Leica LasX.

During the observation, we constantly supplied M9(Glc+a.a.) medium and $0.5\unit{wt\%}$ bovine serum albumin (BSA)
 from the medium inlet (Supplementary~Fig.~2d) at the rate of $0.7\unit{ml~hr^{-1}}$
(flow speed in the drain was approximately $1\unit{mm~sec^{-1}}$).
Cells were observed by phase contrast microscopy and recorded at the time interval of $3\unit{min}$.
The velocity field of the coherent flow was obtained by particle image velocimetry, using MatPIV (MATLAB toolbox).
The stream-wise component of the velocity field (Supplementary~Fig.~2f) was averaged over the span-wise direction, and also over the time period of $150\unit{min}$.

\subsection{Cell growth measurement in U-shape traps in the EMPS}
Here the choice of the strain, the medium, the culture condition and the instruments was the same as those for the measurement with the PDMS-based device, unless otherwise stipulated.
Here, the cultured bacterial suspension was diluted to $\mathrm{OD}=0.04$ before it was inoculated on the coverslip of the EMPS.
As sketched in Supplementary~Fig.~2a, the substrate consisted of drain channels ($100\unit{\mu{}m}$ wide, $7\unit{mm}$ long, $13\unit{\mu{}m}$ deep) and U-shape traps ($30\unit{\mu{}m}$ wide, $80\unit{\mu{}m}$ long, $1.0\unit{\mu{}m}$ deep), which were prepared by the methods described in ref.~\cite{Hashimoto2016}.
When the bilayer membrane was attached to the substrate, care was taken not to cover the two ends of the drain channel to use, so that cells in the drain could escape from it.
After the assembly of the device with the bacterial suspension, it was fixed on the microscope stage inside the incubation box maintained at $37\degc$.
To fill the device with medium, we injected fresh medium stored at $37\degc$ from the inlet (Supplementary~Fig.~1),
 at the rate of $60\unit{ml~hr^{-1}}$ for $5\unit{min}$ by a syringe pump (NE-1000, New Era Pump Systems).
 
During the observation, the flow rate of the M9(Glc+a.a. and BSA) medium was set to be $2\unit{ml~hr^{-1}}$ (approximately $0.2\unit{mm~sec^{-1}}$ above the membrane), except that it was increased to $60\unit{ml~hr^{-1}}$ (approximately $6\unit{mm~sec^{-1}}$ above the membrane) at a constant interval, in order to remove the cells expelled from the trap efficiently.
The time interval of this flush was $60\unit{min}$ before the observed trap was filled with cells, and $30\unit{min}$ thereafter.

\subsection{Evaluation of the rhodamine exchange efficiency in the EMPS}
\label{sec:rhodamine}
We used a non-motile mutant strain W3110 $\Delta$fliC $\Delta$flu $\Delta$fimA.
Before the time-lapse observation,
 we inoculated the strain from a glycerol stock into $2\unit{ml}$ M9(Glc+a.a.) medium in a test tube.
After shaking it overnight at $37\degc$, we transferred $20\unit{\mu{}l}$ of the incubated suspension to $2\unit{ml}$ fresh M9(Glc+a.a.) medium and cultured it until the OD at $600\unit{nm}$ wavelength reached $0.1$-$0.5$.
The bacterial suspension was finally diluted to $\mathrm{OD}=0.1$ before it was inoculated on the coverslip.

Before starting the observations, we injected a PBS solution with $10\unit{\mu{}M}$ rhodamine to adsorb fluorescent dye on the surface of the coverslip, then removed non-adsorbed rhodamine molecules by injecting a pure PBS buffer.
We repeated this medium exchange several times.
For the observation, we used a laser-scanning confocal microscope, Nikon A1R-HD25, equipped with a 100x (N.A. 1.49) oil immersion objective and operated by Nikon NIS-Elements.
The resolution along the Z-axis was $0.12\unit{\mu{}m}$, and cross-sectional images were taken over a height of $10\unit{\mu{}m}$, at the time interval of $1.8\unit{sec}$.
In the presence of bacterial cells, we first monitored the fluorescent intensity while we switched the medium to supply from PBS without rhodamine to that containing $10\unit{\mu{}M}$ rhodamine, at the flow rate of $6\unit{mm~sec^{-1}}$ (approximately $60\unit{ml~hr^{-1}}$ above the membrane) (Supplementary~Fig.~3b,d).
The medium switch was performed by exchanging the syringe connected to the device.
After this observation, followed by a time interval of a few minutes without flow of the solution in the device, we started to monitor the fluorescent intensity while we switched the medium from the PBS with rhodamine to that without rhodamine, at the flow rate of $6\unit{mm~sec^{-1}}$ (Supplementary~Fig.~3c,e).

The data shown in Supplementary~Fig.3b-e represent the fluorescent intensity with the background subtracted.
We defined the background at each time as the spatially averaged intensity inside the coverslip where rhodamine molecules cannot enter, or more specifically, inside the area of about 10 pixels around the bottom right region in Supplementary~movies~5,6.
For evaluating the spatial profile of the fluorescent intensity, the location of the substrate bottom was detected by image analysis in each frame, in order to avoid the influence of vibrations caused by the high flow rate used here.
Such vibrations have been removed in Movies~S5 and S6 by using the detected location.
The spatial average of intensity in the well (Supplementary~Fig.~3d,e, blue curves) was taken in a square ROI of height 5 pixels ($0.6\unit{\mu{}m}$) from the substrate bottom, and width 200 pixels ($24\unit{\mu{}m}$) along the y-axis, around the center of the well. 
The spatial average of intensity in the membrane (Supplementary~Fig.~3d,e, red curves) was taken in a linear ROI of length 200 pixels ($24\unit{\mu{}m}$) along the y-axis, located at $4.8\unit{\mu{}m}$ above the substrate bottom, around the center.

\subsection{Observation of the bacterial reductive division}
We used a wild-type strain MG1655.
Before the time-lapse observation,
 we inoculated the strain from a glycerol stock into $2\unit{ml}$ growth medium in a test tube.
The same medium as for the main observation was used (LB broth, M9(Glc+a.a.) or M9(Glc)).
After shaking it overnight at $37\degc$, we transferred $20\unit{\mu{}l}$ of the incubated suspension to $2\unit{ml}$ fresh medium and cultured it until the OD at $600\unit{nm}$ wavelength reached $0.1$-$0.5$.
The bacterial suspension was finally diluted to $\mathrm{OD}=0.05$ before it was inoculated on the coverslip.

For this experiment, we used a substrate with wells of $55\unit{\mu{}m}$ diameter and $0.8\unit{\mu{}m}$ depth.
The well diameter was chosen so that all cells in the well can be recorded.
The device was placed on the microscope stage, in the incubation box maintained at $37\degc$.
The microscope we used was Leica DMi8, equipped with a 100x (N.A. 1.30) oil immersion objective and operated by Leica LasX.
To fill the device with growth medium, we injected fresh medium stored at $37\degc$ from the inlet (Supplementary~Fig.~1),
 at the rate of $60\unit{ml~hr^{-1}}$ for $5\unit{min}$ by a syringe pump (NE-1000, New Era Pump Systems).

In the beginning of the observation, growth medium was constantly supplied at the rate of $2\unit{ml~hr^{-1}}$
 (flow speed approximately $0.2\unit{mm~sec^{-1}}$ above the membrane).
When a microcolony composed of approximately 100 cells appeared, 
 we quickly switched the medium to a non-nutritious buffer (PBS or M9 medium with $\alpha$-methyl-D-glucoside ($\alpha$MG), see Supplementary~Table~1) stored at $37\degc$, by exchanging the syringe.
The flow rate was set to be $60\unit{ml~hr^{-1}}$ for the first 5 minutes, then returned to $2\unit{ml~hr^{-1}}$.
Throughout the experiment, the device and the media were always in the microscope incubation box, maintained at $37\degc$.
Cells were observed by phase contrast microscopy and recorded at the time interval of $5\unit{min}$.
The data for obtaining each distribution are taken from several wells (stated in the figure captions) in a single experiment.

The cell volumes were evaluated as follows.
We determined the major axis and the minor axis of each cell, manually, by using a painting software.
By measuring the axis lengths, we obtained the set of the length $L_i$ and the width $w_i$ for all cells (indexed by $i$).
We estimated the uncertainty in manual segmentation at $\pm 0.15\unit{\mu{}m}$.
However, the measurement of the individual cell widths is less accurate than that of the lengths, essentially because the width depends on the choice of the section of the cell.
To estimate the cell volume, therefore, we neglected the fluctuation of the width among the cells as follows.
We measured the width $w_i$ at the center of each cell and took the ensemble average $\expct{w_i}$.
Together with the cell length $L_i$, we obtained the volume of this cell, $v_i$, by $v_i = \frac{4\pi}{3}\( \frac{\expct{w_i}}{2}\)^3 + \pi\( \frac{\expct{w_i}}{2}\)^2(L_i-\expct{w_i})$.
Note that the scale invariance holds for the length distribution as well (Supplementary~Fig.~6), which suggests that neglecting the width fluctuation does not affect the main finding of the paper.

Finally, let us note that there may be some technical limitation specific to the combination of the LB medium and the EMPS.
When we used LB medium in the growth phase, the bacteria continued growing, albeit very slowly, even long time after the medium was switched to PBS (see Supplementary~Movie~7), while they stopped growing completely in all cases where we used chemically defined medium before the switch (Supplementary~Fig.~4 and Supplementary~Movie~8-11).
This may be because some nutrient molecules specific to LB might remain on the well surface or inside the membrane.
However, since we are focusing on the earlier stage of the starvation process, in which the typical cell sizes change most significantly, we believe that this remaining slow growth in the case LB $\to$ PBS does not affect our main results.
More quantitatively, from Fig.~2b inset, we can evaluate the rate of this remaining cell growth observed in the case LB $\to$ PBS to be nearly $10^{-4}\unit{min^{-1}}$ or eventually even less (Fig.~2b inset). Because the corresponding time scale $\gtrsim 10^4\unit{min}$ is much longer than the time scales relevant to the scale invariance we found, which are around $100\unit{min}$, there is a clear scale separation, from which we can expect that the remaining slow cell growth will not affect our main finding.
We also noticed relatively poor reproducibility of experiments using LB medium (Supplementary~Fig.~7), which may be attributed to its chemical undefinedness \cite{Sezonov2007}, since the experiments using defined media recorded good reproducibility (see the same figure).

\section*{Supplementary Note 2: Experiments for performance evaluation of EMPS}
We first examined the flatness of the observation area using motile bacteria.
In the setup with circular wells of $110\unit{\mu{}m}$ diameter and $1.1\unit{\mu{}m}$ depth,
 although a cellulose membrane alone is bent and adheres to the bottom of the well (Supplementary~Fig.~1c,d and Supplementary~Movie~1),
 our PET-cellulose bilayer membrane keeps flat enough so that bacteria can freely swim in the shallow well
 (Supplementary~Fig.~1e,f and Supplementary~Movie~2).
The EMPS can realize such observations for a long time 
 with little hydrodynamic perturbation by medium flow
 and no mechanical stress.
We also check whether the additional PET membrane may affect the growth condition of cells, by using {\sl E. coli} MG1655 and M9 medium with glucose and amino acids (Glc+a.a.).
We find that the doubling time of the cell population is 59 $\pm$ 10 min, which is comparable to that in the previous system without the PET membrane \cite{Inoue2001, Wakamoto2005, Hashimoto2016}.
Therefore, our bilayer membrane can still exchange medium efficiently.

We then evaluate the spatial uniformity of the culture condition in the EMPS.
To this end, we design U-shape traps with an open end, for both the EMPS and for the conventional PDMS-based device (Supplementary~Fig.~2).
For comparison, and for our need to use large chambers, here the size of the traps in both systems was set to be nearly the same.
With this geometry, nutrients are supplied via diffusion
 from the open end in the PDMS-based system, while nutritious medium is directly and uniformly delivered through the membrane above the trap in the EMPS.
When we culture {\sl E. coli} MG1655 in M9(Glc+a.a.),
 the trap is eventually filled with cells, and they exhibit coherent flow toward the open end due to the cell growth and proliferation (Supplementary~Movie~3,4).
To evaluate the uniformity of the cell growth,
 we measure the velocity field of the cell flow by particle image velocimetry (PIV).
The velocity component along the stream-wise direction clearly shows that the velocity profile is stable for the EMPS over long time periods, while it gradually decreases for the PDMS-based device (Supplementary~Fig.~2f,g).
The cell growth rate is then obtained by the spatial derivative of the velocity field.
The result shows that the growth rate is indeed uniform and kept constant for the EMPS,
 while for the PDMS device it is heterogeneous, being higher near the open end,
 and it decreases as time elapses (Supplementary~Fig.~2f,g).
The growth rate decays at the distance of roughly $30\unit{\mu{}m}$ from the open end.
This observation is consistent with the nutrient depletion length that we evaluate at $32\unit{\mu{}m}$ by following Mather et al.'s calculation \cite{Mather2010}, for which we used the diffusion constant of glucose \cite{MARUCCI2006} and the division time of 60 min.
Such heterogeneity is not seen in the EMPS.
While cell growth regulation pathways may also be influenced by such factors as mechanical pressure caused by cell elongation \cite{Volfson2008, Boyer2011, Si2015, Chu2018},
 quorum sensing \cite{Bruger2016, Haeaar2018}, etc., 
 our results indicate that the EMPS can indeed realize a uniform and stable culture condition while the same medium is kept supplied.
The EMPS has thus potential applications in a wide range of problems with dense cellular populations.

Similarly to other microfluidic devices, we can also switch the culture condition by changing the medium to supply.
A characteristic advantage of the EMPS is that we can switch the medium without worrying about hydrodynamic perturbations or the intrusion of small bubbles.
Here we evaluate how efficiently the medium in the well is exchanged.
In the presence of non-motile {\sl E. coli} W3110 $\Delta$fliC $\Delta$flu $\Delta$fimA,
 we switch the medium to supply from phosphate buffered saline (PBS) to
 PBS solution of rhodamine fluorescent dye,
 and monitor the fluorescent signal in a cross-section of the device
 by a confocal microscope (Supplementary~Fig.~3a, from right to left).
The result, with background substracted (see Supplementary experimental methods), shows that the medium inside the well is exchanged uniformly in space (Supplementary~Fig.~3b)
 and that it is almost completed within $2$-$4\unit{min}$ (Supplementary~Fig.~3d).
We also change the medium from PBS with rhodamine to that without rhodamine
 (Supplementary~Fig.~3a, from left to right).
The exchange then took longer time, $\simeq 5\unit{min}$, presumably because of
 adsorption of rhodamine on the substrate and membrane (see Supplementary~Fig.~3a).
In any case, the time to take for exchanging medium is much shorter than the timescale of the bacterial cell cycle.
Our observations also indicate that the membrane is indeed kept flat above the well (Supplementary~Fig.~3a)
 and that the Brownian motion of non-motile cells is hardly affected by relatively strong medium flow above the membrane (estimated at roughly $6\unit{mm~sec^{-1}}$
 ) induced when switching the medium (Supplementary~Movie~5,6).
We therefore conclude that the EMPS is indeed able to change the growth condition for cells under observation uniformly, without noticeable fluid flow perturbations.

\section*{Supplementary Note 3: Supplementary simulation methods}
\setcounter{subsection}{0}
\subsection{Derivation of Eqs.~(5) and (6) for $\lambda(t)$ and $\mu_i(t)$}
Here we describe how we obtained the functional forms of $\lambda(t)$ and $\mu_i(t)$, Eqs.~(5) and (6), in our CH model for the bacterial reductive division (before the simplification described in ``Violation of the scale invariance''). 
We consider the situation where growth medium is switched to non-nutritious buffer at $t=0$;
 therefore, $t$ denotes time passed since the switch to the non-nutritious condition.
First, on the basis of the Monod equation \cite{Monod1949},
 we assume that the growth rate $\lambda(t)$ can be expressed as a function of 
 the concentration of substrates, $S(t)$, inside each cell:
\begin{equation}
    \lambda(t) = A'\frac{S(t)}{B'+S(t)},
    \label{eq:monod}
\end{equation}
 where $A'$ and $B'$ are constant coefficients.
We consider that substrates in each cell are simply diluted by volume growth
 and consumed at a constant rate $c$, as follows:
\begin{equation}
    \frac{dS(t)}{dt} = - \lambda(t)S(t) - cS(t).
    \label{eq:substrate}
\end{equation}
There is no uptake of substrates because of the non-nutritious condition considered here.
Note that although the consumption rate $c$ may depend on the growth rate, we assume it to be a constant for simplicity.
When a cell divides, we assume that the two daughter cells take the same concentration $S(t)$; therefore the cell division does not affect the time evolution of $S(t)$.
By combining \eqref{eq:monod} and \eqref{eq:substrate},
 we obtain the following differential equation for $S(t)$:
\begin{equation}
    \frac{dS(t)}{dt} = - \frac{S(t)(A'S(t)+cB'+cS(t))}{B'+S(t)}.
    \label{eq:substrate2}
\end{equation}
We can solve this equation:
\begin{equation}
    \frac{1}{c}\ln S + \( \frac{1}{A'+c}-\frac{1}{c}\)\ln \((A'+c)S+cB'\) = -t + \ln D,
    \label{eq:substrate3}
\end{equation}
where $D$ is a constant due to integration.
To obtain an approximate analytical solution for the sake of simplicity,
 we assume $A'\gg c$ and neglect $\frac{1}{A'+c}$, which is justified later.
We then analyticaly solve this problem as
\begin{equation}
    S(t) = \frac{DcB'}{e^{ct}-D(A'+c)},
    \label{eq:substrate3}
\end{equation}
 
Substituting it to \eqref{eq:monod}, we obtain
\begin{equation}
    \lambda(t) = \frac{DcA'}{e^{ct}-DA'} = \lambda_0\frac{1-A}{e^{ct}-A},
    \label{eq:lambda2}
\end{equation}
 with $A=DA'$ and $\lambda_0=\lambda(0)= DcA'/(1-DA')$ being the growth rate
 in the growth phase, before the onset of starvation.
By fitting \eqref{eq:lambda2} to our experimental data, we obtain $c\approx 0.0059-0.011\unit{min^{-1}}$.
Since $A'\approx 0.035\unit{min^{-1}}$ assuming that the minimum doubling time for {\sl E. coli} is 20 min,
 the assumption used in the derivation is reasonable.

Next, we determine the functional form of the cycle progression speed $\mu_i(t)$.
Here, we propose a functional form that comforms with the adder principle for the C+D period in steady environments, as assumed in Witz \textit{et al.}'s model \cite{Witz2019}.
We assume that division occurs when a given amount of relevant molecules, such as DNA, is produced.
Therefore, using the concentration of such relevant molecules denoted by $[\mathrm{RM}^\mathrm{W}_\mathrm{C+D}]$, we can express the cycle progression speed as follows:
\begin{align}
    \mu_i(t) & = \frac{d\([\mathrm{RM}^\mathrm{W}_\mathrm{C+D}]v_i(t)\)}{dt} \notag \\
    & = v_i(t)\frac{d[\mathrm{RM}^\mathrm{W}_\mathrm{C+D}]}{dt} + [\mathrm{RM}^\mathrm{W}_\mathrm{C+D}]\frac{dv_i(t)}{dt} \notag \\
    & = v_i(t)\frac{d[\mathrm{RM}^\mathrm{W}_\mathrm{C+D}]}{dt} + [\mathrm{RM}^\mathrm{W}_\mathrm{C+D}]\lambda(t)v_i(t),
    \label{eq:RMV}
\end{align}
 where $v_i$ is the size of each cell indexed by $i$.
Note that the cell division does not affect this expression either, because we assume that the two daughter cells take the same concentration of intracellular molecules, too.
We assume that such relevant molecules are synthesized from substrates through enzyme catalyses, according to the Michaelis-Menten equation with the effect of dilution:
\begin{equation}
    \frac{d[\mathrm{RM}^\mathrm{W}_\mathrm{C+D}]}{dt} = \frac{\nu_0[\mathrm{S}_\mathrm{C+D}]}{K+[\mathrm{S}_\mathrm{C+D}]} - \lambda(t)[\mathrm{RM}^\mathrm{W}_\mathrm{C+D}],
    \label{eq:Michaelis-Menten}
\end{equation}
 with constants $\nu_0,K$ and $[\mathrm{S_{C+D}}]$ being the concentration of the corresponding substrates.
Combining \eqref{eq:RMV} and \eqref{eq:Michaelis-Menten}, the progression speed is obtained as
\begin{equation}
    \mu_i(t) = \frac{\nu_0[\mathrm{S}_\mathrm{C+D}]}{K+[\mathrm{S}_\mathrm{C+D}]}v_i(t).
    \label{eq:muRM}
\end{equation}
The time dependence of $[\mathrm{S}_\mathrm{C+D}]$ can be evaluated by considering dilution due to volume growth, degradation and consumption:
\begin{equation}
       \frac{d[\mathrm{S}_\mathrm{C+D}]}{dt} = -\lambda(t)[\mathrm{S}_\mathrm{C+D}] - E[\mathrm{S}_\mathrm{C+D}],
       \label{eq:Sdiff}
\end{equation}
 with a constant rate $E$.
Here we assumed that \eqref{eq:Sdiff} is independent of $\numori$, based on the experimental report that the duration of the C+D period was independent of $\numori$ in the steady environments \cite{SI2017}.
Then, with \eqref{eq:lambda2}, the differential equation becomes
\begin{equation}
    \frac{d[\mathrm{S}_\mathrm{C+D}]}{dt} = -\lambda_0\frac{1-A}{e^{ct}-A}[\mathrm{S}_\mathrm{C+D}] - E[\mathrm{S}_\mathrm{C+D}].
    \label{eq:Sdiff2}
\end{equation}
The solution to this equation is
\begin{equation}
    [\mathrm{S}_\mathrm{C+D}] = c_0'\frac{e^{-[E-\lambda_0(1-A)/A]t}}{(e^{ct}-A)^{\lambda_0(1-A)/Ac}}.
    \label{eq:S}
\end{equation}
 with a constant $c_0'$.
Since this is essentially an exponential function, here we replace it by the following simple exponential decay
\begin{equation}
 [\mathrm{S}_\mathrm{C+D}]\approx c_0\exp(-t/\tau),
 \label{eq:Sexp}
\end{equation}
 with the initial concentration $c_0$ and the decay time scale $\tau$.
As a result, we obtain
\begin{equation}
    \mu_i(t) = \nu_0\frac{1}{k\exp(t/\tau)+1}v_i(t),
    \label{eq:munu_0}
\end{equation}
 with $k = K/c_0$.
$\nu_0$ can be determined by considering the progression speed in the exponential phase, which corresponds to setting $t=0$ here.
By taking the ensemble average on both sides, we obtain
\begin{equation}
    \nu_0 = \frac{\mu_0(k+1)}{v_0},
    \label{eq:nu_0}
\end{equation}
 with the mean progression speed of the C+D period $\mu_0=\expct{\mu_i(0)}$ and the mean cell size $v_0=\expct{v_i(0)}$ in the exponential phase.
Finally, we obtain the following equation [Eq.(6) in the main paper] for cycle progression speed in the starvation process:
\begin{equation}
    \mu_i(t) = \frac{\mu_0}{v_0}\frac{k+1}{k\exp(t/\tau)+1}v_i(t).
    \label{eq:mu2}
\end{equation}

For steady environments, \eqref{eq:mu2} implies that the progression speed in this case is given by
\begin{equation}
    \mu_i^\mathrm{s}(t) = \frac{\mu_0}{v_0}v_i^\mathrm{s}(t),
    \label{eq:mu0}
\end{equation}
 where the superscript ``s'' indicates that those quantities are for steady environments.
Now, suppose a cell $i$ initiated the C+D period at time $t_I$. Then this cell divides when its $\XCD_i(t) = \int_{t_I}^t \mu_i(t) dt$ reaches a constant threshold $\XCDth$. Therefore, the division time $t_D$ satisfies the following equation:
\begin{align}
    \XCDth & = \int^{t_D}_{t_I} \mu_i^\mathrm{s}(t) dt \notag \\
    & = \frac{\mu_0}{v_0} \int^{t_D}_{t_I} v_i^\mathrm{s}(t) dt \notag \\
    & = \frac{\mu_0}{v_0\lambda_0} \int^{t_D}_{T_I} \frac{dv_i^\mathrm{s}(t)}{dt} dt \notag \\
    & = \frac{\mu_0}{v_0\lambda_0} \[ v_i^\mathrm{s}(t_D) - v_i^\mathrm{s}(t_I) \].
    \label{eq:CDadder}
\end{align}
This corresponds to the adder principle during the C+D period assumed in Witz \textit{et al.}'s model \cite{Witz2019}.

\subsection{Extension of the model by Ho and Amir}

Ho and Amir's CH model \cite{Ho2015} assumed the ``timer'' principle for the C+D period in steady environments,
 so that each cell divides after a constant time has passed since replication is initiated.
Most parts of the model are identical to those of the model based on Witz \textit{et al.}'s one \cite{Witz2019}, except the functional form of $\mu_i(t)$.
Here, to obtain a functional form that conforms with the timer principle, we consider that assembly processes of relevant molecules such as deoxynucleotide triphosphates control the cycle progression speed.
We then consider that the progression speed $\mu_i(t)$ is given through the Hill equation, which usually describes the binding probability of a receptor and a ligand, with cooperative effect taken into account.
Specifically, with $[\mathrm{RM}^\mathrm{HA}_\mathrm{C+D}]$ being the concentration of relevant molecules considered here, we use
\begin{equation}
    \mu_i(t) \propto \frac{[\mathrm{RM}^\mathrm{HA}_\mathrm{C+D}]^n}{K'^n + [\mathrm{RM}^\mathrm{HA}_\mathrm{C+D}]^n},
    \label{eq:Hill}
\end{equation}
 where $n$ is the Hill coefficient,
 and $K'$ is the equilibrium constant of the (collective) assembly process.
Then we assume that $[\mathrm{RM}^\mathrm{HA}_\mathrm{C+D}]$ evolves in the same way as $[\mathrm{S}_\mathrm{C+D}]$ in \eqref{eq:Sdiff2}.
As a result, we obtain
\begin{equation}
 [\mathrm{RM}^\mathrm{HA}_\mathrm{C+D}]\approx c^\mathrm{HA}_0\exp(-t/\tau_\mathrm{HA}'),
 \label{eq:RMexp}
\end{equation}
 with the initial concentration $c^\mathrm{HA}_0$ and the decay time scale $\tau_\mathrm{HA}'$.
This gives
\begin{equation}
    \mu_i(t) = \mu_0\frac{k_\mathrm{HA}+1}{k_\mathrm{HA}\exp(t/\tau_\mathrm{HA})+1},
    \label{eq:mu2_HA}
\end{equation}
 with $k_\mathrm{HA} = (K'/c_0)^n$, $\tau_\mathrm{HA} = \tau_\mathrm{HA}'/n$, and $\mu_0$ being the cycle progression speed in the growth phase before the onset of starvation.
Note that the formula of $\mu_i(t)$ without $v_i$ dependence assures the timer principle for the C+D period in steady environments.
To obtain the numerical results in Supplementary~Fig.~9g,h, we set $k_\mathrm{HA} = 1$ and $\tau_\mathrm{HA}=60\unit{min}$.
For investing the violation of the scale invariance by changing the relaxation time $\tau_\mu$, $\mu_i(t)$ was assumed to be
\begin{equation}
    \mu_i(t) = \mu_0\exp(-t/\tau_\mu).
    \label{eq:mu2_HA_violation}
\end{equation}
The other paramaters are identical to those of the model based on Witz \textit{et al.}'s one.

\subsection{Simulation methods}

The parameters used in the simulations were evaluated as follows. 
First, from the observations of the exponential growth phase, we determined the growth rate $\lambda_0$ and the mean cell size $v_0$ directly.
This allowed us to set the cycle progression speed $\mu_0$ too, by using the relation $\mu_0^{-1}\simeq (1.3\lambda_0^{-0.84}+42)$ proposed by Wallden \textit{et al.} \cite{Wallden2016} (the values of $\lambda_0$ and $\mu_0$ are in the unit of $\mathrm{min^{-1}}$ here).
Concerning the volume threshold for initiating the replication, we found such a value of $\delta_\mean$ (or $\vth_\mean$) that reproduced the experimentally observed mean cell volume in the growth phase.
The standard deviation $\delta_\std$ (or $\vth_\std$) was set to be $10\%$ of the mean $\delta_\mean$ ($\vth_\mean$), based on the relation on the initiation volume found by Wallden \textit{et al.} \cite{Wallden2016}.
They also measured the fluctuations of the time length of the C+D period; this led us to estimate $\XCDth_\std$ at 5\% of $\expct{\XCDth}$, i.e., $\XCDth_\std=0.05$.
On the septum positions, we measured their fluctuations and found little difference in $x^\mathrm{sep}_\std$ among the different growth conditions we used, and also in the non-nutritious case (Supplementary~Fig.~8).
We therefore used a single value $x^\mathrm{sep}_\std=0.0325$ for all simulations.
Note that, without this stochastic asymmetric division, the cell size distribution exhibited periodic oscillations, presumably because cellular states between siblings were strongly correlated then.

In the following, we describe how the remaining parameters were evaluated and how the simulations were carried out for each set of the simulations presented in this work.

\subsubsection{Methods for the results that reproduced the experimental observations}

We evaluated the time-dependent rates $\lambda(t)$ and $\mu_i(t)$ as follows.
The growth rate $\lambda(t)$ can be determined independently of the cell divisions, because the total volume $V_\mathrm{tot}(t) = \sum_i v_i(t)$ grows as $V_\mathrm{tot}(t) = V_\mathrm{tot}(0) \exp(\int^t_0 \lambda(t)dt)$.
With $\lambda(t)$ given by Eq.~(5), we compared $V_\mathrm{tot}(t)$ with experimental data and determined the values of $A$ and $c$ (Fig.~2b).
Finally, only $k$ and $\tau$ in Eq.~(6) remained as free parameters.
We tuned them so that the mean cell volume $V(t)$ and the number of the cells $n(t)$ observed in the simulations reproduced those from the experiments (Fig.~2c).
The parameter values determined thereby are summarized in Table~1, for the simulations for LB$\to$PBS and M9(Glc+a.a.)$\to$PBS.

We started the simulations from 10 cells with
 volumes in the range of $0.07$-$1.13\unit{\mu{}m^3}$, randomly generated from the uniform distribution.
The cells grew in the exponential phase
 (with the constant growth rate $\lambda_0$ and the cycle progression speed $\mu_i = (\mu_0/v_0)v_i$)
 until the number of cells reached 100,000.
We then randomly picked up 10 cells from this ``precultured'' sample and grew them until the number of cells exceeded 500.
Those cells were kept growing for 1000 minutes to sufficiently mix cell cycle progressions in the population.
During this process, we kept the number of cells constant, by eliminating one of the daughter cells after each division.
We then used them as the initial population of each simulation.
To precisely compare the number of cells in simulations with the experimentally obtained population $n_\mathrm{exp}(t)$ (Fig.~2c and Supplementary~Fig.~4b),
 the numerically obtained population $n_\mathrm{sim}(t)$ is rescaled by multiplying $n_\mathrm{exp}(0)/n_\mathrm{sim}(0)$. 

\subsubsection{Methods for the results on violation of the scale invariance}

The functional forms of $\lambda(t)$ and $\mu_i(t)$ were given by Eqs.~(10) and (11), with variable parameters $\tau_\lambda$ and $\tau_\mu$.
For $\mu_0$, we determined it from $\lambda_0$ using the empirical relation reported by Wallden \textit{et al.} \cite{Wallden2016}.
For $v_0$, we set its value self-consistently, so that the mean cell volume $\expct{v_i(0)}$ obtained numerically in the exponential phase with $\mu_i = (\mu_0/v_0)v_i$ falls within 1\% error from the given value of $v_0$.
As a result, we obtained $v_0 = 1.9,2.5,3.2,4.0,5.0$ $\mu\unit{m}^3$ for $\lambda = 0.01, 0.015, 0.02, 0.025, 0.03\unit{min^{-1}}$, respectively. 
These values satisfy the growth law, i.e., the mean cell size increases exponentially with the growth rate \cite{TAHERIARAGHI2015}.
The other parameters were fixed at $\delta_\mean = 0.25 \unit{\mu m^3}$, $\delta_\std = 0.025 \unit{\mu m^3}$, $\XCDth_\std = 0.05$, and $x^\mathrm{sep}_\std = 0.0325$.
We started the simulations from 50 cells with
 volumes in the range of $0.07$-$1.13\unit{\mu{}m^3}$, randomly generated from the uniform distribution.
The cells grew in the exponential phase until the number of cells reached 500,000.
We then randomly picked up 50 cells and grew them until the number of cells exceeded 50,000.
Those cells were kept growing for 1,000 minutes to sufficiently mix cell cycle progressions in the population, with the number of cells kept constant by eliminating one of the daughter cells produced by division.
Using them as the initial population (at $t=0$), we started the simulations for $t \geq 0$, with the number of cells still kept constant by the same method.
Strictly, this situation with a constant number of cells is different from the experimental setting, but we confirmed that this change did not influence the validity of the scale invariance and had only a minor effect on the value of CV, at least for the situation shown in Fig.~4d. Therefore, for the results on the violation of the scale invariance (Fig.~5), we carried out simulations with a constant number of cells as described above, to reduce the computation time.

\section*{Supplementary Note 4: Theory}
\setcounter{subsection}{0}
\subsection{Theoretical conditions for the scale invariance}
Here, we describe the detailed derivation of the sufficient condition for the scale invariance of the cell size distribution.
We start from the time evolution equation, Eq.~(7):
\begin{equation}
    \frac{\partial N(v,t)}{\partial t} = -\frac{\partial}{\partial v}[\lambda(t)vN(v,t)] - B(v,t)N(v,t)+4B(2v,t)N(2v,t),
    \label{eq:balanceeq2}
\end{equation}
 where $\lambda(t)$ is the growth rate and $B(v,t)$ is the division rate function (see the main article for its definition).
We assume the scale invariance in the form of Eq.~(2) for the cell size distribution $p(v,t) = N(v,t)/n(t)$, where $n(t) = \int_0^\infty N(v,t)dv$ is the total number of the cells.
For $N(v,t)$, this reads
\begin{equation}
    N(v,t) = \frac{n(t)}{v}F\( \frac{v}{V(t)}\),
    \label{eq:sianswer}
\end{equation}
 where $V(t)$ is the mean cell volume at time $t$, $V(t)=\expct{v}$.
One can then rewrite the time evolution equation \pref{eq:balanceeq2} in terms of the function $F(x)$ with $x=v/V(t)$, as follows:
\begin{equation}
    \frac{\partial n(t)}{\partial t}\frac{1}{v}F\( x\)
     - \frac{n(t)}{V(t)^2}\frac{\partial V(t)}{\partial t}\frac{\partial F(x)}{\partial x} = -\lambda(t)n(t)\frac{1}{V(t)}\frac{\partial F(x)}{\partial x} - B(v,t)\frac{n(t)}{v}F\( x\) + 4B(2v,t)\frac{n(t)}{2v}F\( 2x\).
    \label{eq:balanceeq3}
\end{equation}
If we neglect cell-to-cell fluctuations of the growth rate,
 one can evaluate $\lambda(t)$ from the total biomass growth, as follows:
\begin{align}
    \lambda(t) & = \frac{d(V(t)n(t))}{dt}(V(t)n(t))^{-1} \notag \\
     & = \( V(t)\frac{\partial n(t)}{\partial t} + n(t)\frac{\partial V(t)}{\partial t} \) (n(t)V(t))^{-1} \notag \\
     & = \frac{1}{n(t)}\frac{\partial n(t)}{\partial t} + \frac{1}{V(t)}\frac{\partial V(t)}{\partial t}.
    \label{eq:biomassgrowth}
\end{align}
From Eqs.~\pref{eq:balanceeq3} and \pref{eq:biomassgrowth}, we obtain
\begin{equation}
    \frac{\partial n(t)}{\partial t}\frac{1}{v}F(x) =
       -\frac{1}{V(t)}\frac{\partial n(t)}{\partial t}\frac{\partial F(x)}{\partial x}
       -\frac{B(v,t)}{v}n(t)F(x) + 4\frac{B(2v,t)}{2v}n(t)F(2x).
    \label{eq:balanceeq4}
\end{equation}
Now, the time derivative of $n(t)$ can be calculated as
\begin{align}
    \frac{\partial n(t)}{\partial t} = & \int^\infty_0 dv \frac{\partial N(v,t)}{\partial t} \notag \\
     = & -\int^\infty_0 dv B(v,t)N(v,t) + 4\int^\infty_0 dv' B(2v',t)N(2v',t) \notag \\
     = & n(t)\int^\infty_0 dv \frac{B(v,t)}{v}F\( \frac{v}{V(t)} \) \notag \\
     = & n(t)\overline{B}(t),
    \label{eq:balanceeq5}
\end{align}
 where $\overline{B}(t) = \int^\infty_0 dv B(v,t)p(v,t) = \int^\infty_0 dv B(v,t)v^{-1} F(v/V(t))$.
Substituting it to \eqref{eq:balanceeq4}, we finally obtain the following self-consistent equation for $F(x)$ (Eq.~(8)):
\begin{equation}
    F(x) = -x\frac{\partial F(x)}{\partial x} - \frac{B(v,t)}{\overline{B}(t)}F(x)
     + 2\frac{B(2v,t)}{\overline{B}(t)}F(2x).
    \label{eq:Fconsistent2}
\end{equation}

For the scale invariance, \eqref{eq:Fconsistent2} should hold at any time $t$.
In other words, the coefficient $B(v,t)/\overline{B}(t)$ should be independent of both $t$ and $V(t)$.
Note here that $B(v,t)/\overline{B}(t)$ can be rewritten as
\begin{equation}
    \frac{\overline{B}(t)}{B(v,t)}
     = \int^\infty_0 dv' \frac{B(v',t)}{B(v,t)}\frac{1}{v'}F\(\frac{v'}{V(t)}\)
     = \int^\infty_0 dx' \frac{B(x'V(t),t)}{B(xV(t),t)}\frac{1}{x'}F(x').
    \label{eq:bb}
\end{equation}
Therefore, $B(v,t)/\overline{B}(t)$ is time-independent if $\frac{B(x'V(t),t)}{B(xV(t),t)}$ does not depend on $V(t)$, being a function of two dimensionless variables $x$ and $x'$ only, as follows:
\begin{equation}
    \frac{B(x'V(t),t)}{B(xV(t),t)} = \beta(x',x).
    \label{eq:sc1}
\end{equation}
This condition can be rewritten as follows.
For a constant $x_0$, we can define $B_t(t)$ by
\begin{equation}
    B(x_0V(t),t) = B_t(t)
    \label{eq:bt}
\end{equation}
 and $B_v(x)$ by
\begin{equation}
    \beta(x,x_0) = B_v(x).
    \label{eq:bm}
\end{equation}
Then, the division rate can be expressed as $B(xV(t),t)=B_v(x)B_t(t)$ for any $x$ and $t$.
This gives the sufficient condition we presented in the main text, Eq.~(9),
\begin{equation}
    B(v,t) = B_v\(\frac{v}{V(t)}\)B_t(t).
    \label{eq:separability2}
\end{equation}

\subsection{Test of the derived conditions for $B(v,t)$ and $F(x)$}
We tested the sufficient condition for $B(v,t)$ (\eqref{eq:separability2}) and the resulting self-consistent equation for $F(x)$ (\eqref{eq:Fconsistent2}) with numerical data we obtained from our CH model (Fig.~4e).
We evaluated the division rate $B(v,t)$ in the simulations for LB $\to$ PBS by
\begin{equation}
B(v,t) = \frac{\text{\# of division events of cells of volume between $v$ and $v+\Delta v$, between time $t$ and $t+\Delta t$}}{\text{\# of cells of volume between $v$ and $v+\Delta v$ at time $t$}}.
\label{eq:DRinmodel}
\end{equation}
Here, $\Delta v$ was set to be approximately $0.2 \times V(t)$ and $\Delta t$ to be approximately $20$-$30\unit{min}$
 for each time point, respectively.
The value of $B(v,0)$ was determined by counting all division events in the exponential growth phase ($t<0$).
The ratio $B_t(0)/B_t(t)$ can be evaluated by $B_t(0)/B_t(t) = \int B(xV(0),0)dx / \int B(xV(t),t) dx$.
We found that the curves $B(v,t)B_t(0)/B_t(t)$ taken at different $t$ overlap reasonably well (Fig.~4e, Inset), which support the variable separability condition of the division rate, \eqref{eq:separability2}.

Given the functional form of $B(v,t)$ that we obtained numerically, we can also test the self-consistent equation for $F(x)$, \eqref{eq:Fconsistent2}.
Here we remind that the ratio $B(v,t)/\overline{B}(t)$ in \eqref{eq:Fconsistent2} can be expressed as \eqref{eq:bb}, and $\frac{B(x'V(t),t)}{B(xV(t),t)}$ is time-independent (cf. \eqref{eq:sc1}).
Therefore, 
\begin{equation}
    \frac{\overline{B}(t)}{B(v,t)} = \int^\infty_0 dx' \frac{B(x'V(0),0)}{B(xV(0),0)}\frac{1}{x'}F(x')
\end{equation}
 with $B(xV(0),0) = B(v,0) = B(v,t)B_t(0)/B_t(t)$.
Since we already confirmed the time independence of $B(v,t)B_t(0)/B_t(t)$ (Fig.~4e, Inset), we took the average of this quantity obtained at $t=0,30,60,90\unit{min}$.
Since the observed range of $v$ is finite and $F(x)$ almost vanishes for $x \gtrsim 2$,
 we evaluated the integral over $x'$ in the range $0 \leq x' \leq 2$.

With the $\overline{B}(t)/B(v,t)$ evaluated thereby, we substituted $F(x)$ obtained by the simulations for LB$\to$PBS
 to \eqref{eq:Fconsistent2} (Supplementary~Fig.~10a),
 using the time average of $F(x)$ (the dashed line in Fig.~4d) and $F(x)=0$ for $x\ge 2$.
We also tested \eqref{eq:Fconsistent2} with $F(x)$ obtained in the experiment for LB$\to$PBS,
 in the same way as for the model (Supplementary~Fig.~10b).
In both cases, the right-hand side (rhs) of \eqref{eq:Fconsistent2}
 differs significantly from the observed form of $F(x)$.

\subsection{Theory with septum fluctuations}
As a possible improvement of our theory, here we take into account septum fluctuations.
We define the kernel function $q(v|\nu)$ which represents the probability
 that a mother cell of volume $\nu$ produces daughter cells of volume $v$ and $\nu-v$ (therefore, $q(v|\nu) = q(\nu-v|\nu)$).
The time evolution equation of $N(v,t)$ is then written as
\begin{equation}
    \frac{\partial N(v,t)}{\partial t} = -\frac{\partial(\lambda(t)vN(v,t))}{\partial v} - B(v,t)N(v,t)
     + \int^\infty_v q(v|\nu)B(\nu,t)N(\nu,t)\frac{\nu}{v} d\nu.
    \label{eq:asbalanceeq}
\end{equation}
Now, assuming that the fluctuations of the septum position are Gaussian for simplicity, we have
\begin{equation}
    q(v|\nu) = 2\times \sqrt{\frac{1}{2\pi (\xi\nu)^2}}\exp\( -\frac{(v-\nu/2)^2}{2(\xi\nu)^2} \),
    \label{eq:kernel}
\end{equation}
 where the coefficient $2$ corresponds to the two daughter cells produced from a single division event.
Here, following experimental observations (Supplementary~Fig.~8, see also \cite{TAHERIARAGHI2015}),
 we assumed that the standard deviation is proportional to the volume of the mother cell, $\nu$, with coefficient $\xi=0.0325$.

Under this modification, we obtain the self-consistent equation for $F(x)$ as follows.
One can again calculate the time derivative of $n(t)$ as
\begin{align}
\frac{\partial n(t)}{\partial t} = & \int^\infty_0 dv \frac{\partial N(v,t)}{\partial t} \notag \\
= & -n(t)\int^\infty_0 dv B(v,t)p(v,t)
+ n(t)\sqrt{\frac{2}{\pi\xi^2}}\int^\infty_0 dv' \int^\infty_{v'} d\nu \frac{B(\nu,t)}{v'}p(\nu,t) \exp\( -\frac{(v'/\nu-1/2)^2}{2\xi^2}\) \notag \\
= & n(t)\overline{B_2}(t),
\label{eq:asbalanceeq2}
\end{align}
 where $\overline{B_2}(t)$ is defined by
\begin{equation}
    \overline{B_2}(t) = -\int^\infty_0 dv B(v,t)p(v,t)
     + \sqrt{\frac{2}{\pi\xi^2}}\int^\infty_0 dv' \int^\infty_{v'} d\nu \frac{B(\nu,t)}{v'}p(\nu,t) \exp\( -\frac{(v'/\nu-1/2)^2}{2\xi^2}\).
    \label{eq:asb}
\end{equation}
With $x=v/V(t)$ and $y=\nu/V(t)$,
 one can obtain the self-consistent equation for $F(x)$ as
\begin{equation}
    F(x) = -x\frac{\partial F(x)}{\partial x} - \frac{B(V(t)x,t)}{\overline{B_2}(t)}F(x)
     + \sqrt{\frac{2}{\pi\xi^2}}\int^\infty_x dy \frac{B(V(t)y,t)}{\overline{B_2}(t)}y^{-1}\exp\( -\frac{(x/y-1/2)^2}{2\xi^2}\)F(y).
    \label{eq:asFconsistent}
\end{equation}
 with
\begin{align}
\frac{\overline{B_2}(t)}{B(V(t)x,t)}
= & \int^\infty_0 dx' \( -\frac{B_v(x')}{B_v(x)}\frac{F(x')}{x'} + \sqrt{\frac{2}{\pi\xi^2}}\frac{1}{x'}\int^\infty_{x'} dy' \frac{B_v(y')}{B_v(x)} \exp\( -\frac{(x'/y'-1/2)^2}{2\xi^2}\)\frac{F(y')}{y'} \), \\
\frac{\overline{B_2}(t)}{B(V(t)y,t)}
= & \int^\infty_0 dx' \( -\frac{B_v(x')}{B_v(y)}\frac{F(x')}{x'} + \sqrt{\frac{2}{\pi\xi^2}}\frac{1}{x'}\int^\infty_{x'} dy' \frac{B_v(y')}{B_v(y)} \exp\( -\frac{(x'/y'-1/2)^2}{2\xi^2}\)\frac{F(y')}{y'} \).
\label{eq:asbb}
\end{align}
We tested this for both experimentally and numerically observed $F(x)$,
 but the modification did not improve the results (Supplementary~Fig.~10).

\newpage
\section*{Supplementary tables and figures}

\begin{table}[h]\centering
    \caption{Culture conditions, ingredients and osmotic pressure.}
    \begin{tabular}{llr}
        Medium & Ingredients & Concentration \tabularnewline
        (Osmotic pressure) & &  \tabularnewline
        \hline
        LB broth (Millar) & Tryptone & 1 wt\% \tabularnewline
        (0.38 Osm kg${}^{-1}$) & Sodium Chloride & 1 wt\% \tabularnewline
        & Yeast extract & 0.5 wt\% \tabularnewline
        \hline
        M9(Glc+a.a.) medium & Disodium Phosphate (Anhydrous) & 0.68 wt\% \tabularnewline
        (0.21 Osm kg${}^{-1}$) & Monopotassium Phosphate & 0.3 wt\% \tabularnewline
        & Sodium Chloride & 0.05 wt\% \tabularnewline
        & Ammonium Chloride & 0.1 wt\% \tabularnewline
        & Magnesium Phosphate & 2 mM \tabularnewline
        & Calcium Chloride & 0.1 mM \tabularnewline
        & Glucose [Glc] & 0.2 wt\% \tabularnewline
        & MEM Amino Acids solution (M5550, Sigma) [a.a.] & 1 wt\% \tabularnewline
        \hline
        M9(Glc) medium & Disodium Phosphate (Anhydrous) & 0.68 wt\% \tabularnewline
        (0.22 Osm kg${}^{-1}$) & Monopotassium Phosphate & 0.3 wt\% \tabularnewline
        & Sodium Chloride & 0.05 wt\% \tabularnewline
        & Ammonium Chloride & 0.1 wt\% \tabularnewline
        & Magnesium Phosphate & 2 mM \tabularnewline
        & Calcium Chloride & 0.1 mM \tabularnewline
        & Glucose [Glc] & 0.2 wt\% \tabularnewline
        \hline
        M9(Glyc) medium & Disodium Phosphate (Anhydrous) & 0.68 wt\% \tabularnewline
        & Monopotassium Phosphate & 0.3 wt\% \tabularnewline
        & Sodium Chloride & 0.05 wt\% \tabularnewline
        & Ammonium Chloride & 0.1 wt\% \tabularnewline
        & Magnesium Phosphate & 2 mM \tabularnewline
        & Calcium Chloride & 0.1 mM \tabularnewline
        & Glycerol [Glyc] & 0.4 wt\% \tabularnewline
        \hline
        M9($\alpha$MG) medium & Disodium Phosphate (Anhydrous) & 0.68 wt\% \tabularnewline
        (0.21 Osm kg${}^{-1}$) & Monopotassium Phosphate & 0.3 wt\% \tabularnewline
        & Sodium Chloride & 0.05 wt\% \tabularnewline
        & Ammonium Chloride & 0.1 wt\% \tabularnewline
        & Magnesium Phosphate & 2 mM \tabularnewline
        & Calcium Chloride & 0.1 mM \tabularnewline
        & Alpha-methyl-D-glucopyranoside. [$\alpha$MG] & 0.2 wt\% \tabularnewline
        \hline
        TB medium & Tryptone & 1 wt\% \tabularnewline
        (0.21 Osm kg${}^{-1}$) & Sodium Chloride & 0.5 wt\% \tabularnewline
        \hline
        PBS(-) & Potassium Dihydrogenphosphate & 0.02 wt\% \tabularnewline
        (0.27 [Osm kg${}^{-1}$]) & Disodium phosphate (Anhydrous) & 0.115 wt\% \tabularnewline
        & Potassium Chloride & 0.02 wt\% \tabularnewline
        & Sodium Chloride & 0.8 wt\% \tabularnewline
        \hline
    \end{tabular}
\end{table}

\newpage

\begin{figure}[p]
    \centering
    \includegraphics[width=0.8\textwidth]{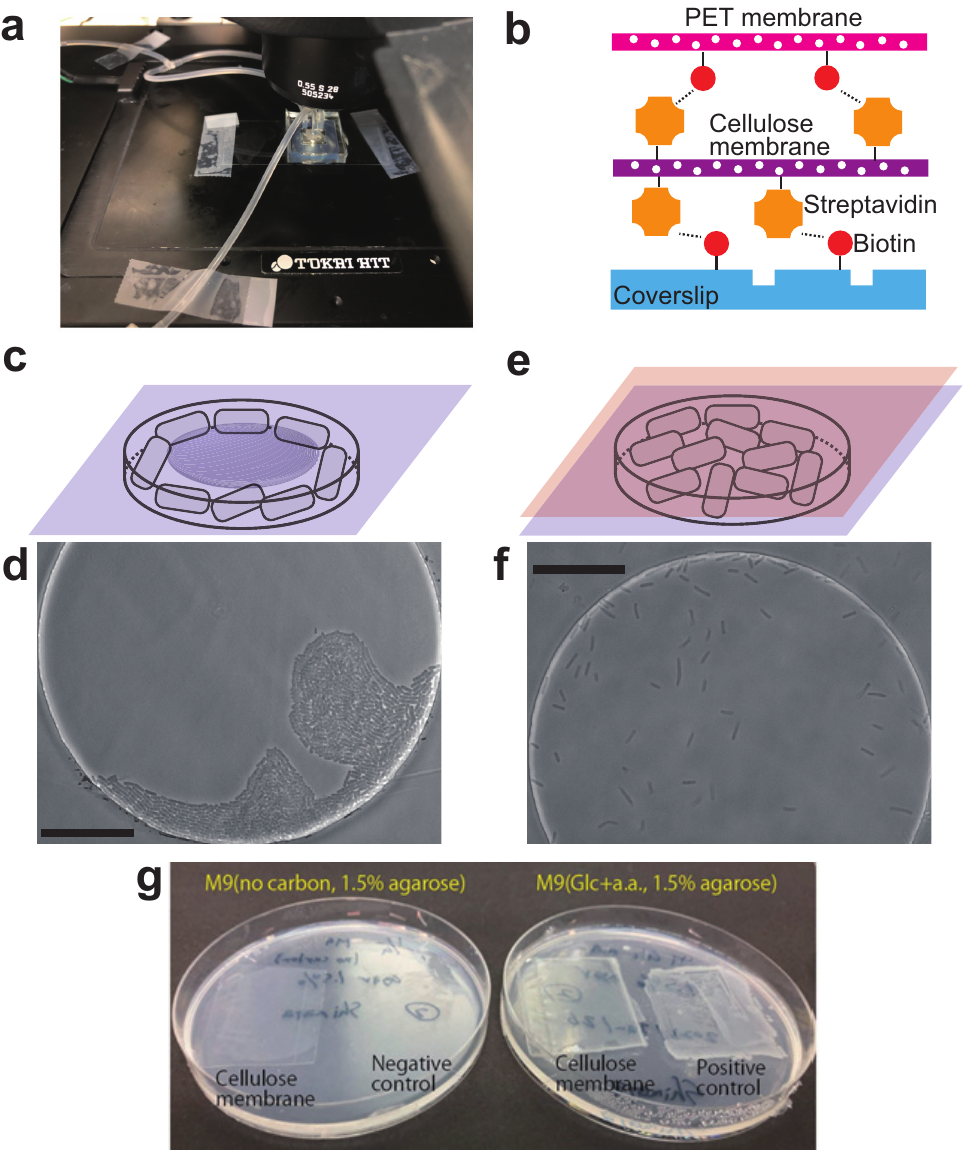}
    \caption{Supplementary figures on the setup of the EMPS.
    {\bf a} Photograph of the device mounted on the microscope stage.
    The coverslip and tubes are fixed to the stage by mending tapes.
    {\bf b} Illustration of the chemical bonding between a bilayer membrane and a glass coverslip.
    A streptavidin-decorated cellulose membrane is sandwiched by a biotin-coated PET membrane and coverslip.
    {\bf c} Sketch of growth of bacterial cells inside a circular well
     covered only by a cellulose membrane.
    Since the membrane bends and presses cell beneath, cells do not swim but form clusters, extending toward the wall.
    {\bf d} Photograph of motile {\sl E. coli} RP437 in a well (diameter $110\unit{\mu{}m}$, depth $1.1\unit{\mu{}m}$) covered only by a cellulose membrane. 
    TB medium was constantly supplied at $37\degc$ (see Materials and Methods for details).
    Despite the motility, cells were confined near the wall and unable to swim freely (see also Supplementary~Movie~1).
    The scale bar is $30\unit{\mu{}m}$.
    {\bf e} Sketch of growth of cells inside a well covered by a PET-cellulose bilayer membrane.
    The rigid bilayer membrane is sustained without bending, leaving a sufficient gap beneath for cells.
    {\bf f} Photograph of motile {\sl E. coli} RP437 in a well (same diameter and depth as in  ({\bf c}, bottom) covered by a bilayer membrane. Cells were able to swim freely (see also Supplementary~Movie~2).
    Growth conditions are same as in ({\bf c}, bottom).
    {\bf g} {\sl E. coli} MG1655 inoculated on cellulose membranes and agar plates.
    Square cellulose membranes were placed on top of the agar plates.
    Colonies were formed in a rectangular shape on the M9(Glc+a.a.) agar plate, both with and without the cellulose membrane, while no colony was formed on the cellulose membrane in the case without carbon sources in the agar plate.
    This indicates that cellulose cannot be metabolized by the {\sl E. coli} strain MG1655 used in our experiments on the scale invariance.
    }
\end{figure}

\begin{figure}[p]
    \centering
    \includegraphics[width=0.9\textwidth]{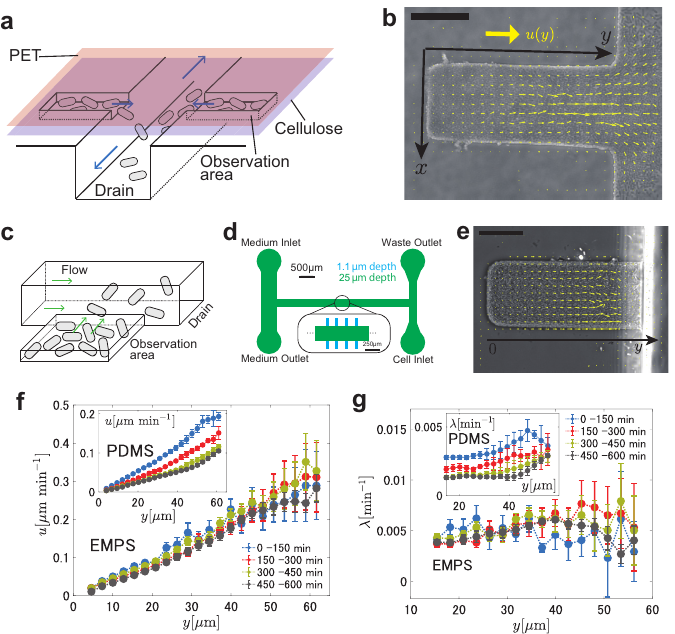}
    \caption{Supplementary figures on the cell growth measurement.
    {\bf a} Sketch of the design of microchannels.
    Non-motile cells are trapped in the shallow observation area.
    Cells in the trap can escape to the deep drain channel.
    {\bf b} Top view of the trap ($30\unit{\mu{}m}$ wide, $80\unit{\mu{}m}$ long, $1.0\unit{\mu{}m}$ deep)
    filled with {\sl E. coli} W3110 $\Delta$fliC $\Delta$flu $\Delta$fimA (see also Supplementary~Movie~4).
   The scale bar is $25\unit{\mu{}m}$.
   The yellow arrows represent the velocity field of flow driven by cell proliferation,
   measured by particle image velocimetry (PIV).
   Coherent flow of cells driven by cell proliferation is directed toward the drain ($15\unit{\mu{}m}$ deep).
    {\bf c} Sketch of the design of the microchannels in the PDMS-based device.
    Medium flow in the drain channel removes cells expelled from the observation area (trap).
    Nutrient is supplied to the cells inside the trap via diffusion from the drain channel.
    {\bf d} The design of the PDMS-based device.
    The green region corresponds to the drain channel, and the blue regions are the U-shape traps.
    {\bf e} Top view of the trap ($30\unit{\mu{}m}$ wide, $88\unit{\mu{}m}$ long, $1.1\unit{\mu{}m}$ deep) in the PDMS-based device.
    The trap is filled with {\sl E. coli} W3110 $\Delta$fliC $\Delta$flu $\Delta$fimA.
    The scale bar is $25\unit{\mu{}m}$.
    See also Supplementary~Movie~3.
    {\bf f} The stream-wise ($y$) component of the velocity field averaged over the span-wise ($x$) direction, $u(y)=\expct{u(x,y)}_x$, measured in different time periods.
    The data were taken from a single trap.
    $t=0$ is the time at which the trap is filled with cells.
    Error bars represent the standard error, which we roughly evaluate by picking up 5 data points along the span-wise direction, at intervals of roughly $5\unit{\mu{}m}$, half of the size of the PIV window.
    The open end is located near $y \approx 81\unit{\mu{}m}$.
    (Inset) The same quantity measured for a PDMS-based device. The open end is at $y \approx 88\unit{\mu{}m}$. Time dependence is clearly observed.
    {\bf g} The growth rate profiles $\lambda$, evaluated by $\lambda(y)=\expct{\diff{u(x,y)}{y}}_x$, for the EMPS (main panel) and the PDMS-based device (inset).
    Error bars represent the standard error of the ensemble.
    }
\end{figure}

\begin{figure}[p]
    \centering
    \includegraphics[width=\hsize]{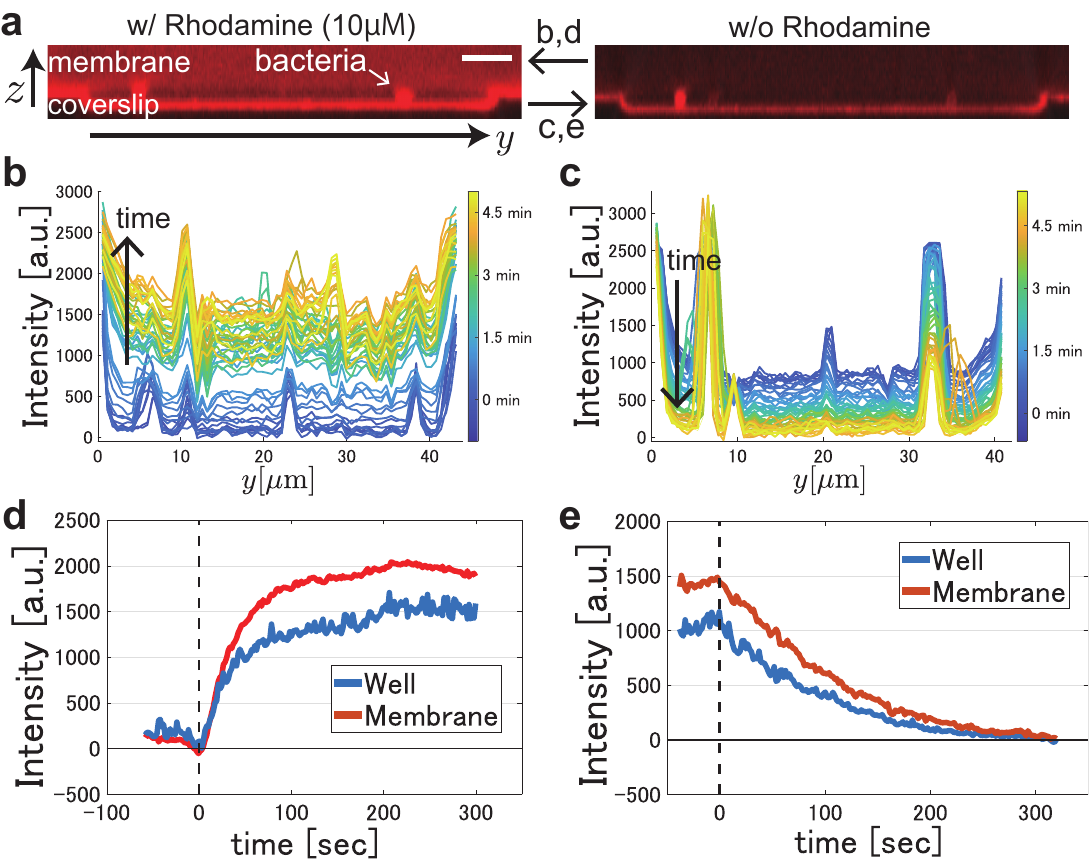}
    \caption{
    Direct observations of medium exchange in the EMPS.
    {\bf a} Cross-sectional images of the device taken by confocal microscopy.
    The scale bar is $5\unit{\mu{}m}$.
    Medium flows from the front side to the rear side of the images.
    (left) A snapshot of the device filled with a PBS solution with $10\unit{\mu{}M}$ rhodamine.
    Bacterial cells and the bilayer membrane are dyed and visualized.
    (right) A snapshot of the device filled with PBS without rhodamine.
    The surface of the coverslip and the cells still exhibit fluorescence because of adsorption of rhodamine.
    {\bf b,c} Time evolution of the spatial profile of the fluorescent intensity with the background subtracted,
    when the medium is switched to the rhodamine solution ({\bf b}, see also Supplementary~Movie~5)
    and to the PBS without rhodamine ({\bf c}, see also Supplementary~Movie~6).
    The intensity averaged over 5 pixels ($0.6\unit{\mu{}m}$) from the substrate bottom is shown.
    The peaks seen in the profiles are due to bacterial cells, walls or dust.
    {\bf d,e} Time series of the spatially averaged fluorescence intensity with the background subtracted,
    when the medium is switched to the rhodamine solution ({\bf d}) and to the PBS without rhodamine ({\bf e}).
    During the experiment, medium flowed above the membrane at a constant speed of approximately $6\unit{mm~sec^{-1}}$.
    $t=0$ is the time at which the rhodamine solution entered the device (black dashed line).
    See Supplementary experimental methods for details of the experiment and analysis.
    }
\end{figure}

\begin{figure}[p]
    \centering
    \includegraphics[width=\textwidth]{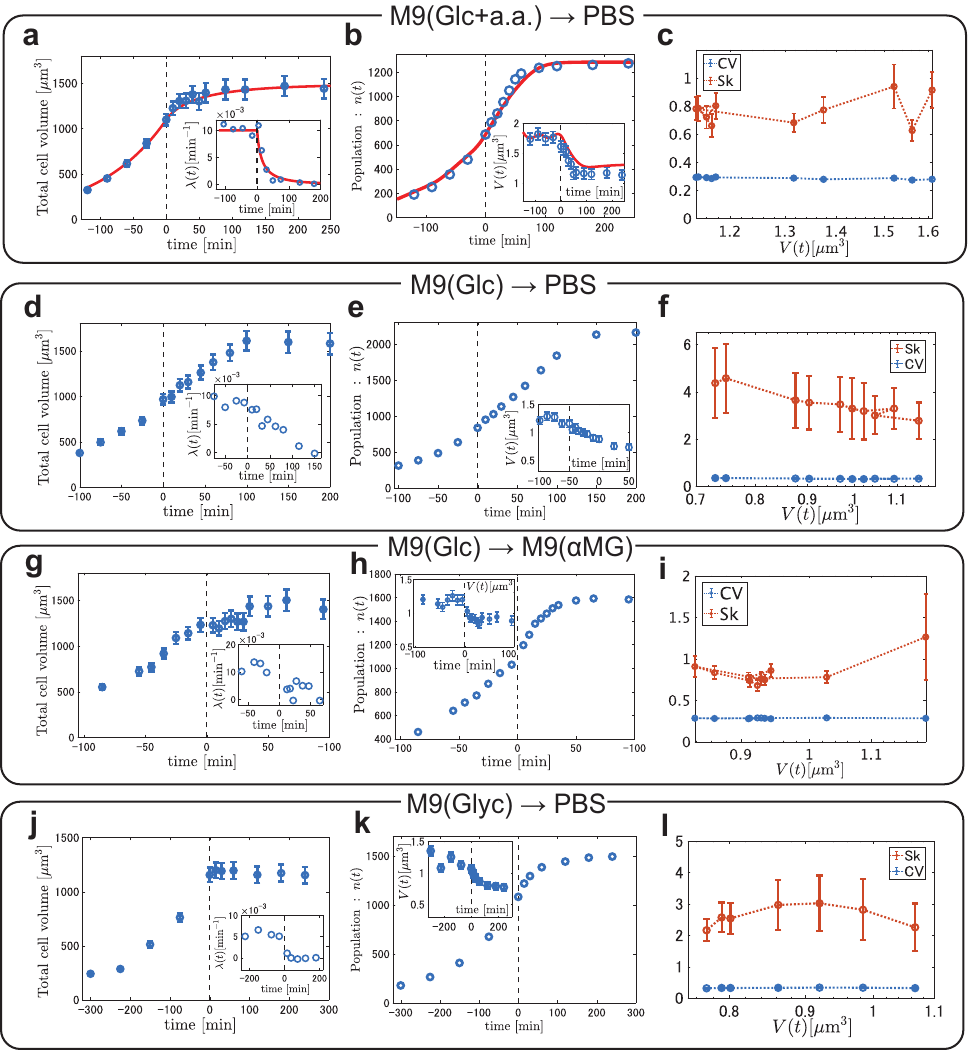}
    \caption{Supplemental results from the observations of reductive division.
    $t=0$ is the time at which the non-nutritious buffer entered the device (black dashed line).
    {\bf a-c} Results for the case of M9(Glc+a.a.) $\to$ PBS.
    The data were collected from 17 wells recorded in a single experiment.
    See also Supplementary~Movie~8.
    The total cell volume $V_\mathrm{tot}(t)$ ({\bf a}), the growth rate $\lambda(t)$ ({\bf a}, Inset), the number of the cells $n(t)$ ({\bf b}), the mean cell volume $V(t)$ ({\bf b}, Inset), from the experiment (blue) and simulation (red).
    The error bars indicate segmentation uncertainty in the image analysis (see Methods).
    CV and Sk [Eq.(3)] against $V(t)=\expct{v}$, where the error bars were estimated by the bootstrap method with 1000 realizations ({\bf c}).
    {\bf d-f} Results for the case of M9(Glc) $\to$ PBS.
    The data were collected from 26 wells recorded in a single experiment.
    See also Supplementary~Movie~9.
    {\bf g-i} Results for the case of M9(Glc) $\to$ M9($\alpha$MG).
    The data were collected from 26 wells recorded in a single experiment.
    See also Supplementary~Movie~10.
    {\bf j-l} Results for the case of M9(Glyc) $\to$ PBS.
    The data were collected from 19 wells recorded in a single experiment.
    See also Supplementary~Movie~11.
    Note that oscillatory behavior of Sk is seen in some of the results in this figure. 
    In fact, we observed similar behavior in our simulations presented in Supplementary~Fig.~S9, which disappeared if we kept a constant number of cells by eliminating one of the two daughter cells after each cell division. 
    This led us to believe that those oscillations are due to strong correlation of cell cycles between siblings.
    }
\end{figure}

\begin{figure}[p]
    \centering
    \includegraphics[width=0.7\textwidth]{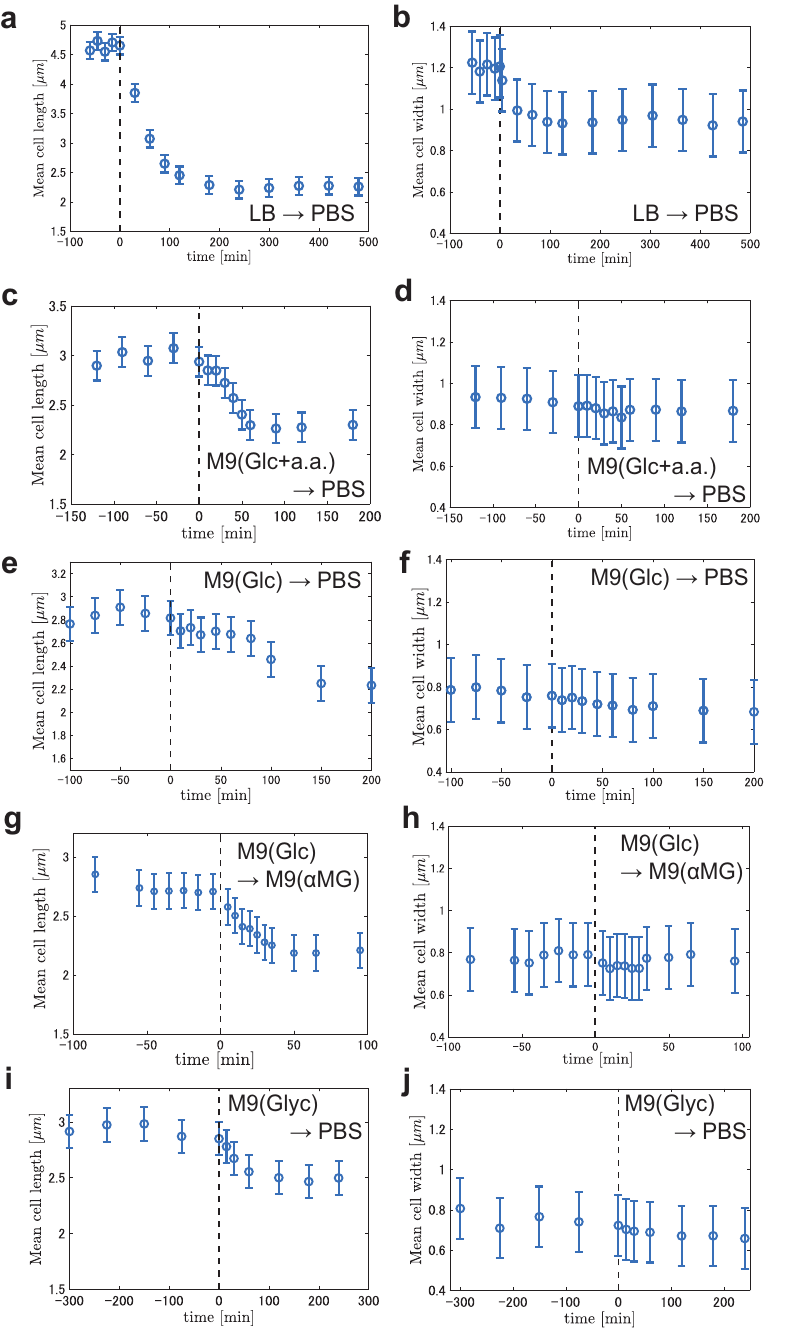}
    \caption{Time series of the mean cell length and the mean cell width during the starvation process.
    $t=0$ is the time at which the non-nutritious buffer entered the device (black dashed line).
    {\bf a,b} Results for the case of LB $\to$ PBS.
    The data were taken from 26 wells recorded in a single experiment.
    Time series of the mean cell length({\bf a}) and the mean cell width ({\bf b}).
    The error bars indicate segmentation uncertainty in the image analysis (see Methods).
    {\bf c,d} Results for the case of M9(Glc+a.a.) $\to$ PBS.
    The data were collected from 17 wells recorded in a single experiment.
    {\bf e,f} Results for the case of M9(Glc) $\to$ PBS.
    The data were collected from 26 wells recorded in a single experiment.
    {\bf g,h} Results for the case of M9(Glc) $\to$ M9($\alpha$MG).
    The data were collected from 26 wells recorded in a single experiment.
    {\bf i,j} Results for the case of M9(Glyc) $\to$ PBS.
    The data were collected from 19 wells recorded in a single experiment.
    }
\end{figure}

\begin{figure}[h]
    \centering
    \includegraphics[width=\textwidth]{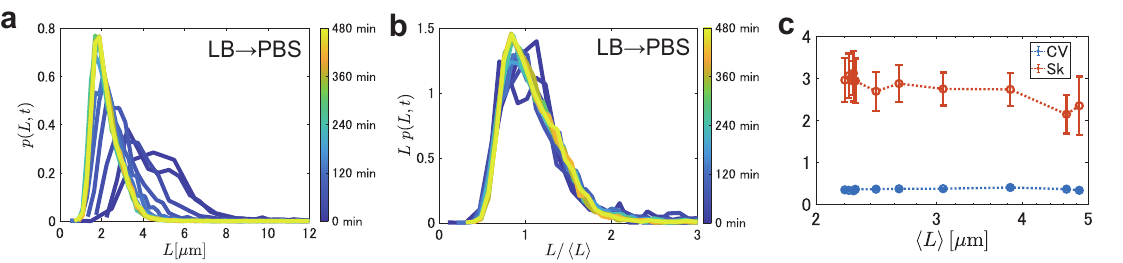}
    \caption{Results on the cell length $L$.
    {\bf a}
    Time evolution of the cell length distributions during starvation
    in the case of LB $\to$ PBS at
     $t=0,5,30,60,90,120,180,240,300,360,420,480\unit{min}$ from right to left.
    The data were taken from 26 wells recorded in a single experiment.
    {\bf b}
    Rescaling of the data in ({\bf d}) against the mean cell length $\expct{L}$.
    {\bf c}
    CV and Sk [Eq.~(3)].
    }
\end{figure}

\newpage

\begin{figure}[h]
    \centering
    \includegraphics[width=\textwidth]{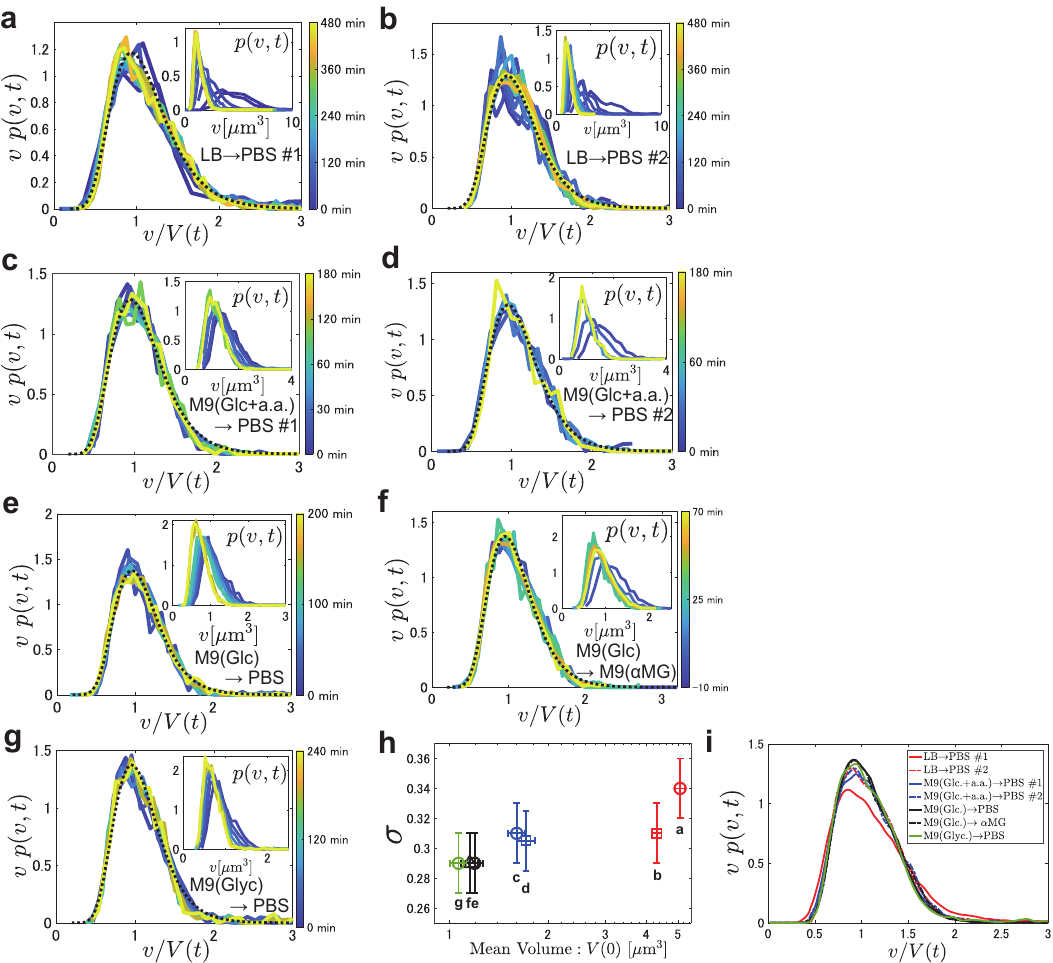}
    \caption{Comparison of the distributions obtained from different experimental conditions.
    {\bf a} The cell size distributions in the case of LB $\to$ PBS at
     $t=0,5,30,60,90,120,180,240,300,360,420,480\unit{min}$.
    The dashed line represents the fitted log-normal distribution ($\sigma=0.34(2)$).
    The data were taken from 26 wells recorded in a single experiment.
    {\bf b} The distributions obtained from a biological replicate for the case LB $\to$ PBS at
     $t=0,5,30,60,90,120,180,240,300,360,420,480\unit{min}$ with the log-normal distribution ($\sigma=0.31(2)$).
    The data were taken from 7 wells recorded in a single experiment.
    {\bf c} The distributions in the case of M9(Glc+a.a.) $\to$ PBS at
     $t=0,10,20,30,40,50,60,90,120,180\unit{min}$ with the log-normal distribution ($\sigma=0.31(2)$).
    The data were taken from 17 wells recorded in a single experiment.
    {\bf d} The distributions obtained from a biological replicate for the case M9(Glc+a.a.) $\to$ PBS at
    $t=0,20,40,60,180\unit{min}$ with the log-normal distribution ($\sigma=0.305(20)$).
    The data were taken from 21 wells recorded in a single experiment.
    {\bf e} The distributions in the case of M9(Glc) $\to$ PBS at
    $t=0,10,20,30,40,50,60,80,100,150,200\unit{min}$ with the log-normal distribution ($\sigma=0.29(2)$).
    The data were taken from 26 wells recorded in a single experiment.
    {\bf f} The distributions in the case of M9(Glc) $\to$ M9($\alpha$MG) at
    $t=-5,5,10,20,25,30,35,50,65\unit{min}$ with the log-normal distribution ($\sigma=0.29(2)$).
    The data were taken from 26 wells recorded in a single experiment.
    {\bf g} The distributions in the case of M9(Glyc) $\to$ PBS at
    $t=0,15,30,60,120,180,240 \unit{min}$ with the log-normal distribution ($\sigma=0.29(2)$).
    The data were taken from 19 wells recorded in a single experiment.
    {\bf h} Relation between $\sigma$ of the fitted log-normal distribution and the mean cell volume before starvation for the different conditions studied in this work.
    The horizontal axis is shown in the log scale.
    {\bf i} Time-averaged $vp(v,t)$ for each condition.
    }
\end{figure}

\newpage

\begin{figure}[p]
    \centering
    \includegraphics[width=0.95\textwidth]{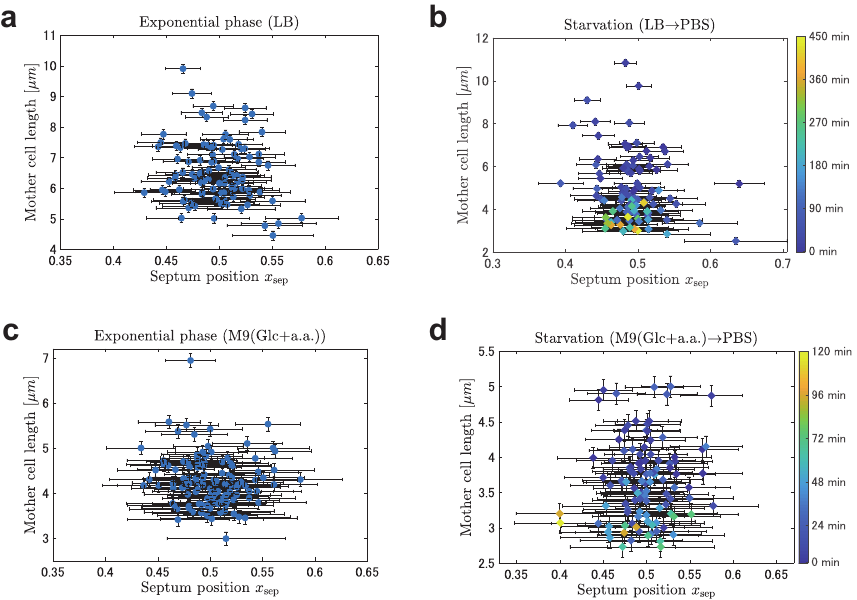}
    \caption{Experimentally measured cell-to-cell fluctuations of the septum position.
    Scatter plots of the mother cell length against the septum position $x^\mathrm{sep}$ are shown for four different cases.
    The data were taken from more than 100 cell division events chosen randomly for each case.
    The error bars indicate segmentation error in the image analysis.
    There is no visible correlation between the mother cell size
     and the standard deviation of the septum position.
    {\bf a} Scatter plot for the exponential growth phase in LB broth. 
    The standard deviation of the septum position is $x^\mathrm{sep}_\mathrm{std}=0.028$.
    {\bf b} Scatter plot during the starvation process for LB $\to$ PBS.
    The color represents the time passed since PBS entered the device.
    The standard deviation of the septum position is $x^\mathrm{sep}_\mathrm{std}=0.036$.
    {\bf c} Scatter plot for the exponential growth phase in M9(Glc+a.a) medium.
    The standard deviation of the septum position is $x^\mathrm{sep}_\mathrm{std}=0.029$.
    {\bf d} Scatter plot during the starvation process for M9(Glc+a.a.) $\to$ PBS.
    The standard deviation of the septum position is $x^\mathrm{sep}_\mathrm{std}=0.032$.
    }
\end{figure}

\begin{figure}[p]
    \centering
    \includegraphics[width=\textwidth]{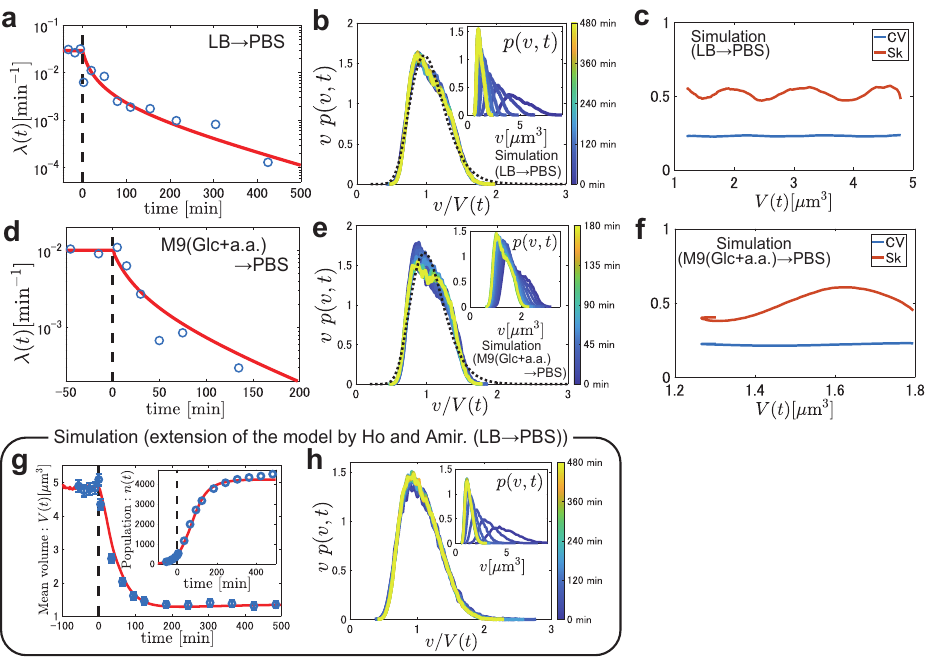}
    \caption{Supplementary figures on the simulation results. The results in panels {\bf a} to {\bf f} were obtained with the model based on that by Witz et al.
    {\bf a} The growth rate obtained by the simulations (red curves) and the experiments (blue symbols) for the case of LB $\to$ PBS.
    The same data are shown in Fig.~2b in linear scale.
    {\bf b} Overlapping of the rescaled cell size distributions during starvation
    obtained by the simulations for LB $\to$ PBS (the same figure as Fig.~4d).
    The dashed line represents the fitted log-normal distribution ($\sigma=0.25(2)$).
    (Inset) The non-rescaled cell size distributions
    at $t=0,5,30,60,90,120,180,240,300,360,420,480\unit{min}$ from right to left.
    {\bf c} CV and Sk [Eq.~(3)] against $V(t)=\expct{v}$
    obtained by the simulations for LB $\to$ PBS.
    The error bars estimated by the bootstrap method with 1000 realizations were smaller than the symbol size.
    The oscillation of Sk disappeared if we kept a constant number of cells, by eliminating one of the two daughter cells after each cell division.
    This led us to believe that those oscillations are due to strong correlation of cell cycles between siblings.
    The important observation here is that the values of Sk do not increase or decrease monotonically even in the presence of such an oscillation.
    {\bf d} The growth rate obtained by the simulations (red curves) and the experiments (blue symbols) for the case of M9(Glc+a.a.) $\to$ PBS.
    The same data are shown in Supplementary~Fig.~2b in linear scale.
    {\bf e} Overlapping of the rescaled cell size distributions during starvation
    obtained by the simulations for M9(Glc+a.a.) $\to$ PBS.
    (Inset) The non-rescaled cell size distributions
    at $t=0,10,20,30,40,50,60,90,120,180\unit{min}$ from right to left.
    {\bf f} CV and Sk [Eq.~(3)] against $V(t)=\expct{v}$
    obtained by the simulations for M9(Glc+a.a.) $\to$ PBS.
    The error bars estimated by the bootstrap method with 1000 realizations were smaller than the symbol size.
    Similarly to the result in panel {\bf c}, we believe that the oscillation of Sk is due to correlation between siblings.
    {\bf g} Comparison of the experiments (LB $\to$ PBS) and simulations based on the model by Ho and Amir (LB $\to$ PBS).
    {\bf h} Overlapping of the rescaled cell size distributions during starvation
    obtained by the simulations based on the model by Ho and Amir (LB $\to$ PBS).
    (Inset) The non-rescaled cell size distributions
    at $t=0,5,30,60,90,120,180,240,300,360,420,480\unit{min}$ from right to left.
    }
\end{figure}

\begin{figure}[h]
    \centering
    \includegraphics[width=\hsize]{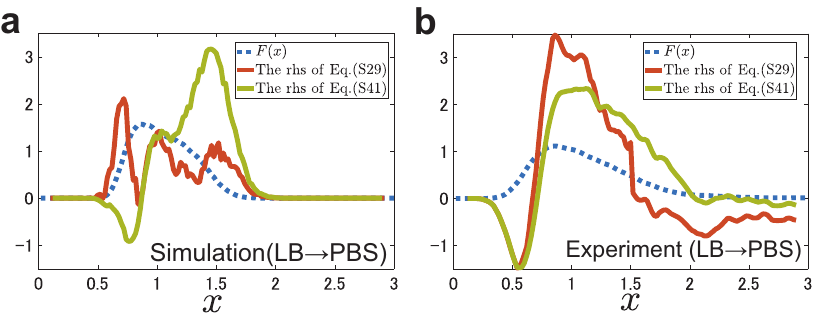}
    \caption{
    Test of the self-consistent equations derived in Supplementary Theory. 
    The functional form of $F(x)$, obtained numerically ({\bf a}) or experimentally ({\bf b}) for the case LB $\to$ PBS, is compared with the right-hand side (rhs) of the self-consistent equations derived in Supplementary Theory, Eqs.~(\ref{eq:Fconsistent2}) and (\ref{eq:asFconsistent}).
    The effect of septum fluctuations is neglected in \eqref{eq:Fconsistent2} and considered in \eqref{eq:asFconsistent}.
    $F(x)$ is obtained as the time average of the instantaneous data.
    }
\end{figure}

\begin{figure}[h]
    \centering
    \includegraphics[width=0.9\textwidth]{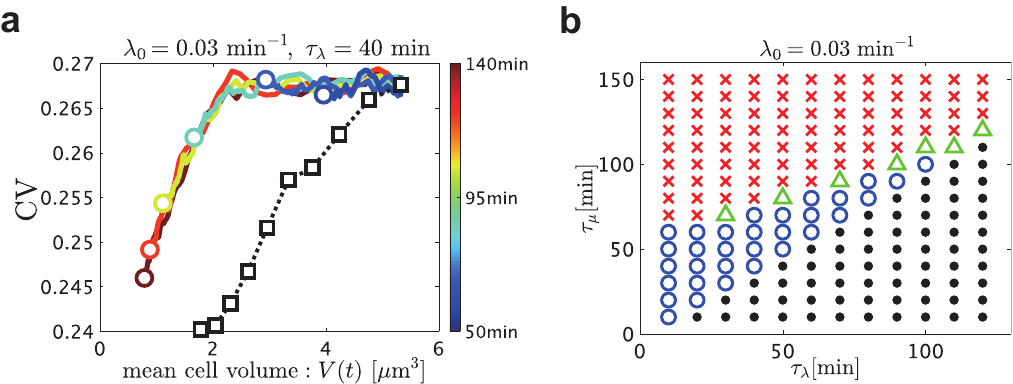}
    \caption{
    Results on the range of validity of the scale invariance based on the model by Ho and Amir.
    The initial growth rate is fixed at $\lambda(0)=\lambda_0=0.03\unit{min^{-1}}$ unless otherwise stipulated.
    {\bf a} Trajectories in the $\expct{v}$-CV space for different $\tau_\mu$ (from $50$ to $140\unit{min^{-1}}$), with $\tau_\lambda=40\unit{min}$ fixed.
    The endpoint of each trajectory is indicated by a colored open circle.
    The black squares represent the states in steady growth conditions with the growth rate $\lambda_0$ ranging from $0.0075$ to $0.03\unit{min^{-1}}$.
    {\bf b} Phase diagram. $\times$: the scale invariance breaks down.
     Blue $\circ$: the scale invariance holds.
     Green $\triangle$: near the boundary.
     Black dots: the scale invariance holds but the mean volume increases.
    }
\end{figure}

\clearpage

\section*{Supplementary References}
\bibliography{EMPSpaperSI}